\documentclass[seceqn,MSNbibl,number,citesort,dvips]{arxbj}
\usepackage{upgreek}
\usepackage{graphicx}

\aid{0}
\volume{20}
\issue{4}
\pubyear{2014}
\firstpage{1879}
\lastpage{1929}
\doi{10.3150/13-BEJ545} 

\makeatletter

\newcommand{\rright}{\right}
\newcommand{\lleft}{\left}
\newcommand{\rrVert}{\Vert}
\newcommand{\rrvert}{\vert}
\newcommand{\llVert}{\Vert}
\newcommand{\llvert}{\vert}

\newcommand{\Prob}{\mathsf{P}}
\newcommand{\Real}{\mathbb{R}}
\newcommand{\mB}{\mathcal{B}}
\newcommand{\mP}{\mathcal{P}}
\newcommand{\mK}{\mathcal{K}}
\newcommand{\mL}{\mathcal{L}}
\newcommand{\mS}{\mathcal{S}}
\newcommand{\mI}{\mathcal{I}}
\newcommand{\mH}{\mathcal{H}}
\newcommand{\mG}{\mathcal{G}}
\newcommand{\mF}{\mathcal{F}}
\newcommand{\proj}{\star}
\newcommand{\compl}{\mathsf{c}}

\newtheorem{Teorema}{Theorem}[section]
\newproclaim{Definicion}{Definition}[section]
\newtheorem{Lema}{Lemma}[section]
\newtheorem{Proposicion}{Proposition}[section]
\newtheorem{Corolario}{Corollary}[section]
\newtheorem{Assumption}{\em A.}
\newremark{Nota}{Remark}[section]
\newremark{Ejemplo}{Example}[section]

\newcommand{\eqref}[1]{(\ref{#1})}
\renewcommand{\emptyset}{\varnothing}

\def\sfrac#1#2{#1/#2}
\def\vfrac#1#2{(#1)/#2}
\def\afrac#1#2{#1/(#2)}

\makeatother

\begin{document}
\begin{frontmatter}

\title{Particle-kernel estimation of the filter density in state-space models}
\runtitle{Particle-kernel density estimation}

\begin{aug}
\author[1]{\inits{D.}\fnms{Dan} \snm{Crisan}\thanksref{1}\ead[label=e1]{d.crisan@imperial.ac.uk}} \and
\author[2]{\inits{J.}\fnms{Joaqu\'{\i}n} \snm{M\'{\i}guez}\corref{}\thanksref{2}\ead[label=e2]{joaquin.miguez@uc3m.es}}
\address[1]{Department of Mathematics, Imperial College London,
Huxley Building, 180 Queen's Gate, London SW7 2BZ, UK. \printead{e1}}
\address[2]{Department of Signal Theory \& Communications, Universidad
Carlos III de Madrid, Avenida de la Universidad 30, 28911 Legan\'es
(Madrid), Spain.
\printead{e2}}
\end{aug}

\received{\smonth{12} \syear{2012}}
\revised{\smonth{7} \syear{2013}}

%
\begin{abstract}
Sequential Monte Carlo (SMC) methods, also known as particle filters,
are simulation-based recursive algorithms for the approximation of the
a posteriori probability measures generated by state-space
dynamical models. At any given time $t$, a SMC method produces a set of
samples over the state space of the system of interest (often termed
``particles'') that is used to build a discrete and random
approximation of the posterior probability distribution of the state
variables, conditional on a sequence of available observations. One
potential application of the methodology is the estimation of the
densities associated to the sequence of a posteriori
distributions. While practitioners have rather freely applied such
density approximations in the past, the issue has received less
attention from a theoretical perspective. In this paper, we address the
problem of constructing kernel-based estimates of the posterior
probability density function and its derivatives, and obtain asymptotic
convergence results for the estimation errors. In particular, we find
convergence rates for the approximation errors that hold uniformly on
the state space and guarantee that the error vanishes almost surely as
the number of particles in the filter grows. Based on this uniform
convergence result, we first show how to build continuous measures that
converge almost surely (with known rate) toward the posterior measure
and then address a few applications. The latter include maximum a posteriori
 estimation of the system state using the approximate
derivatives of the posterior density and the approximation of
functionals of it, for example, Shannon's entropy.
\end{abstract}

%

\end{frontmatter}

\section{Introduction}\label{intro}

\subsection{Background} \label{ssIntro-Background}

Consider two random sequences, $\{X_t\}_{t \ge0}$ and $\{Y_t\}_{t\ge
1}$, possibly multidimensional, where $X_t$ represents the unobserved
state of a system of interest and $Y_t$ is a related observation. Very
often, the dependence between the two sequences is given by a Markov
state-space model and the posterior probability measure that
characterizes the random variable $X_t$ conditional on the observations
$\{Y_s, 1\le s\le t\}$ is usually termed the ``filtering measure'',
denoted as $\pi_t$ in the sequel. If the model is linear and Gaussian,
$\pi_t$ is also Gaussian and can be computed exactly using a set of
recursive equations known as the Kalman filter \cite{Kalman60}.
However, if $X_t$ takes values in a continuous space and the model is
nonlinear or non-Gaussian, the exact filter is intractable and
numerical approximation techniques are necessary. The class of
sequential Monte Carlo (SMC) methods, also known as particle filters
\cite{Gordon93,Kitagawa96,Liu98,Doucet00,Doucet01b}, has become a very
popular tool for this purpose. Particle filters generate discrete
random measures (constructed from random samples in the state space)
that can be naturally used to approximate integrals with respect to
(w.r.t.) the filtering measure.

The asymptotic convergence of SMC algorithms has been well studied
during the past two decades. The first formal results appeared in \cite
{DelMoral96a,DelMoral96b}, while the analysis in \cite{Crisan99}
already took into account the branching (resampling) step indispensable
in most practical applications. Currently, there is a broad knowledge
about the convergence of particle filters in some of the forms commonly
used in practical applications; see \cite
{DelMoral00,Crisan01,DelMoral04,LeGland04,Bain08,Heine08} and the
references therein. Most of these results are aimed to show that
integrals of real functions w.r.t. $\pi_t$ can be accurately
approximated by weighted sums when the particle filter is run with a
sufficiently large number of random samples (commonly referred to as
particles). More recently, other types of convergence have been
investigated. For instance, the convergence of particle approximations
of maximum a posteriori (MAP) estimates of sequences has also been
proved. Convergence in probability can be shown using random
genealogical trees (see \cite{Najim06} and \cite{DelMoral04}) while
almost sure convergence can also be guaranteed by extending the
analysis in \cite{Crisan00} (see~\cite{Miguez13b}).\looseness=1

In most cases of interest, the filtering measure has a density, denoted
$p_t$, w.r.t. a dominating measure (usually Lebesgue's) and
practitioners have freely used various estimators of this function.
Less attention has been devoted to this problem from a theoretical
perspective, though. Note that the samples generated by the particle
filter are not drawn directly from $p_t$: they can only be considered
as \emph{approximate samples}, in the sense that they can be used to
estimate the value of integrals w.r.t. the measure $\pi_t$. As a
consequence, the convergence of a kernel density estimate of $p_t$
built from the output of a particle filter cannot be justified directly
using the classical theory of kernel density estimation, which is
concerned with samples drawn directly from the distribution of interest
(see, e.g., \cite{Devroye85,Silverman86,Wand95,Simonoff96}).

The estimation of $p_t$ is of interest by itself, since it naturally
enables the computation of confidence regions, as well as MAP and
maximum likelihood estimators, but also because it leads to the
approximation of $\pi_t$ by a continuous (instead of discrete) random
measure. The convergence of continuous approximations of the filtering
measure in total variation distance has been investigated in the
context of regularized particle filters \cite{LeGland04} as well as
for accept/reject and auxiliary particle filters \cite{Kunsch05}.
\subsection{Contributions}

In this paper, we analyze the approximation of $p_t$ constructed as the
sum of properly scaled kernel functions located at the particle
positions. Kernel methods \cite{Silverman86,Wand95} are the most
widely used techniques for the nonparametric estimation of probability
density functions (pdfs) and, therefore, it seems natural to analyze
their convergence when applied to the approximate samples generated by
particle filters.

The pdf estimators we analyze are based on generic kernel functions
which are only required to satisfy mild standard conditions
(essentially the same as in classical density estimation theory \cite
{Silverman86}). We describe how to build approximations in
arbitrary-dimensional spaces $\Real^d$, $d \ge1$, and then analyze
their convergence as the number of particles is increased and the
bandwidth of the kernels is decreased. In particular, we obtain
point-wise convergence rates for the absolute approximation errors,
both of $p_t$ and its derivatives\footnote{Let us note here that the
approximation of derivatives of the filter has received attention
recently, related to problems of parameter estimation in state-space
systems \cite{Dean11,DelMoral11}. In the latter context, the filtering
pdf is made to depend explicitly on a parameter vector $\theta=(\theta
_1, \ldots, \theta_d)$, and the interest is in the computation of the
partial derivatives $\partial p_t / \partial\theta_i$ in order to
implement, for example, maximum likelihood estimation algorithms \cite
{DelMoral11}. In this paper, however, we consider derivatives with
respect to the state variables in $X_t=(X_{1,t}, \ldots, X_{d_x,t})$,
that is, $\partial p_t / \partial x_{i,t}$.} (provided they exist). The
latter results can be extended to deduce uniform (instead of
point-wise) convergence rates, again both for $p_t$ and its
derivatives. Specifically, we provide explicit bounds for the supremum
of the approximation error and prove that it converges almost surely
(a.s.) toward 0 as the number of particles is increased. Our analysis
is different from the standard methods in kernel density estimation.
The latter address the bias and variance of the estimators using
approximations based on Taylor series (see, e.g., \cite{Silverman86}, Chapter~4 or \cite{Wand95}, Chapter~4) or Edgeworth expansions
\cite{Hall01}, which enable the asymptotic approximation of the mean
integrated square error (MISE) of the density estimate and yield
expressions involving the number of samples and the kernel bandwidth.
We directly obtain convergence rates for various estimation errors (not
only the MISE), given in terms of a single index that links the number
of samples and the kernel bandwidth. This link is briefly discussed in
Section~\ref{ssSummary}.

The uniform (on the support of $p_t$) convergence result can be
exploited in a number of ways. For instance, if we let $p_t^N$ be the
approximation of $p_t$ with $N$ particles, then we can obtain a
continuous approximation of the filtering measure $\pi_t(\mathrm{d}x)$ as
$\breve\pi_t^N(\mathrm{d}x) = p_t^N(x)\,\mathrm{d}x$, prove that $\breve\pi_t^N$
converges to $\pi_t$ a.s. in total variation distance (as $N
\rightarrow
\infty$) and provide explicit convergence rates. A similar kind of
analysis also leads to the calculation of convergence rates for the
MISE of the particle-kernel density estimator $p_t^N$. Additionally, we
prove that the (random) integrated square error (ISE) of a truncated
version of $p_t^N$ converges to 0 a.s. and provide convergence rates. A
comparison of these results with the standard asymptotic approximation
of the MISE for kernel estimators built from i.i.d. samples is
presented at the end of Section~\ref{ssISE}.

The convergence in total variation distance of a continuous
approximation of the filtering measure $\pi_t$ was also addressed in
\cite{LeGland04} and \cite{Kunsch05}. Compared to these earlier
contributions, our analysis guarantees the almost sure convergence of
the (random) total variation distance toward 0, with explicit rates,
rather than the convergence of its expected value (as in \cite
{LeGland04}) or its convergence in probability (as in \cite
{Kunsch05}). Also, our assumptions on the Markov kernel of the state
process $\{ X_t \}_{t\ge0}$ and the conditional densities of $\{
Y_t|X_t \}_{t \ge1}$ are relatively mild and simple to check. In
particular, our results also hold for light-tailed Markov kernels
(e.g., Gaussian), unlike Theorems 2 and 3 in \cite{Kunsch05}.

The last part of the paper is devoted to some applications of the
density approximation method and the uniform convergence result. We
first consider the problem of MAP estimation. We refer here to the
maximization of the filtering density, a problem different from that of
MAP estimation in the path space addressed, for example, in \cite
{Godsill01,Najim06,Miguez13b}. We first prove that the maxima of the
approximation of the filtering density actually converge,
asymptotically, to the maxima of the true function $p_t$ and then show
some simulation results that illustrate the use of gradient algorithms
on the estimated density function. 

The second application we describe is the approximation of functionals
of $p_t$. We provide first a generic result that guarantees the almost
sure convergence of such approximations for bounded and Lipschitz
continuous functionals. Then, we address the problem of approximating
Shannon entropies \cite{Cover91}, which is of practical interest in
various machine learning and signal processing problems. The $\log$
function is neither bounded nor Lipschitz continuous and, therefore,
the latter generic result does not apply to the computation of
entropies. We specifically address this problem resorting to a new
result on the convergence of the particle approximations of integrals
of the form $\int f(x) \pi_t(\mathrm{d}x)$ when the test function $f$ is
possibly unbounded. Let us remark that a large majority of the results
in the literature \cite
{DelMoral00,Crisan01,DelMoral04,LeGland04,Bain08} refer exclusively to
the approximation of integrals of bounded functions. Only recently, the
convergence of approximate integrals of unbounded test functions has
been proved \cite{Hu08}, albeit for a modified particle filter and
assuming that the product of the test and the likelihood functions
\emph{is} bounded. Here, we prove the almost sure convergence of the
approximations of integrals of unbounded functions for the standard
particle filter, placing only integrability assumptions on the test
function. From this result, we deduce the almost sure convergence
toward 0 of the errors in the approximation of Shannon entropies for
densities with a compact support. A numerical illustration is given.

\subsection{Organization of the paper}

The rest of the paper is organized as follows. Section~\ref{sBackground} contains background material, including a summary of
notation, a description of Markov state space models and the standard
particle (\emph{bootstrap}) filter. A new lemma that establishes the
convergence of the particle approximation of posterior expectations of
unbounded test functions is also introduced in Section~\ref{sBackground}. The construction of particle-kernel approximations of
the filtering density and its derivatives is described in Section~\ref{sApproximation}, where we also review some basics of kernel density
estimation and the most relevant results in \cite{LeGland04} and \cite
{Kunsch05} for density estimation with particle filters. Our formal
results on the convergence of the particle-kernel density estimators
and the smooth approximation of the filtering measure are introduced in
Section~\ref{sConvergence}. This includes the point-wise and uniform
approximations of $p_t(x)$, the convergence in total variation distance
of the smooth measures $\breve\pi_t^N$ and convergence rates for the
mean integrated square error and the (random) integrated square error
of $p_t^N$ and its truncated version, respectively. In Section~\ref{sApplications}, we discuss applications of the particle-kernel
estimator of $p_t$ and its derivatives. In particular, we consider the
problem of (marginal) MAP estimation of the state variables and the
approximation of functionals of the filtering density, including
Shannon's entropy. Finally, brief conclusions are presented in Section~\ref{sConclusions}.

\section{Particle filtering} \label{sBackground}

\subsection{Notation}
We first introduce some common notations to be used through the paper,
broadly classified\vspace*{-2pt} by topics. Below, $\Real$ denotes the real line,
while for an integer $d\ge1$, $\Real^d=\overbrace{\Real\times
\cdots\times\Real}^{d\ \mathrm{times}}$

\begin{itemize}
\item Measures and integrals.
\begin{itemize}
\item$\mB(\Real^d)$ is the $\sigma$-algebra of Borel subsets of
$\Real^d$.
\item$\mP(\Real^d)$ is the set of probability measures over $\mB
(\Real^d)$.
\item$(f,\mu) \triangleq\int f(x) \mu(\mathrm{d}x)$ is the integral of a real
function $f\dvtx \Real^d \rightarrow\Real$ w.r.t. a measure $\mu\in\mP
(\Real^d)$.
\item Take a measure $\mu\in\mP(\Real^d)$, a Borel set $A \in\mB
(\Real^d)$ and a real function $f \dvtx  \Real^d \rightarrow\Real^+$. The
projective product $f \proj\mu$ is a measure, absolutely continuous
w.r.t. $\mu$ and proportional to $f$, constructed as
%
\begin{equation}
(f \proj\mu) (A) = \frac{
\int_A f(x) \mu(\mathrm{d}x)
}{
(f,\mu)
}.
\end{equation}
\end{itemize}

\item Functions.
\begin{itemize}
\item The supremum norm of a real function $f\dvtx \Real^d \rightarrow
\Real$ is
denoted as $\| f \|_\infty= \sup_{x\in\Real^d} | f(x) |$.
\item$B(\Real^d)$ is the set of bounded real functions over $\Real
^d$, that is, $f \in B(\Real^d)$ if, and only if, $\| f \|_\infty<
\infty$.
\item$C_b(\Real^d)$ is the set of continuous and bounded real
functions over $\Real^d$.
\end{itemize}

\item Sets.
\begin{itemize}
\item Given a probability measure $\mu\in\mP(\Real^d)$, a Borel set
$A \in\mB(\Real^d)$ and the indicator function
\[
I_A(x) = \lleft\{ %
\begin{array} {l@{\qquad}l} 1, &\mbox{if } x
\in A,
\\
0, &\mbox{otherwise}, \end{array} %
\rright.
\]
$\mu(A) = (I_A,\mu) = \int_A \mu(\mathrm{d}x)$ is the probability of $A$.
\item The Lebesgue measure of a set $A \in\mB(\Real^d)$ is denoted
$\mL(A)$.
\item For a set $A \in\Real^d$, $A^\compl= \Real^d \backslash A$
denotes its complement.
\end{itemize}

\item Sequences, vectors and random variables (r.v.).
\begin{itemize}
\item We use a subscript notation for sequences, $x_{t_1:t_2}
\triangleq\{
x_{t_1}, \ldots, x_{t_2} \}$.
\item For an element $x=(x_1,\ldots,x_d) \in\Real^d$ of an Euclidean
space, its norm is denoted as $\| x \| = \sqrt{ x_1^2+\cdots+x_d^2 }$.
\item The $L_p$ norm of a real r.v. $Z$, with $p \ge1$, is written as
$\| Z \|_p \triangleq E[ |Z|^p ]^{1/p}$, where $E[\cdot]$ denotes expectation.
\end{itemize}
\end{itemize}

\subsection{Filtering in discrete-time, state-space Markov models}

Consider two random sequences, $\{ X_t \}_{t \ge0}$ and $\{ Y_t \}_{t
\ge1}$, taking values in $\Real^{d_x}$ and $\Real^{d_y}$,
respectively. The common probability measure for the pair $ ( \{
X_t\}_{t\ge0}, \{Y_t\}_{t\ge1}  )$ is denoted $\Prob$, and we
assume that it is absolutely continuous w.r.t. the Lebesgue measure. We
refer to the first sequence as the state process and we assume that it
is an inhomogeneous Markov chain governed by an initial probability
measure $\tau_0 \in\mP(\Real^{d_x})$ and a sequence of transition
kernels $\tau_t \dvtx  \mB(\Real^{d_x}) \times\Real^{d_x} \rightarrow[0,1]$,
defined as
%
\begin{equation}
\tau_t(A|x_{t-1}) \triangleq\Prob \{ X_t
\in A | X_{t-1}=x_{t-1} \},
\end{equation}
where $A \in\mB(\Real^{d_x})$ is a Borel set. The sequence $\{ Y_t \}
_{t \ge1}$ is termed the observation process. Each r.v. $Y_t$ is
assumed to be conditionally independent of other observations given the
state $X_t$, meaning that
%
\begin{equation}
\Prob \bigl\{ Y_t \in A | X_{0:t} = x_{0:t},
\{ Y_k = y_k \}_{k \ne t} \bigr\} = \Prob \{
Y_t \in A | X_t = x_t \}
\end{equation}
for any $A \in\mB(\Real^{d_y})$. Additionally, we assume that every
probability measure $\gamma_t \in\mP(\Real^{d_y})$ in the sequence
%
\begin{equation}
\gamma_t(A|x_t) \triangleq\Prob \{ Y_t
\in A | X_t=x_t \} ,\qquad  A \in\mB\bigl(
\Real^{d_x}\bigr), t=1,2,\ldots,
\end{equation}
has a positive density w.r.t. the Lebesgue measure. We denote this
density as $g_t(y|x)$, hence we write $\gamma_t(A|x_t) = \int_A g_t(y|x_t)\,\mathrm{d}y$.

The filtering problem consists in the computation of the posterior
probability measure of the state $X_t$ given a sequence of observations
up to time $t$. Specifically, for a fixed observation record $Y_{1:T} =
y_{1:T}$, $T < \infty$, we seek the measures $\pi_t \in\mP(\Real
^{d_x})$ given by
%
\begin{equation}
\pi_t(A) \triangleq\Prob \{ X_t \in A |
Y_{1:t}=y_{1:t} \}, \qquad t=0,1,\ldots, T,
\end{equation}
where $A \in\mB(\Real^{d_x})$. For many practical problems, the
interest actually lies in the computation of integrals of the form
$(f,\pi_t)$. Note that, for $t=0$, we recover the prior measure, that
is, $\pi_0=\tau_0$.

\subsection{Particle filters} \label{ssPFs}

The sequence of measures $\{ \pi_t \}_{t \ge1}$ can be numerically
approximated using particle filtering. Particle filters are numerical
methods based on the recursive decomposition \cite{Bain08}
%
\begin{equation}
\pi_t = g_t^{y_{t}} \proj\tau_t
\pi_{t-1},
\end{equation}
where $g_t^{y_{t}}\dvtx \Real^{d_x}\rightarrow\Real^+$ is the function
defined as
$g_t^{y_{t}}(x) \triangleq g_t(y_t|x)$, $\proj$ denotes the projective
product and $\xi_t \triangleq\tau_t \pi_{t-1}$ is the (predictive)
probability measure
%
\begin{equation}
\xi_t(A) = \tau_t \pi_{t-1}(A) = \int
\tau_t(A|x) \pi_{t-1}(\mathrm{d}x), \qquad A \in\mB\bigl(
\Real^{d_x}\bigr). \label{eqSuccessive}
\end{equation}
Specifically, the simplest particle filter, often called `standard
particle filter' or `bootstrap filter' \cite{Gordon93} (see also \cite
{Doucet01}), can be described as follows.
\begin{enumerate}[2.]
\item[1.] \textit{Initialization.} At time $t=0$, draw $N$ i.i.d. samples
from the initial distribution $\tau_0 \equiv\pi_0$, denoted
$x_0^{(n)}$, $n=1,\ldots,N$.
\item[2.] \textit{Recursive step.} Let $\Omega_{t-1}^N=\{ x_{t-1}^{(n)} \}
_{n=1,\ldots,N}$ be the particles (samples) generated at time $t-1$. At
time $t$, proceed with the two steps below.
\begin{longlist}[(b)]
\item[(a)] For $n=1,\ldots,N$, draw a sample $\bar x_t^{(n)}$ from the
probability distribution $\tau_t(\cdot|x_{t-1}^{(n)})$ and compute
the normalized weight
%
\begin{equation}
w_t^{(n)} = \frac{
g_t^{y_{t}}(\bar x_t^{(n)})
}{
\sum_{k=1}^N g_t^{y_{t}}(\bar x_t^{(k)})
}.
\end{equation}

\item[(b)] For $n=1,\ldots,N$, let $x_t^{(n)}=\bar x_t^{(k)}$ with probability
$w_t^{(k)}$, $k \in\{1,\ldots,N\}$.
\end{longlist}
\end{enumerate}
Step 2(b) is referred to as resampling or selection. In the form
stated here, it reduces to the so-called multinomial resampling
algorithm \cite{Doucet00,Douc05} but the convergence of the algorithm
can be easily proved for various other schemes (see, e.g., the
treatment of the resampling\vspace*{-1pt} step in \cite{Crisan01}). Using the
samples in $\Omega_t^N=\{ x_t^{(n)} \}_{n=1,\ldots,N}$, we construct
a random approximation of $\pi_t$, namely
%
\begin{equation}
\pi_t^N(\mathrm{d}x_t) = \frac{1}{N} \sum
_{n=1}^N \delta_{x_t^{(n)}}(\mathrm{d}x_t),
\end{equation}
where $\delta_{x_t^{(n)}}$ is the delta unit-measure located at
$X_t=x_t^{(n)}$. For any integrable\vspace*{-1pt} function $f$ in the state space, it
is straightforward to approximate the integral $(f,\pi_t)$ as
%
\begin{equation}\label{eqApprox_of_f}
(f,\pi_t) \approx\bigl(f,\pi_t^N\bigr) =
\frac{1}{N} \sum_{n=1}^N f
\bigl(x_t^{(n)}\bigr).
\end{equation}

The convergence of particle filters has been analyzed in a number of
different ways. Most of the results to be described in this paper rely
only on the convergence of the $L_p$ norm of the approximation errors
$(f,\pi_t^N)-(f,\pi_t)$ for bounded functions. Additionally, we
establish the a.s. convergence toward 0 of the approximation errors for
a class of possibly unbounded functions. Specifically, let $f$ be a
real function over the state space and introduce the notation
\[
\tau_t(f) (x) = \int f(z) \tau_t(\mathrm{d}z|x)
\]
for conciseness. Note that $\tau_t(f) \dvtx  \Real^{d_x} \rightarrow\Real
$ is
also a real function over the state space. We define the following
class of functions.
%
\begin{Definicion} \label{defFtp}
$\mathsf{F}_T^p$ is a family of functions $f \dvtx  \Real^{d_x}
\rightarrow\Real
$ that satisfy:
\begin{longlist}[(ii)]
\item[(i)] $(f^p,\pi_t) < \infty$ for $t=0, \ldots, T$, and
\item[(ii)] if $f \in\mathsf{F}_T^p$ then $\tau_t(f^p) \in\mathsf
{F}_T^p$ for $t=1, \ldots, T$.
\end{longlist}
\end{Definicion}
%
The set $\mathsf{F}_T^p$ includes functions that are $p$-integrable
w.r.t. $\pi_t$, $0 \le t \le T$, and remain $p$-integrable when
sequentially transformed by the kernels $\tau_t$, $1 \le t \le T$.
Note that if $p \le q$ then $F_T^q \subseteq F_T^p$. It turns out that
if $f \in\mathsf{F}_T^p$ for some $p \ge4$, then the error of the
particle approximations vanishes for large $N$ at every time step. This
is precisely stated by the following proposition.
%
%
\begin{Proposicion} \label{propConvPF}
Assume that the sequence of observations $Y_{1:T}=y_{1:T}$ is fixed,
with $T$ being some large but finite time horizon, $g_t^{y_{t}} \in
B(\Real^{d_x})$ and $g_t^{y_t} > 0$ (in particular, $(g_t^{y_{t}},\xi
_t) > 0$) for every $t=1,2,\ldots,T$. The following results hold.

\begin{enumerate}[(b)]
\item[(a)] For any $f \in B(\Real^{d_x})$ and any $p \ge1$,
%
\begin{equation}
\bigl\llVert \bigl(f,\pi_t^N\bigr) - (f,
\pi_t) \bigr\rrVert _p 
\le \frac{
c_t \| f \|_\infty
}{
\sqrt{N}
}
\end{equation}
for $t=0,1,\ldots, T$, where $c_t$ is a constant independent of $N$,
$\| f \|_\infty= \sup_{x \in\Real^{d_x}} |f(x)|$ and the
expectation is taken over all possible realizations of the random
measure $\pi_t^N$. In particular,
\[
\lim_{N\rightarrow\infty} \bigl| \bigl(f,\pi_t^N\bigr)
- (f,\pi_t) \bigr| = 0\qquad  \mbox{a.s. for } 0 \le t \le T.
\]


\item[(b)] If $f \in\mathsf{F}_T^4$ , then
$
\lim_{N\rightarrow\infty} \llvert
(f,\pi_t^N) - (f,\pi_t)
\rrvert  = 0 \mbox{ a.s. for } 0 \le t \le T$.
\end{enumerate}
\end{Proposicion}
%

See Appendix~\ref{apProofLemma 1} for a proof.
%
%
%
\begin{Nota}
Part (a) of Proposition~\ref{propConvPF} is fairly standard. A similar
proposition was already proved in \cite{DelMoral00}, albeit under
additional assumptions on the state-space model. Bounds for $p=2$ and
$p=4$ can also be found in a number of references (see, e.g., \cite
{Crisan01,Crisan02,DelMoral04}). Part (b) establishes the almost sure
convergence for the approximate integrals of unbounded functions (e.g.,
for the approximation of the posterior mean) as long as they are
``sufficiently integrable''. A similar result can be found in \cite
{Hu08}, including convergence rates. However, the analysis in \cite
{Hu08} is carried out for a modified particle filtering algorithm, that
involves a rejection test on the generated particles, and cannot be
applied to the standard particle filter presented in this section.
\end{Nota}
%

\section{Particle-kernel approximation of the filtering density}
\label{sApproximation}

In the sequel, we will be concerned with the family of Markov
state-space models for which the posterior probability measures $\{ \pi
_t \}_{t \ge1}$ are absolutely continuous w.r.t. the Lebesgue measure
and, therefore, there exist pdfs $p_t\dvtx \Real^{d_x} \rightarrow
[0,+\infty)$,
$t=1, 2, \ldots\,$, such that $\pi_t(A) = \int_A p_t(x)\,\mathrm{d}x$ for any $A
\in\mB(\Real^{d_x})$. The density $p_t$ is referred to as the
filtering pdf at time $t$. In this section, we briefly review the basic
methodology for kernel density estimation and then describe the
construction of sequences of approximations of $p_t$ using the
particles generated by a particle filter and a generic kernel function.
The section concludes with the discussion on the relationship between
the complexity of the particle filter (i.e., the number of particles
$N$) and the choice of kernel bandwidth for the density estimators.

\subsection{Kernel density estimators}

In order to build an approximation of the function $p_t(x)$ using a
sample of size $N$, $\{ x_t^{(n)} \}_{i=1,\ldots,N}$, we resort to the
classical kernel approach commonly used in density estimation \cite
{Silverman86,Wand95,Simonoff96}. Specifically, given a kernel function
$\phi\dvtx \Real^{d_x}\rightarrow\Real^+$, we build a regularized density
function of the form
%
\begin{equation}
p_t^N(x) = \frac{1}{N} \sum
_{n=1}^N \phi\bigl(x-x_t^{(n)}
\bigr). \label{eqDefinition_of_KDE}
\end{equation}
In the classical theory, the kernel function $\phi$ is often taken to
be a nonnegative and symmetric probability density function with zero
mean and finite second order moment. Specifically, the following
assumptions are commonly made \cite{Silverman86} and we abide by them
in this paper.
%
\begin{Assumption} \label{asPositivePdf}
The kernel $\phi$ is a pdf w.r.t. the Lebesgue measure. In particular,
$\phi(x) \ge0\ \forall x\in\Real^{d_x}$ and $\int\phi(x)\,\mathrm{d}x=1$.
\end{Assumption}
%
%
\begin{Assumption} \label{asSecondOrder}
The probability distribution with density $\phi$ has a finite second
order moment, that is, $c_2 = \int\|x\|^2\phi(x)\,\mathrm{d}x < \infty$.
\end{Assumption}
%

Given a function $\phi$ satisfying A.\ref{asPositivePdf} 
and A.\ref{asSecondOrder} it is possible to define a family of
rescaled kernels
%
\begin{equation}
\phi_{\sfrac{1}{h}}(x)=h^{-d_x}\phi \bigl(h^{-1}x \bigr),
\label{eqNotationKernel}
\end{equation}
where $h>0$ is often referred to as the bandwidth of the kernel
function. Both the kernel and the bandwidth can be optimized to
minimize the mean integrated square error (MISE) between the
regularized density and the target densities \cite{Wand95}.
Specifically, the MISE is defined as
%
\begin{equation}
\mathit{MISE} \equiv\int E \Biggl[ \Biggl( p_t(x) -
\frac{1}{N} \sum_{n=1}^N
\phi_{\sfrac{1} {h}}\bigl(x-x_t^{(n)}\bigr)
\Biggr)^2 \Biggr] \,\mathrm{d}x, \label{eqMISEdef}
\end{equation}
where the expectation is taken over the random sample. Although the
MISE given in equation \eqref{eqMISEdef} is intractable in general,
asymptotic approximations (as $N \rightarrow\infty$) are known \cite
{Wand95}. Moreover, if we assume that $x_t^{(1)},\ldots,x_t^{(N)}$ are
i.i.d. and drawn exactly from $p_t(x)$ (beware that this is \emph{not}
the case in the particle filtering framework, though), then the MISE is
minimized by the Epanechnikov kernel \cite{Silverman86}
%
\begin{equation}\label{eqEpanechnikov}
\phi_\mathrm{E}(x) = \lleft\{ %
\begin{array} {l@{\qquad}l}
\displaystyle \frac{
d_x+2
}{
2v_{d_x}
} \bigl(1-\|x\|^2\bigr), &\mbox{if } \|x\|<1,
\\
0, &\mbox{otherwise}, \end{array} %
\rright.
\end{equation}
where $v_{d_x}$ is the volume of the unit sphere in $\Real^{d_x}$. If,
additionally, $p_t(x)$ is Gaussian with unit covariance matrix, then
the scaling of $\phi_\mathrm{E}$ that yields the minimum MISE is given by the
bandwidth \cite{Wand95}
\[
h_{\mathrm{opt}} = \bigl[ 8v_{d_x}^{-1}(d_x+4)
(2\sqrt{\pi})^{d_x} \bigr]^{\afrac{1}{d_x+4}} N^{-\afrac{1}{d_x+4}}.
\]

In our case, $p_t(x)$ is not known (it is known not to be Gaussian in
general, though) and the random sample $x_t^{(1)},\ldots,x_t^{(N)}$ is not
drawn from $p_t(x)$, so the standard results of \cite
{Silverman86,Wand95,Simonoff96} and others cannot be applied directly
and a specific analysis is needed \cite{LeGland04,Kunsch05}.

In \cite{LeGland04}, two regularized particle filtering algorithms
were studied, each of them yielding a different kernel estimator of
$p_t$. Using the notation in the present paper, they can be written as
%
\begin{equation}\label{nonumber}
p_{t,\mathrm{pre}}^N(x) \propto\frac{1}{N} \sum
_{n=1}^N g_t(y_t|x)
\phi _{\sfrac{1}{h}}\bigl(x-\bar x_t^{(n)}\bigr)
\end{equation}
for the \emph{pre-regularized particle filter}, and
\[
p_{t,\mathrm{post}}^N(x) = \sum_{n=1}^{N}
w_{t}^{(n)} \phi_{\sfrac{1}{h}}\bigl(x - \bar
x_t^{(n)}\bigr)
\]
for the post-regularized particle filter. Note that $p_{t,\mathrm{pre}}^N(x)$ is
an unnormalized approximation of $p_t(x)$ (the normalization constant
cannot be computed in general). For the post-regularized density
estimator, it can be shown that under certain regularity assumptions
(\cite{LeGland04}, Theorem~6.15)
\[
E \biggl[ \int\bigl\llvert p_t(x) - p_{t,\mathrm{post}}^N(x)
\bigr\rrvert \,\mathrm{d}x \bigl| Y_{1:t} \biggr] \rightarrow0\qquad  \mbox{a.s.}
\]
(where the expectation is taken w.r.t. $p_{t,\mathrm{post}}^N$) when $N
\rightarrow
\infty$ and $h \rightarrow0$ jointly. Specifically, the mean total variation
decreases as $\mathrm{O}(N^{-{\sfrac{1}{2}}} + h^2)$. A similar result can be
shown for $p_{t,\mathrm{post}}^N$ (\cite{LeGland04}, Theorem~6.9).
%
%
\begin{Nota}
Although we use the same notation for the particles, $\bar x_t^{(i)}$,
$i=1,\ldots, N$, as in Section~\ref{ssPFs}, the sampling/resampling
schemes in the pre-regularized and post-regularized particle filters
are different from the basic `bootstrap' filter \cite
{Musso01,LeGland04}. The pre-regularized filter, in particular,
involves the use of a rejection sampler.
\end{Nota}
%
%
\begin{Nota}
The convergence results in \cite{LeGland04} for the post-regularized
density estimator $p_{t,\mathrm{post}}^N$ hold true when the following
assumptions on the state-space model are guaranteed.
\begin{itemize}
\item The transition kernel $R_t(x_{t-1},A) = \int_A g_t^{Y_t}(x) \tau
_t(\mathrm{d}x|x_{t-1})$ is mixing (\cite{LeGland04}, Definition~3.2).
\item The likelihood satisfies $\sup_{u \in\mathsf{W}^{2,1}} \frac{
g_t^{Y_t} u }{ \| u \|_{2,1} } < \infty$, where $\mathsf{W}^{2,1}$ is
the Sobolev space of\vspace*{-1pt} functions defined on $\Real^{d_x}$ which,
together with their derivatives up to order 2, are integrable with
respect to the Lebesgue measure, and $\| \cdot\|_{2,1}$ is the
corresponding norm.
\item The measure $\tau_t(\mathrm{d}x|x_{t-1})$ is absolutely continuous w.r.t.
the Lebesgue measure, with density $\tau_t^{x_{t-1}}(x) \in\mathsf
{W}^{2,1}$ and $\sup_{x_{t-1} \in\Real^{d_x}} \| \tau_t^{x_{t-1}} \|
_{2,1} < \infty$.
\end{itemize}
\end{Nota}
%

Assuming that $\tau_t = \tau$ for every $t \ge1$ (hence, the Markov
state process is homogeneous), the analysis in \cite{Kunsch05} targets
the convergence in total variation distance of the continuous measure
$\rho_t^N(x)\,\mathrm{d}x$, where the density estimator $\rho_t^N$ is defined as
\[
\rho_t^N(x) = c_t \sum
_{n=1}^N g_t(y_t|x)
\tau^{x_{t-1}^{(n)}}(x)
\]
with normalization constant $c_t =  ( \sum_{n=1}^N \int
g_t(y_t|x)\tau^{x_{t-1}^{(n)}}(x)\,\mathrm{d}x  )^{-1}$. This is similar to
the pre-regularized approximation $p_{t,\mathrm{pre}}^N$ but using the Markov
kernel of the model, $\tau$, for smoothing, instead of the generic
kernel $\phi_{\sfrac{1}{h}}$. Although in most problems it is possible
to draw from $\tau^{x_{t-1}}$, it is often not possible to evaluate it
and, in such cases, the approximation $\rho_t^N$ is not practical.
Also note that $\rho_t^N$ is \emph{not} a kernel density estimator of
$p_t$ in the classical form of equation \eqref{eqDefinition_of_KDE}. The
sample of size $N$ from which the approximation is constructed
corresponds to the variable $X_{t-1}$, rather than $X_t$, and smoothing
is achieved by way of a prediction step (using the Markov kernel $\tau
$). It is \emph{not} possible, in general, to write $\rho_t^N(x)
\propto\sum_{n=1}^N g_t(y_t|x) \phi_{\sfrac{1}{h}}(x-x_{t-1}^{(n)})$
for some kernel function $\phi$. Under regularity assumptions on $g_t$
and $\tau$, it is proved in \cite{Kunsch05}, Theorem~2, that
%
\begin{equation}
\Prob \biggl\{ \int\bigl|\rho_t^N(x) - p_t(x)
\bigr| \,\mathrm{d}x > \epsilon \biggr\} \le c_1\exp\{-c_2N\}, \qquad t \ge1,
\end{equation}
for any $\epsilon>0$ and some constants $c_1, c_2>0$.
%
%
\begin{Nota}
The regularity assumptions on the state-space model in \cite{Kunsch05}, Theorem~2, are the following.
\begin{enumerate}[(b)]
\item[(a)] There are pdfs $\{ b_t \}_{t \ge1}$ and two constants $0
< c_\tau< C_\tau< \infty$ such that
\[
c_\tau b_t(x) \le\tau^{x_{t-1}}(x) \le
C_\tau b_t(x) \qquad \mbox{for all } x, t.
\]
\item[(b)] The likelihood $g_t$ satisfies that $\sup_{ t \ge1; x, x'
\in\Real^{d_x}; y \in\Real^{d_y} } \frac{ g_t(y|x) }{ g_t(y|x') }
< \infty$. 
\end{enumerate}
The assumption in (a) excludes, for example, models of the form $X_t =
h(X_{t-1}) + V_t$ where the function $h\dvtx \Real^{d_x}\rightarrow\Real^{d_x}$
is not bounded or the noise process $V_t$ is Gaussian (\cite{Kunsch05}, Section~4.2). The assumption in (b) is also stronger than required
for Proposition~\ref{propConvPF} to hold true.
\end{Nota}
%

\subsection{Approximation of the filtering density and its
derivatives} \label{ssApproximations}

We investigate particle-kernel approximations of $p_t$ constructed from
a kernel function $\phi$ and the samples $x_t^{(n)}$, $n=1,\ldots,N$,
generated by the particle filter. Instead of restricting our attention
to procedures based on a single kernel, however, we consider a sequence
of functions $\phi_k\dvtx \Real^{d_x} \rightarrow\Real^+$, $k \in
\mathbb{N}$,
defined according to the notation in equation (\ref{eqNotationKernel}),
that is, $\phi_k(x) = k^{d_x}\phi(kx)$. If $\phi$ complies with
A.\ref{asPositivePdf} 
and A.\ref{asSecondOrder}, then we have similar properties for $\phi
_k$. Trivially, $\phi_k(x) \ge0\ \forall x \in\Real^{d_x}$, and it
is also straightforward to check that $\int\phi_k(x)\,\mathrm{d}x = 1$.
Moreover, if we apply the change of variable $y=kx$ and note that $\mathrm{d}y =
k^{d_x} \,\mathrm{d}x$, then
\[
\int\| x \|^2 \phi_k(x) \,\mathrm{d}x = \frac{1}{k^2} \int\|y
\|^2 \phi(y)\,\mathrm{d}y = \frac{c_2}{k^2}
\]
from A.\ref{asSecondOrder}.

The approximation of $p_t$ generated by the particles $x_t^{(n)}$,
$n=1,\ldots,N$, and the $k$th kernel, $\phi_k$, is denoted as $p_t^{k}$
and has the form
\[
p_t^{k}(x) \triangleq\frac{1}{N} \sum
_{n=1}^N \phi_k\bigl(x-x_t^{(n)}
\bigr) = \bigl(\phi _k^x,\pi_t^N
\bigr),
\]
where $\phi_k^x(x') \triangleq\phi_k(x-x')$. Beware that, in our notation,
we skip the dependence of $p_t^k$ on the number of particles, $N$, for
the sake of simplicity. In Section~\ref{ssAS_convergence}, we will
assume a certain relationship between $N$ and $k$ that will be carried
on through the rest of the paper and justifies the omission in the
notation. Let us also remark that we do not construct $p_t^k$ in order
to approximate integrals w.r.t. the filtering measure (this is more
efficiently achieved using equation \eqref{eqApprox_of_f}). Instead, we aim
at applications where an explicit approximation of the density $p_t$ is
necessary. Some examples are considered in Section~\ref{sApplications}.

In order to investigate the approximation of derivatives of $p_t$, let
us consider the multi-index $\alpha= (\alpha_1, \alpha_2, \ldots,
\alpha_{d_x}) \in\mathbb{N}^* \times\mathbb{N}^* \times\cdots
\times\mathbb{N}^*$, where $\mathbb{N}^* = \mathbb{N} \cup\{0\}$,
and introduce the partial derivative operator $D^\alpha$ defined as
\[
D^\alpha h \triangleq\frac{
\partial^{\alpha_1} \cdots\partial^{\alpha_{d_x}} h
}{
\partial x_1^{\alpha_1} \,\cdots\,\partial x_{d_x}^{\alpha_{d_x}}
}
\]
for any (sufficiently differentiable) function $h\dvtx \Real^{d_x}
\rightarrow
\Real$. The order of the derivative $D^\alpha h$ is denoted as
$|\alpha|=\sum_{i=1}^{d_x} \alpha_i$. We are interested in the
approximation of functions $D^\alpha p_t(x)$ which are continuous, as
explicitly given below.
%
\begin{Assumption}\label{asLipschitz}
For every $x$ in the domain of $p_t(x)$, $D^\alpha p_t(x)$ exists and
is Lipschitz continuous, that is, there exists a constant $c_{\alpha
,t}>0$ such that
\[
\bigl\llvert D^\alpha p_t(x-z) - D^\alpha
p_t(x) \bigr\rrvert \le c_{\alpha,t} \| z \|
\]
for all $x,z \in\Real^{d_x}$.
\end{Assumption}
%
%
\begin{Nota}
It is possible to check whether A.\ref{asLipschitz} holds by
inspecting the transition kernel $\tau_t$ and the likelihood function
$g_t^{y_t}$. For example, assume that $\tau_t(\mathrm{d}x|x')$ has an
associated density w.r.t. the Lebesgue measure, denoted $\tau_t^{x'}$.
A sufficient condition for $D^\alpha p_t$ to be Lipschitz is that both
$g_t^{y_t}$ and $\tau_t^{x'}$ be bounded with bounded derivatives up
to order $1+|\alpha|$. Specifically, it is sufficient that $g_t^{y_t}
\in B(\Real^{d_x})$ and, for any $\beta=(\beta_1, \ldots, \beta
_{d_x})$ such that $0 \le|\beta| \le1+|\alpha|$, $D^\beta g_t^{y_t}
\in B(\Real^{d_x})$ and there exist constants $c_\beta$, independent
of $x$ and $x'$, such that $D^\beta\tau_t^{x'} \le c_\beta$.
\end{Nota}
%

For the same $\alpha$, we also impose the following condition on the
kernel $\phi$.
%
\begin{Assumption}\label{asDPhi_k_in_Cb}
$D^\alpha\phi\in C_b(\Real^{d_x})$, that is, $D^\alpha\phi$ is a
continuous and bounded function. In particular, $\| D^\alpha\phi\|
_\infty= \sup_{x\in\Real^{d_x}} |D^\alpha\phi(x)| < \infty$.
\end{Assumption}
%
%
\begin{Nota} \label{remD_fi_k}
Trivially, if $D^\alpha\phi\in C_b(\Real^{d_x})$ then $D^\alpha\phi
_k \in C_b(\Real^{d_x})$ for any finite $k$. In particular,
$
\| D^\alpha\phi_k \|_\infty= k^{d_x + |\alpha|} \| D^\alpha\phi\|
_\infty$.
\end{Nota}
%

The approximation of $D^\alpha p_t$ computed from the samples
$x_t^{(n)}$, $n=1,\ldots,N$, and the $k$th kernel, $\phi_k$, has the form
%
\begin{equation}\label{eqApproxDp}
D^\alpha p_t^k(x) = \frac{1}{N} \sum
_{n=1}^N D^\alpha\phi
_k^x\bigl(x_t^{(n)}\bigr) =
\bigl(D^\alpha\phi_k^x,\pi_t^N
\bigr).
\end{equation}

\subsection{Complexity of the particle filter and choice of kernel
bandwidth} \label{ssSummary}

In the sequel, we will be concerned with the convergence of the
sequence of approximations $\{D^\alpha p_t^k\}_{k \ge1}$ under the
generic assumptions A.\ref{asPositivePdf}--A.\ref{asDPhi_k_in_Cb}.
The convergence results introduced in Sections~\ref{sConvergence} and
\ref{sApplications} are given either as limits, for $k \rightarrow
\infty$,
or as error bounds that decrease with $k$.

Recall, however, that $p_t^k(x) = (\phi_k^x,\pi_t^N)$, that is, the
density estimator $p_t^k$ depends both on the kernel bandwidth $h=\frac
{1}{k}$ and the number of particles $N$. A distinctive feature of the
analysis in Sections~\ref{sConvergence} and \ref{sApplications} is
that it links both indices by way of the inequality
$
N \ge k^{2(d_x+|\alpha|+1)}$,
where $|\alpha|=\sum_{i=1}^{d_x} \alpha_i$ is the order of the
derivative $D^\alpha$. For $\alpha=(0,\ldots,0)$, $D^\alpha
p_t^k=p_t^k$ and
%
\begin{equation}
N \ge k^{2(d_x+1)}. \label{eqN_vs_k}
\end{equation}
Obviously, $k \rightarrow\infty$ implies that $N \rightarrow\infty$
and $h \rightarrow0$.

This connection is useful to provide simple bounds for the
approximation errors, but also because it yields guidance for the
numerical implementation of the density estimators. In particular, for
$|\alpha|=0$ and a fixed kernel bandwidth $h=\frac{1}{k}$, the
inequality in \eqref{eqN_vs_k} determines the minimum number of
particles $N$ that are needed in the particle filter in order to
guarantee that convergence, at the rates given by the Theorems of
Sections~\ref{sConvergence} and \ref{sApplications}, holds. A lesser
number of samples (i.e., some $N < k^{2(d_x+1)}$) would result in an
under-smoothed density $p_t^k(x)$ with a bigger approximation error.

If the computational complexity of the particle filter is limited by
practical considerations, then $N$ is given and the error bounds to be
introduced only hold when $k \le N^{\afrac{1}{2(d_x+1)}}$ or,
equivalently, when the kernel bandwidth is lower-bounded as $h = \frac
{1}{k} \ge N^{-\afrac{1}{2(d_x+1)}}$. A smaller bandwidth would, again,
result in an under-smoothed approximation $p_t^k(x)$. On the other
hand, since over-smoothing also increases the approximation error of
kernel density estimators \cite{Silverman86}, it is convenient to
choose the smallest possible bandwidth $h$. For given $N$, we should
therefore select\footnote{In practice, an adaptive choice of the
kernel bandwidth (see, e.g., \cite{Brewer00,Zhang06}) is generally
more efficient. In this paper, however, we restrict our attention to
fixed-bandwidth kernels.} $h=h(N)=N^{-\afrac{1}{2(d_x+1)}}$. 

\section{Convergence of the approximations} \label{sConvergence}

Starting from Proposition~\ref{propConvPF}, we prove that the kernel
approximations of the filtering pdf, $p_t^k(x)$, and its derivates
converge a.s. for every $x$ in the domain of $p_t$, both point-wise and
uniformly on $\Real^{d_x}$. We also prove that the smoothed
approximating measure $\breve\pi_t^{N(k)}(\mathrm{d}x)=p_t^k(x)\,\mathrm{d}x$ converges
to $\pi_t$ in total variation distance and that the integrated square
error of a sequence of truncated density estimators converges
quadratically (in $k$) toward 0 a.s. Explicit convergence rates for the
approximations are given.

\subsection{Almost sure convergence} \label{ssAS_convergence}

In this section, we obtain convergence rates for the particle-kernel
approximation $D^\alpha p_t^k(x)$ of equation \eqref{eqApproxDp}. Depending
on the support of the density $p_t(x)$, these rates may be point-wise
or uniform (for all $x$). In both cases, convergence is attained a.s.
based on the following auxiliary result.
%
\begin{Lema} \label{lmU}
Let $ \{ \theta^k  \}_{k \in\mathbb{N}}$ be a sequence of
nonnegative random variables such that, for $p \ge2$,
%
\begin{equation}\label{eqAss_lema_U}
E \bigl[ \bigl( \theta^k \bigr)^p \bigr] \le
\frac{
c
}{
k^{p-\nu}
},
\end{equation}
where $c>0$ and $0\le\nu<1$ are constant w.r.t. $k$. Then, there
exists a nonnegative and a.s. finite random variable $U^\varepsilon$,
independent of $k$, such that
%
\begin{equation}\label{eqStatement_lemmaU}
\theta^k \le\frac{
U^\varepsilon
}{
k^{1-\varepsilon}
},
\end{equation}
where $\frac{1+\nu}{p}<\varepsilon<1$ is also a constant w.r.t. $k$.
\end{Lema}
%
\begin{pf} See Appendix~\ref{apLema_U}.
\end{pf}
%
\begin{Nota}
In Lemma~\ref{lmU}, if the inequality \eqref{eqAss_lema_U} holds for
all $p\ge2$ then the constant $\varepsilon$ in \eqref
{eqStatement_lemmaU} can be made arbitrarily small, that is, we can
choose $0 < \varepsilon<1$.
\end{Nota}
%

Using Lemma~\ref{lmU}, it is possible to prove that $D^\alpha p_t^k(x)
\rightarrow D^\alpha p_t(x)$ a.s. and obtain explicit convergence
rates. In
order to establish a connection between the sequence of kernels $\phi
_k(x)$, $k \in\mathbb{N}$, and the sequence of measure approximations
$\pi_t^N$, $N \in\mathbb{N}$, we define the number of particles to
be a function of the kernel index and denote it as $N(k)$. To be
specific, for a given multi-index $\alpha$, we assume that $N(k) \ge
k^{2(d_x+|\alpha|+1)}$. 
In this way, all the convergence rates to be presented in this paper
are primarily given in terms of the kernel index $k$. We first show
that $D^\alpha p_t^k \rightarrow D^\alpha p_t$ point-wise for $x \in
\Real^{d_x}$.
%
%
\begin{Teorema} \label{thConvergenceDp}
Under assumptions A.\ref{asPositivePdf}, A.\ref{asSecondOrder},
A.\ref{asLipschitz}, A.\ref{asDPhi_k_in_Cb} and $N(k) \ge k^{2 (
d_x + |\alpha| + 1 )}$, the inequality
%
\begin{equation}\label{eqIneqDpU}
\bigl\llvert D^\alpha p_t^k(x) -
D^\alpha p_t(x) \bigr\rrvert \le\frac{
V^{x,\alpha,\varepsilon}
}{
k^{1-\varepsilon}
}
\end{equation}
holds true, with $V^{x,\alpha,\varepsilon}$ an a.s. finite,
nonnegative random variable and a constant $0<\varepsilon<1$. In particular,
%
\begin{equation}\label{eqConvergenceDpas}
\lim_{k\rightarrow\infty} \bigl\llvert D^\alpha
p_t^k(x) - D^\alpha p_t(x) \bigr
\rrvert = 0 \qquad \mbox{a.s.}
\end{equation}
\end{Teorema}
%
\begin{pf} Let us construct an approximation of $p_t(x)$
using the kernel $\phi_k$ and the true filtering measure $\pi_t$,
namely, $\tilde p_t^k(x) = (\phi_k^x,\pi_t)$. Since $\pi
_t(\mathrm{d}x)=p_t(x)\,\mathrm{d}x$, the approximation $\tilde p_t^k$ is actually a
convolution integral and can be written in two alternative ways using
the commutative property, namely
%
\begin{equation}\label{eqCommutative}
\tilde p_t^k(x) = \int\phi_k(x-z)p_t(z)\,\mathrm{d}z
= \int\phi _k(z)p_t(x-z)\,\mathrm{d}z.
\end{equation}
Let us now consider the derivative $D^\alpha p_t$. If we apply the
operator $D^\alpha$ to $\tilde p_t^k$ in (\ref{eqCommutative}), we
readily obtain
\[
D^\alpha\tilde p_t^k(x) = \int
\phi_k(z) D^\alpha p_t(x-z) \,\mathrm{d}z
\]
and, using the latter expression, we find an upper bound for the error
$| { D^\alpha\tilde p_t }^k(x) - D^\alpha p_t(x) |$. In particular,
%
\begin{eqnarray}
\bigl\llvert { D^\alpha\tilde p_t }^k(x) -
D^\alpha p_t(x) \bigr\rrvert &=& \biggl\llvert \int
\phi_k(z) D^\alpha p_t(x-z)\,\mathrm{d}z -
D^\alpha p_t(x) \biggr\rrvert
\nonumber
\\
\label{eqUseConmutative}&\le& \int\phi_k(z) \bigl\llvert D^\alpha
p_t(x-z) - D^\alpha p_t(x) \bigr\rrvert \,\mathrm{d}z
\\
\label{eqUseLipschitz}&\le& c_{\alpha,t} \int\phi_k(z)\| z \| \,\mathrm{d}z
\\
\label{eqUseJensen}&\le& c_{\alpha,t} \sqrt{ \int\phi_k(z)\| z \|^2 \,\mathrm{d}z
}
\\
\label{eqLastBound}&=& c_{\alpha,t} \frac{
\sqrt{c_{2}}
}{
k
},
\end{eqnarray}
where equation (\ref{eqUseConmutative}) follows from A.\ref{asPositivePdf}
(namely, $\phi\ge0$), (\ref{eqUseLipschitz}) is obtained from the
Lipschitz assumption A.\ref{asLipschitz}, (\ref{eqUseJensen}) follows
from Jensen's inequality and, finally, the bound in (\ref
{eqLastBound}) is obtained from assumption A.\ref{asSecondOrder}. Note
that $c_{\alpha,t}$ and $c_2$ are constants with respect to both $x$
and $k$. As a consequence of (\ref{eqLastBound}),
\[
\lim_{k\rightarrow\infty} D^\alpha\tilde p_t^k(x)
= D^\alpha p_t(x).
\]

Consider now the approximation, with $N(k)$ particles, $D^\alpha p_t^k
= (D^\alpha\phi_k^x,\pi_t^{N(k)})$ of the integral $(D^\alpha\phi
_k^x,\pi_t)$. From Proposition~\ref{propConvPF} and assumption A.\ref
{asDPhi_k_in_Cb}, we obtain
%
\begin{eqnarray}\label{eqApplicLemma}
\bigl\llVert D^\alpha p_t^k(x) -
D^\alpha\tilde p_t^k(x) \bigr\rrVert
_p &=& \bigl\llVert \bigl(D^\alpha\phi_k^x,
\pi_t^{N(k)}\bigr) - \bigl(D^\alpha
\phi_k^x,\pi_t\bigr) \bigr\rrVert
_p
\nonumber
\\[-8pt]\\[-8pt]
&\le& \frac{
\bar c_t k^{d_x + |\alpha|} \| D^\alpha\phi\|_\infty
}{
\sqrt{N(k)}
}, \nonumber
\end{eqnarray}
where we have used Remark~\ref{remD_fi_k} and the constant $\bar c_t$
is independent of $N(k)$ and $x$.

A straightforward application of the triangle inequality now yields
%
\begin{equation}\label{eqTriangleIneqDp}
\bigl\llVert D^\alpha p_t^k(x) -
D^\alpha p_t(x) \bigr\rrVert _p \le\bigl
\llVert D^\alpha p_t^k(x) - D^\alpha
\tilde p_t^k(x) \bigr\rrVert _p + \bigl
\llVert D^\alpha\tilde p_t^k(x) -
D^\alpha p_t(x) \bigr\rrVert _p.
\end{equation}
The first term on the right-hand side of (\ref{eqTriangleIneqDp}) can
be bounded using (\ref{eqApplicLemma}), while the second term also has
an upper bound given by\footnote{Note that $
\llVert
D^\alpha\tilde p_t^k(x) - D^\alpha p_t(x)
\rrVert _p = \llvert
D^\alpha\tilde p_t^k(x) - D^\alpha p_t(x)
\rrvert
$ because ${D^\alpha\tilde p_t}^{k}(x)$ does not depend on~$\pi
_t^{N(k)}$.} (\ref{eqLastBound}). Taking both bounds together, we
arrive at
%
\begin{equation}\label{eqLpConvergenceDp}
\bigl\llVert D^\alpha p_t^k(x) -
D^\alpha p_t(x) \bigr\rrVert _p \le
\frac{
\bar c_t k^{d_x+|\alpha|} \| D^\alpha\phi\|_\infty
}{
\sqrt{N(k)}
} + \frac{
c_{\alpha,t}\sqrt{c_{2}}
}{
k
} \le\frac{\bar c_{\alpha,t}}{k},
\end{equation}
where the second inequality follows from the assumption $N(k) \ge
k^{2 ( d_x + |\alpha| + 1  )}$ and $\bar c_{\alpha,t} =
\bar c_t \| D^\alpha\phi\|_\infty+ c_{\alpha,t}\sqrt{c_{2,\alpha
}}$ is a constant.

The inequality \eqref{eqLpConvergenceDp} immediately yields
%
\begin{equation}\label{eqMISE1}
E \bigl[ \bigl\llvert D^\alpha p_t^k(x) -
D^\alpha p_t(x) \bigr\rrvert ^p \bigr] \le
\frac{
\bar c_{\alpha,t}^p
}{
k^p
}
\end{equation}
and we can apply Lemma~\ref{lmU}, with $\theta^k = \llvert  D^\alpha
p_t^k(x) - D^\alpha p_t(x) \rrvert $, $\nu= 0$ and arbitrarily large
$p \ge2$, to obtain
%
\begin{equation}\label{eqIneqV}
\bigl\llvert D^\alpha p_t^k(x) -
D^\alpha p_t(x) \bigr\rrvert \le\frac{
V^{\alpha,x,\varepsilon}
}{
k^{1-\varepsilon}
},
\end{equation}
where $V^{\alpha,x,\varepsilon}$ is a nonnegative and a.s. finite
random variable and $0 < \varepsilon< 1$ is a constant, both of them
independent of $k$. The limit in equation \eqref{eqConvergenceDpas} follows
immediately from the inequality \eqref{eqIneqV}.
\end{pf}

%
\begin{Nota}
The convergence rate for the approximation error $\| D^\alpha p_t^k(x)
- D^\alpha p_t(x) \|_p$ given by inequality \eqref{eqLpConvergenceDp}
can be improved if we place additional assumptions on the filter
density and the kernel, and increase the number of particles $N(k)$. In
particular, if in addition to A.\ref{asPositivePdf}--A.\ref
{asDPhi_k_in_Cb} we assume that
\begin{itemize}
\item$p_t(x)$ has continuous and bounded derivatives up to order
$|\alpha|+2$,
\item the kernel satisfies
$
\int z_i\phi(z)\,\mathrm{d}z = 0 
$, for $i=1, \ldots, d_x$, and\vspace*{1pt}
\item$N(k) \ge k^{2(d_x+|\alpha|+2)}$,
\end{itemize}
then it can be shown, using the multivariate version of Taylor's
theorem, that
\[
\bigl\| D^\alpha p_t^k(x) - D^\alpha
p_t(x) \bigr\|_p \le\frac{\bar C_{\alpha,t}}{k^2}
\]
for some constant $\bar C_{\alpha,t}$ independent of $k$. A specific
result that relies on these extended assumptions is given in Theorem~\ref{thMISE2} (see Section~\ref{ssISE}).
\end{Nota}
%
%
\begin{Nota}
The constant $\bar c_{\alpha,t}$ of equation \eqref{eqLpConvergenceDp} is
independent of the index $k$ and the point $x \in\Real^{d_x}$. The
random variable $V^{\alpha,x,\varepsilon}$ is also independent of the
kernel index $k$, as explicitly given by Lemma~\ref{lmU}. However, it
may depend on the multi-index $\alpha$, the dimension of the state
space $d_x$ and the point $x$ where the derivative of the density is
approximated, hence the notation.
\end{Nota}
%
%
\begin{Nota} \label{rmConvergenceASofDp}
For $\alpha=(0,\ldots,0)=\mathbf{0}$, the inequality (\ref
{eqIneqDpU}) implies that we can construct a particle
approximation of $p_t(x)$ that converges point-wise. In particular,
$D^\mathbf{0} p_t(x) = p_t(x)$ and $D^\mathbf{0} p_t^{k}(x) =
p_t^{k}(x) = (\phi_k^x,\pi_t^{N(k)})$, hence equation (\ref
{eqConvergenceDpas}) becomes
%
\begin{equation}\label{eqLimitASp}
\lim_{k\rightarrow\infty} \bigl| p_t^{k}(x) -
p_t(x) \bigr| = 0 \qquad \mbox{a.s.}
\end{equation}
for every $x\in\Real^{d_x}$.
\end{Nota}
%
%
\begin{Nota}\label{rmStandardAss}
The proof of Theorem~\ref{thConvergenceDp} does not demand that the
assumptions A.\ref{asLipschitz}, A.\ref{asDPhi_k_in_Cb} and $N(k) \ge
k^{2 ( d_x + |\alpha| + 1 )}$ hold for every possible
$\alpha$, but only for the particular derivative we need to
approximate. For instance, if we only aim to approximate $p_t(x)$
(i.e., $\alpha=\mathbf{0}$), assumption A.\ref{asSecondOrder}
implies that the distribution with density $\phi$ must have a finite
second order moment, assumption A.\ref{asLipschitz} means that $p_t$
must be Lipschitz, assumption A.\ref{asDPhi_k_in_Cb} implies that the
basic kernel function $\phi$ must be continuous and bounded, and it
suffices that the number of particles satisfies the inequality $N(k)
\ge k^{2(d_x+1)}$.

Most of the results to be given in the remaining of this paper are
conditional on the assumptions A.\ref{asPositivePdf}, A.\ref
{asSecondOrder}, A.\ref{asLipschitz}, A.\ref{asDPhi_k_in_Cb} and
$N(k) \ge k^{2 ( d_x + |\alpha| + 1 )}$, the same as
Theorem~\ref{thConvergenceDp}. However, they refer only to properties
of $p_t$ and its first order derivatives and, as a consequence, it is
enough to assume that A.\ref{asLipschitz} and A.\ref{asDPhi_k_in_Cb}
hold true for $\alpha= \mathbf{0}$ and $\alpha= \mathbf{1} = (1,
\ldots, 1)$ alone. For the same reason, it suffices to assume $N(k)
\ge k^{2(2d_x + 1)}$.

Through the rest of the paper, we say that the ``standard conditions''
are satisfied when
\begin{itemize}
\item A.\ref{asPositivePdf} and A.\ref{asSecondOrder} hold true;
\item A.\ref{asLipschitz} and A.\ref{asDPhi_k_in_Cb} hold true for,
at least, $\alpha=\mathbf{0}$ and $\alpha= \mathbf{1}$; and
\item$N(k) \ge k^{2(2d_x + 1)}$.
\end{itemize}
\end{Nota}
%

If we restrict $x$ to take values on a sequence of compact subsets of
$\Real^{d_x}$, then we can obtain a convergence rate for the error $|
p_t^k(x) - p_t(x) |$ that is uniform on $x$, instead of point-wise like
in Theorem~\ref{thConvergenceDp}. For the following result, we fix $p
\ge2$ and consider the sequence of hypercubes
\[
\mK_k = [-M_k,+M_k] \times\cdots
\times[-M_k,+M_k] \subset\Real^{d_x},
\]
where $M_k = \frac{1}{2} k^{\afrac{\beta}{d_x p}}$, and $0\le\beta<1$
is a positive constant independent of $k$. Note that, for any fixed $p$
and $\beta> 0$, $\lim_{k\rightarrow\infty} \mK_k = \Real^{d_x}$.
%
\begin{Teorema} \label{thSupK}
If the standard conditions are satisfied, then
\[
\sup_{x \in\mK_k} \bigl\llvert p_t^k(x) -
p_t(x) \bigr\rrvert \le\frac{
U^{\varepsilon}
}{
k^{1-\varepsilon}
},
\]
where $U^\varepsilon\ge0$ is an a.s. finite random variable and
$0<\varepsilon<1$ is a constant, both of them independent of $k$ and
$x$. In particular,
\[
\lim_{k\rightarrow\infty} \sup_{x \in\mK_k} \bigl\llvert
p_t^k(x)-p_t(x) \bigr\rrvert = 0
\qquad \mbox{a.s.}
\]
\end{Teorema}
%
\begin{pf}
For any $x=(x_1,\ldots,x_{d_x}) \in\mK_k$ and a function $f\dvtx \Real
^{d_x}\rightarrow\Real$ continuous, bounded and differentiable,
\[
f(x) - f(0) = \int_{-M_k}^{x_1} \cdots\int
_{-M_k}^{x_{d_x}} D^{\mathbf{1}}f(z) \,\mathrm{d}z-\int
_{-M_k}^{0} \cdots\int_{-M_k}^{0}
D^{\mathbf{1}}f(z) \,\mathrm{d}z.
\]
In particular, for $x_i \in[-M_{k},M_k]$, $i=1,\ldots,d_x$, and the
assumption A.\ref{asDPhi_k_in_Cb} with $\alpha= \mathbf{1}$,
%
\begin{equation}
\bigl\llvert p_t^{k}(x) - p_t(x) \bigr
\rrvert \le2\int_{-M_{k}}^{M_k}\cdots\int
_{-M_{k}}^{M_k} \bigl\llvert D^{\mathbf{1}}p_t^k(z)
- D^{\mathbf{1}}p_t(z) \bigr\rrvert \,\mathrm{d}z + \bigl\llvert
p_t^{k}(0)-p_t(0) \bigr\rrvert
\end{equation}
and, as a consequence,
%
\begin{equation}\label{eqSupError0M}
\sup_{x \in\mK_k} \bigl\llvert p_t^{k}(x) -
p_t(x) \bigr\rrvert \le2A^k+ \bigl\llvert
p_t^{k}(0)-p_t(0) \bigr\rrvert ,
\end{equation}
where
\[
A^{k} = \int_{-M_{k}}^{M_k}\cdots\int
_{-M_{k}}^{M_k} \bigl\llvert D^{\mathbf{1}}p_t^k(z)
- D^{\mathbf{1}}p_t(z) \bigr\rrvert \,\mathrm{d}z.
\]
An application of Jensen's inequality yields, for $p \ge1$,
\[
\biggl( \frac{1}{2^{d_x} M_k^{d_x}} A^{k} \biggr)^p \le
\frac{1}{2^{d_x} M_k^{d_x}} \int_{-M_{k}}^{M_k}\cdots \int
_{-M_{k}}^{M_k} \bigl\llvert D^{\mathbf{1}}p_t^k(z)
- D^{\mathbf{1}}p_t(z) \bigr\rrvert ^p \,\mathrm{d}z,
\]
hence
%
\begin{equation}\label{eqBoundBigA}
\bigl( A^{k} \bigr)^p \le2^{d_x(p-1)}
M_k^{d_x(p-1)} \sum_{\ell=0}^{2^{d_x}-1}
\int_{-M_k}^{M_k}\cdots\int_{-M_k}^{M_k}
\bigl\llvert D^{\mathbf{1}}p_t^k(z) -
D^{\mathbf{1}}p_t(z) \bigr\rrvert ^p \,\mathrm{d}z.
\end{equation}
Since, from inequality (\ref{eqLpConvergenceDp}) in the proof of
Theorem~\ref{thConvergenceDp},
%
\begin{equation}\label{eqSmallBoundAk}
E \bigl[ \bigl\llvert D^{\mathbf{1}}p_t^k
\bigl(s_\ell(z)\bigr) - D^{\mathbf{1}}p_t
\bigl(s_\ell(z)\bigr) \bigr\rrvert ^p \bigr] \le
\frac{
\bar c_{\mathbf{1},t}^p
}{
k^p
},
\end{equation}
we can combine \eqref{eqSmallBoundAk} and \eqref{eqBoundBigA} to
arrive at
\[
E \bigl[ \bigl(A^k\bigr)^p \bigr] \le\frac{
2^{d_x p} M_k^{d_x p} \bar c_{\mathbf{1},t}^p
}{
k^p
}
= \frac{
\bar c_{\mathbf{1},t}^p
}{
k^{p-\beta}
},
\]
where the equality follows from the relationship $M_k = \frac{1}{2}
k^{\afrac{\beta}{d_x p}}$. Using Lemma~\ref{lmU} with $\theta_k =
A^k$, $p \ge2$, $\nu=\beta$ and $c=\bar c_{\mathbf{1},t}^p$, we
obtain a constant $\varepsilon_1 \in ( \frac{1+\beta}{p}, 1
)$ and a nonnegative and a.s. finite random variable
$V^{A,\varepsilon_1}$, both of them independent of $k$, such that
%
\begin{equation}\label{eqRate_part1}
A^k \le\frac{
V^{A,\varepsilon_1}
}{
k^{1-\varepsilon_1}
}.
\end{equation}
Since, from Proposition~\ref{propConvPF},
\[
E \bigl[ \bigl\llvert p_t^k(x) - p_t(x)
\bigr\rrvert ^p \bigr] \le\frac{
\bar c_{\mathbf{0},t}^p
}{
k^p
},
\]
we can apply Lemma~\ref{lmU} again, with $\theta^k = | p_t^k(0) -
p_t(0) |$, $p \ge2$, $\nu=0$ and $c=\bar c_{\mathbf{0},t}^p$ to
obtain that
%
\begin{equation}\label{eqRate_part2}
\bigl\llvert p_t^k(0) - p_t(0) \bigr
\rrvert \le\frac{
V^{p_t(0),\varepsilon_2}
}{
k^{1-\varepsilon_2}
},
\end{equation}
where $\varepsilon_2 \in (\frac{1}{p},1 )$ is a constant
and $V^{p_t(0),\varepsilon_2}$ is a nonnegative and a.s. finite random\vspace*{1pt}
variable, both of them independent of $k$.

If we choose $\varepsilon=\varepsilon_1=\varepsilon_2 \in (
\frac{1+\beta}{p}, 1  )$ and define $U^\varepsilon=
V^{A,\varepsilon_1} + V^{p_t(0),\varepsilon_2}$, then the\vspace*{1pt} combination
of equations \eqref{eqSupError0M}, \eqref{eqRate_part1} and \eqref
{eqRate_part2} yields
\[
\sup_{x \in\mK_k} \bigl\llvert p_t^k(x) -
p_t(x) \bigr\rrvert \le\frac{U^\varepsilon}{k^{1-\varepsilon}},
\]
where $U^\varepsilon$ is a.s. finite. Note that $U^\varepsilon$ and
$\varepsilon$ are independent of $k$. Moreover, we can choose $p$ as
large as we wish and $\beta>0$ as small as needed, hence we can select
$\varepsilon\in(0,1)$.
\end{pf}
%
\begin{Nota}
Assuming that A.\ref{asLipschitz} and A.\ref{asDPhi_k_in_Cb} hold for
the multi-index $\alpha' = \alpha+ \mathbf{1}$, the argument of the
proof of Theorem~\ref{thSupK} can also be adapted to show that
\[
\sup_{x \in\mK_k} \bigl\llvert D^\alpha
p_t^{k}(x) - D^\alpha p_t(x)
\bigr\rrvert \le\frac{
\tilde U^{\varepsilon}
}{
k^{1-\varepsilon}
},
\]
where the constant $0 < \varepsilon< 1$ and the a.s. finite random
variable $\tilde U^{\varepsilon} \ge0$ are independent of $k$.
\end{Nota}
%
%
\begin{Nota} \label{rmFixedK}
Theorem~\ref{thSupK} also holds for a fixed compact subset $\mK
\subset\Real^{d_x}$ instead of the sequence $\mK_1, \mK_2, \ldots\,$.
In particular, the presented proof is easily adapted to a fixed
hypercube $\mK=[-M,+M]\times\cdots\times[-M,+M]$. Therefore,
%
\begin{equation}\label{eqSupFixedK}
\sup_{x\in\mK} \bigl\llvert p_t^k(x) -
p_t(x) \bigr\rrvert \le\frac{
\tilde U^{\varepsilon}
}{
k^{1-\varepsilon}
},
\end{equation}
where the constant $0<\varepsilon<1$ and the a.s. finite random
variable $\tilde U^{\varepsilon} \ge0$ are independent of $k$.
\end{Nota}
%

%

\subsection{Convergence in total variation distance} \label
{ssTV_convergence_1}

The total variation distance (TVD) between two measures $\mu_1, \mu_2
\in\mP(\Real^d)$ on the Borel $\sigma$-algebra $\mB(\Real^d)$ is
defined as
\[
d_{\mathrm{TV}}(\mu_1, \mu_2 ) \triangleq\sup
_{A \in\mB(\Real^d)} \bigl| \mu _1(A) - \mu_2(A) \bigr|.
\]
Correspondingly, a sequence of measures $\mu^n \in\mP(\Real^d)$
converges toward $\mu\in\mP(\Real^d)$ in TVD when $\lim_{n\rightarrow
\infty} d_{\mathrm{TV}}( \mu^n,\mu) = 0$. It can be shown that if $\mu^n$
and $\mu$ have densities w.r.t. the Lebesgue measure, denoted $q^n$
and $q$, respectively, then\vspace*{-1.5pt}
\[
d_{\mathrm{TV}}\bigl(\mu^n, \mu\bigr) = \frac{1}{2} \int
\bigl\llvert q^n(x) - q(x) \bigr\rrvert \,\mathrm{d}x
\]
and, therefore, the sequence $\mu^n$ converges to $\mu$ in TVD if,
and only if,\vspace*{-1.5pt}
%
\begin{equation}\label{eqL1boundForTVD}
\lim_{n\rightarrow\infty} \int\bigl\llvert q^n(x) - q(x) \bigr
\rrvert \,\mathrm{d}x = 0.
\end{equation}

Consider the smooth approximating measures\vspace*{-1.5pt}
\[
\breve\pi_t^{N(k)}(\mathrm{d}x) = p_t^k(x)\,\mathrm{d}x,\qquad
k=1,2,\ldots.
\]
In this section, we show that the sequence $\breve\pi_t^{N(k)}$
converges toward $\pi_t$ in TVD, as $k \rightarrow\infty$, by
proving first
that $\int\llvert  p_t^{k} - p_t \rrvert  \,\mathrm{d}x \rightarrow0$ under the same
assumptions of Theorem~\ref{thSupK}. This result is established by
Theorem~\ref{thConvTV_1} below. The same as in the proof of Theorem~\ref{thSupK}, we consider an increasing sequence of hypercubes $\mK_1
\subset\cdots\subset\mK_k \subset\cdots\subset\Real^{d_x}$,
where $\mK_k = [-M_k,+M_k] \times\cdots\times[-M_k,+M_k]$ and $M_k
= \frac{1}{2} k^{\afrac{\beta}{d_x p}}$, with constants $0<\beta<1$
and $p > 3$. Also, recall that, for a set $A \in\Real^d$, $A^\compl=
\Real^d \backslash A$ denotes its complement and, given a probability
measure $\mu\in{\mathcal P}(\Real^d)$, $\mu(A) = \int_A \mu(\mathrm{d}x)$
is the probability of $A$.
%
\begin{Teorema} \label{thConvTV_1}
If the standard conditions are satisfied and $\pi_t(\mK_k^\compl)
\le\frac{b}{2} k^{-\gamma}$, where $b>0$ and $\gamma>0$ are
arbitrary but constant w.r.t. $k$, then\vspace*{-1.5pt}
\[
\int\bigl\llvert p_t^{k}(x) - p_t(x)
\bigr\rrvert\, \mathrm{d}x < \frac{
Q^{\varepsilon}
}{
k^{\min\{1-\varepsilon,\gamma\}}
},
\]
where $Q^\varepsilon> 0$ is an a.s. finite random variable and
$0<\varepsilon<1$ is a constant, both of them independent of $k$. In
particular,\vspace*{-1.5pt}
\[
\lim_{k\rightarrow\infty} \int\bigl\llvert p_t^{k}(x)
- p_t(x) \bigr\rrvert\, \mathrm{d}x = 0 \qquad \mbox{a.s.}
\]
and, as a consequence,\vspace*{-1.5pt}
\[
\lim_{k \rightarrow\infty} d_{\mathrm{TV}}\bigl( \breve
\pi_t^{N(k)}, \pi_t \bigr) = 0 \qquad \mbox{a.s.}
\]
\end{Teorema}
%
\begin{pf} We start with a trivial decomposition of the
integrated absolute error,\vspace*{-1.5pt}
\begin{eqnarray*}
\int\bigl\llvert p_t^{k}(x) - p_t(x)
\bigr\rrvert \,\mathrm{d}x &=& \int_{\mK_k} \bigl\llvert
p_t^{k}(x) - p_t(x) \bigr\rrvert \,\mathrm{d}x + \int
_{\mK_k^\compl} \bigl\llvert p_t^{k}(x) -
p_t(x) \bigr\rrvert \,\mathrm{d}x
\nonumber
\\
&\le& \int_{\mK_k} \bigl\llvert p_t^{k}(x)
- p_t(x) \bigr\rrvert \,\mathrm{d}x + 2\int_{\mK_k^\compl}
p_t(x)\,\mathrm{d}x
\nonumber
\\
&&{} + \int_{\mK_k^\compl} \bigl( p_t^{k}(x) -
p_t(x) \bigr) \,\mathrm{d}x, 
\end{eqnarray*}
where the equality follows from $\mK_k \cup\mK_k^\compl= \Real
^{d_x}$ and the inequality is obtained from the fact that $p_t$ and
$p_t^{k}$ are nonnegative. Moreover,
\[
\int_{\mK_k^\compl} \bigl(p_t^{k}(x) -
p_t(x) \bigr)\,\mathrm{d}x \le \int_{\mK_k^\compl} \bigl\llvert
p_t^k(x) - p_t(x) \bigr\rrvert \,\mathrm{d}x \le\int
_{\mK_k} \bigl\llvert p_t^{k}(x) -
p_t(x) \bigr\rrvert \,\mathrm{d}x
\]
hence
%
\begin{equation}\label{eqThreeTerms}
\int\bigl\llvert p_t^{k}(x) - p_t(x)\bigr
\rrvert \,\mathrm{d}x \le 2\int_{\mK_k} \bigl\llvert p_t^{k}(x)
- p_t(x) \bigr\rrvert \,\mathrm{d}x + 2\int_{\mK_k^\compl}
p_t(x)\,\mathrm{d}x.
\end{equation}
The first term on the right-hand side \eqref{eqThreeTerms} can be
bounded by
%
\begin{equation}\label{eqIneq1}
\int_{\mK_k} \bigl\llvert p_t^{k}(x)
- p_t(x) \bigr\rrvert\, \mathrm{d}x \le\mL(\mK_k) \sup
_{x \in\mK_k} \bigl\llvert p_t^{k}(x) -
p_t(x) \bigr\rrvert ,
\end{equation}
where $\mL(\mK_k) = ( 2M_k )^{d_x} = k^{\sfrac{\beta}{p}}$ is the
Lebesgue measure of $\mK_k$. From Theorem~\ref{thSupK}, the supremum
in \eqref{eqIneq1} can be bounded as $\sup_{x \in\mK_k} | p_t^k(x)
- p_t(x) | \le V^{\varepsilon_1}/k^{1-\varepsilon_1}$, where
$V^{\varepsilon_1} \ge0$ is\vspace*{-1pt} an a.s. finite random variable and $\frac
{1+\beta}{p} < \varepsilon_1 < 1$ is a constant, both independent of
$k$. Therefore,\vspace*{1pt} the inequality \eqref{eqIneq1} can be extended to yield
%
\begin{equation}\label{eqBoundFirstTerm}
\int_{\mK_k} \bigl\llvert p_t^{k}(x)
- p_t(x) \bigr\rrvert \,\mathrm{d}x \le\frac{
V^{\varepsilon_1}
}{
k^{1-\varepsilon_1-\sfrac{\beta}{p}}
} =
\frac{
V^{\varepsilon}
}{
k^{1-\varepsilon}
},
\end{equation}
where $\varepsilon=\varepsilon_1 + \frac{\beta}{p}$ and
$V^{\varepsilon}=V^{\varepsilon_1}$. If we choose $\varepsilon_1 <
1-\frac{\beta}{p}$, then $\varepsilon\in ( \frac{1+2\beta
}{p}, 1  )$. Note\vspace*{-1pt} that, for $\beta< 1$ and $p>3$, $1-\frac
{\beta}{p}-\frac{1+\beta}{p} > 1-\frac{3}{p} > 0$, hence both
$\varepsilon_1$ and $\varepsilon$ are well defined.

For the second integral in equation \eqref{eqThreeTerms}, note that $\int_{\mK_k^\compl}p_t(x)\,\mathrm{d}x = \pi_t(\mK_k^\compl)$ and, therefore, it
can be bounded directly from the assumptions in the Theorem, that is,
%
\begin{equation}\label{eqBoundSecondTerm}
2\int_{\mK_k^\compl} p_t(x)\,\mathrm{d}x \le b k^{-\gamma},
\end{equation}
where $b>0$ and $\gamma>0$ are constant w.r.t. $k$.
Putting together equations \eqref{eqThreeTerms}, \eqref{eqBoundFirstTerm}
and \eqref{eqBoundSecondTerm} yields the desired result.
\end{pf}
%
\begin{Nota}
The condition $\pi_t(\mK_k^\compl) \le\frac{b}{2} k^{-\gamma}$ in
the statement of Theorem~\ref{thConvTV_1} is satisfied for any $t$ when
\begin{itemize}
\item it is satisfied at time $t=0$, that is, there exists some
constant $b_0$ such that $\pi_0(\mK_k^\compl) \le\frac{b_0}{2}
k^{-\gamma}$,
\item the likelihood is bounded, that is, $g_t^{y_t} \in B(\Real^{d_x})$,
\item and the kernels $\tau_t(\mathrm{d}x|x')$ have sufficiently light tails
for every $t$ and every $x' \in\Real^{d_x}$.
\end{itemize}
The latter can be made more precise using a standard induction
argument. For example, let $\mK_k=[-\frac{1}{2}k^{\afrac{\beta
}{d_xp}}, +\frac{1}{2}k^{\afrac{\beta}{d_xp}}]$ with $p \ge2$ and $0\le
\beta<1$, and assume that for any $x' \in\Real^{d_x}$\vspace*{-1pt} the kernel
$\tau_t$ satisfies that $\tau_t(\mK_k^\compl) \le\frac{b(x')}{2}
k^{-\gamma}$ for some function $b\dvtx \Real^{d_x} \rightarrow(0,\infty
)$. If
$b(x')$ can be upper bounded by a polynomial function, say $b(x') \le
c ( 1+  ( \sum_{i=1}^{d_x} |x_i'|  )^a  )$, for
some constant $c>0$ and degree $a < \frac{d_x p (\gamma- 1)}{\beta
}$, then there exists\vspace*{-1pt} a constant $b_t < \infty$ such that $\pi_t(\mK
_k^\compl) \le\frac{b_t}{2}k^{-\gamma}$.
\end{Nota}
%

\subsection{Integrated square error} \label{ssISE}

A standard figure of merit for the assessment of kernel density
estimators is the mean integrated square error (MISE) \cite
{Silverman86,Simonoff96}. If we assume that both $p_t(x)$ and the
kernel $\phi(x)$ take values on a compact set $\mK$, then it is
relatively simple to prove that the MISE of the sequence of
approximations $D^\alpha p_t^k$ converges toward 0 quadratically with
the index $k$. In particular, we have the following result.
%
%
\begin{Teorema} \label{thMISE}
Assume that A.\ref{asPositivePdf}, A.\ref{asSecondOrder}, A.\ref
{asLipschitz}, A.\ref{asDPhi_k_in_Cb} and $N(k) \ge k^{2 ( d_x +
|\alpha| + 1 )}$ hold true. If both $p_t(x)$ and the kernel
$\phi(x)$ have a compact support set $\mK\subset\Real^{d_x}$, then
\[
\mathit{MISE} \equiv\int_{\mK} E \bigl[ \bigl(
D^\alpha p_t^k(x) - D^\alpha
p_t(x) \bigr)^2 \bigr]\, \mathrm{d}x \le\frac{
c_{\alpha,\mK,t}
}{
k^2
},
\]
where $c_{\alpha,\mK,t} > 0$ is constant w.r.t. $k$.
\end{Teorema}
%
\begin{pf} Since any compact set is contained in a larger
hypercube, we can choose $\mK= [-M,+M] \times\cdots\times[-M,+M]$
without loss of generality. Furthermore, since the assumptions of
Theorem~\ref{thConvergenceDp} are satisfied, we can recall the
inequality in \eqref{eqMISE1}, which, selecting $p=2$, yields
\[
E \bigl[ \bigl( D^\alpha p_t^k(x) -
D^\alpha p_t(x) \bigr)^2 \bigr] \le
\frac{
\bar c_{\alpha,t}^2
}{
k^2
},
\]
where the constant $\bar c_{\alpha,t}^2$ is independent of $k$ and
$x$. Therefore,
\[
\int_\mK E \bigl[ \bigl( D^\alpha
p_t^k(x) - D^\alpha p_t(x)
\bigr)^2 \bigr] \,\mathrm{d}x \le\frac{
\bar c_{\alpha,t}^2
}{
k^2
} \mL(\mK) \le
\frac{
c_{\alpha,\mK,t}
}{
k^2
},
\]
where $\mL(\mK) = (2M)^{d_x}$ is the Lebesgue measure of $\mK$ and
$c_{\alpha,\mK,t} = (2M)^{d_x} \bar c_{\alpha,t}^2$.
\end{pf}

It is also possible to establish a quadratic convergence rate (w.r.t.
$k$) for the integrated square error (ISE) of a sequence of truncated
density approximations. In particular, consider the usual hypercubes
$\mK_k = [-M_k,+M_k] \times\cdots\times[-M_k,+M_k]$ with $M_k =
\frac{1}{2}k^{\afrac{\beta}{d_x p}}$, for some $p > \frac{5}{2}$ and
a constant $0 < \beta< 1$, and define the truncated density estimators
%
\[
p_t^{\top,k}(x) = I_{\mK_k}(x) p_t^k(x)
= \lleft\{ %
\begin{array} {l@{\qquad}l} p_t^k(x), &
\mbox{if } x \in\mK_k,
\\
0, &\mbox{otherwise}. \end{array} %
\rright.
\]
Since $\lim_{k\rightarrow\infty} \mK_k = \Real^{d_x}$, it follows that
$\lim_{k\rightarrow\infty} | p_t^{\top,k}(x) - p_t^k(x) | = 0$ and
we can
make $p_t^{\top,k}$ arbitrarily close to the original approximation.
The theorem below states that $p_t^{\top,k}$ converges a.s. toward
$p_t$, with a quadratic rate.
%
%
\begin{Teorema} \label{thISE}
If the standard conditions are satisfied, $p_t \in B(\Real^{d_x})$ and
$\pi_t(\mK_k^\compl) \le bk^{-\gamma}$, where $b>0$ and $\gamma>0$
are arbitrary but constant w.r.t. $k$, then
\[
\mathit{ISE}\equiv\int \bigl( p_t^{\top,k}(x) -
p_t(x) \bigr)^2 \,\mathrm{d}x \le\frac{
U^\varepsilon
}{
k^{\min\{ 2-\varepsilon, \gamma\}}
},
\]
where $U^\varepsilon\ge0$ is an a.s. finite random variable,
independent of $k$, and $0<\varepsilon<2$ is an arbitrarily small
constant. In particular,
\[
\lim_{k\rightarrow\infty} \int \bigl( p_t^{\top,k}(x) -
p_t(x) \bigr)^2 \,\mathrm{d}x = 0\qquad  \mbox{a.s.}
\]

\end{Teorema}
%
\begin{pf} We start with the trivial decomposition
%
\begin{eqnarray}\label{eqFacil}
\int \bigl( p_t^{\top,k}(x) - p_t(x)
\bigr)^2 \,\mathrm{d}x &=& \int_{\mK_k} \bigl(
p_t^{\top,k}(x) - p_t(x) \bigr)^2
\,\mathrm{d}x
\nonumber
\\[-8pt]\\[-8pt]
&&{} + \int_{\mK_k^\compl} \bigl( p_t^{\top,k}(x) -
p_t(x) \bigr)^2 \,\mathrm{d}x, \nonumber
\end{eqnarray}
where $\mK_k^\compl= \Real^{d_x} \backslash\mK_k$ is the
complement of $\mK_k$, and, expanding the square in the last integral
of equation \eqref{eqFacil}, we obtain
%
\begin{eqnarray}\label{eqDecomposition2}
\int \bigl( p_t^{\top,k}(x) - p_t(x)
\bigr)^2\, \mathrm{d}x &=& \int_{\mK_k} \bigl(
p_t^{\top,k}(x) - p_t(x) \bigr)^2
\,\mathrm{d}x
\nonumber
\\
&& {}+ \int_{\mK_k^\compl} \bigl( p_t(x) -
p_t^{\top,k}(x) \bigr) p_t(x) \,\mathrm{d}x
\\
&&{} + \int_{\mK_k^\compl} \bigl( p_t^{\top,k}(x) -
p_t(x) \bigr) p_t^{\top,k}(x) \,\mathrm{d}x.\nonumber
\end{eqnarray}
In the rest of the proof, we compute upper bounds for each of the
integrals on the right-hand side of equation \eqref{eqDecomposition2}.

%

For the first term in \eqref{eqDecomposition2}, we note that
$p_t^{\top,k}(x) = p_t^k(x)$ for all $x \in\mK_k$, hence
%
\begin{eqnarray}\label{eqBound-a}
\int_{\mK_k} \bigl( p_t^{\top,k}(x) -
p_t(x) \bigr)^2 \,\mathrm{d}x &=& \int_{\mK_k}
\bigl( p_t^k(x) - p_t(x)
\bigr)^2 \,\mathrm{d}x
\nonumber
\\[-8pt]\\[-8pt]
&\le& \mL(\mK_k) \Bigl( \sup_{x \in\mK_k} \bigl\llvert
p_t^{\top,k}(x) - p_t(x) \bigr\rrvert
\Bigr)^2, \nonumber
\end{eqnarray}
where $\mL(\mK_k) = ( 2M_k )^{d_x} = k^{\sfrac{\beta}{p}}$. Using
Theorem~\ref{thSupK}, we obtain an upper bound for the supremum in equation
\eqref{eqBound-a}, namely $\sup_{x \in\mK_k} | p_t^k(x) - p_t(x) |
\le V^{\varepsilon_1}/k^{1-\varepsilon_1}$, where $V^{\varepsilon_1}
\ge0$ is\vspace*{-1pt} an a.s. finite random variable and $\frac{1+\beta}{p} <
\varepsilon_1 < 1$ is a constant. Both $V^{\varepsilon_1}$ and
$\varepsilon_1$ are independent of $k$. We then extend the inequality
in \eqref{eqBound-a} as
%
\begin{equation}\label{eqTerm1RHS}
\int_{\mK_k} \bigl( p_t^{\top,k}(x) -
p_t(x) \bigr)^2 \,\mathrm{d}x \le k^{\sfrac{\beta} {p}}
\frac{
(
V^{\varepsilon_1}
)^2
}{
k^{2-2\varepsilon_1}
} = \frac{
\tilde U^{\varepsilon}
}{
k^{2-\varepsilon}
},
\end{equation}
where $\varepsilon=2\varepsilon_1 + \frac{\beta}{p}$ and $\tilde
U^{\varepsilon}= ( V^{\varepsilon_1}  )^2$. If we choose
$\varepsilon_1 < 1-\frac{\beta}{2p}$, then $\varepsilon\in (
\frac{2+3\beta}{p}, 2  )$. Note\vspace*{-1pt} that, for $\beta< 1$ and $p >
\frac{5}{2}$, $2-\frac{2+3\beta}{p} > 0$, hence $\varepsilon$ is
well defined.

For the second term on the right-hand side of equation \eqref
{eqDecomposition2} we simply note that $p_t^{\top,k}(x)=0$ for all $x
\in\mK_k^\compl$ and $p_t(x) < \| p_t \|_\infty< \infty$, since
$p_t \in B(\Real^{d_x})$. Therefore,
\[
\int_{\mK_k^\compl} \bigl( p_t(x) -
p_t^{\top,k}(x) \bigr) p_t(x)\, \mathrm{d}x \le\|
p_t \|_\infty\int_{\mK_k^\compl}
p_t(x)\,\mathrm{d}x = \| p_t \|_\infty
\pi_t\bigl(\mK_k^\compl\bigr),
\]
and using the assumption $\pi_t(\mK_k^\compl) \le bk^{-\gamma}$ we obtain
%
\begin{equation}\label{eqTerm2RHS}
\int_{\mK_k^\compl} \bigl( p_t(x) -
p_t^{\top,k}(x) \bigr) p_t(x)\, \mathrm{d}x \le
\frac{
b\|p_t\|_\infty
}{
k^{\gamma}
}.
\end{equation}
The third term is trivial. Since $p_t^{\top,k}(x)=0$ for all $x \in
\mK_k^\compl$, it follows that
%
\begin{equation}\label{eqTerm3RHS}
\int_{\mK_k^\compl} \bigl( p_t^{\top,k}(x) -
p_t(x) \bigr) p_t^{\top,k}(x) = 0.
\end{equation}

Substituting equations \eqref{eqTerm1RHS}, \eqref{eqTerm2RHS} and \eqref
{eqTerm3RHS} into equation \eqref{eqDecomposition2} yields
\[
\int \bigl( p_t^{\top,k}(x) - p_t(x)
\bigr)^2 \,\mathrm{d}x \le\frac{\tilde U^\varepsilon}{k^{2-\varepsilon}} + \frac{b\|p_t\|_\infty}{k^\gamma} \le
\frac{
U^\varepsilon
}{
k^{\min\{2-\varepsilon,\gamma\}}
},
\]
where $U^\varepsilon= \tilde U^\varepsilon+ b\|p_t\|_\infty$ and
$0<\varepsilon<2$.
\end{pf}

The classical asymptotic approximation of the MISE (AMISE) for kernel
density estimators built from i.i.d. samples is (see, e.g., \cite
{Duong05} and note that we restrict ourselves to diagonal bandwidth matrices)
%
\begin{equation}\label{eqDefAMISE}
\mathit{AMISE} \equiv h^4 c(\phi,p_o) +
\frac{c(\phi)}{Nh^{d_x}},
\end{equation}
where $h>0$ is the bandwidth parameter, $c(\phi,p_o)>0$ is a constant
that depends on the kernel $\phi$ and the target density (denoted
$p_o$ here and assumed twice differentiable), $c(\phi)>0$ is another
constant depending on $\phi$ alone and $N$ is the number of samples.
If we substitute $h=1/k$ and $N=k^{2d_x+2}$, as given by our analysis,
into the expression above, then we find that the MISE converges
asymptotically as $\frac{ \tilde c(\phi,p_o) }{ k^4 }$, for some
constant $\tilde c(\phi,p_o)>0$. We note, however, that
\begin{itemize}
\item Equation (\ref{eqDefAMISE}) is only an asymptotic approximation of
the MISE, whereas Theorems \ref{thMISE} and \ref{thISE} give actual
upper bounds for the MISE and the ISE that are valid for every $k$;
\item the AMISE of equation (\ref{eqDefAMISE}) is derived under the
assumption that a size $N$ sample drawn from the density $p_t$ is
available \cite{Wand95}, whereas Theorems \ref{thMISE} and \ref
{thISE} hold true for the smoothing of any random measure $\pi_t^N$
that satisfies $\| (f,\pi_t^N) - (f,\pi_t) \|_p \le\frac{c\| f \|
_\infty}{\sqrt{N}}$ for\vspace*{-1pt} some constant $c$ and $f \in B(\Real^{d_x})$.
\end{itemize}

Nevertheless, the convergence rate for the MISE in Theorem~\ref
{thMISE} can be improved if we place some additional assumptions on the
kernel $\phi(x)$, assume that the filter density is sufficiently
smooth and increase the number of particles $N(k)$ in the filter. To be
specific, we consider the approximation of $p_t(x)$ alone for clarity
and make the following assumptions.
\begin{itemize}[a.3]
\item[a.1] The kernel $\phi(x)$ satisfies A.\ref{asPositivePdf}
($\phi>0$, $\int\phi(x)\,\mathrm{d}x = 1$), A.\ref{asSecondOrder} ($\int\| x
\|^2 \phi(x)\,\mathrm{d}x \le C_2 < \infty$ for some constant $C_2$) and it is a
bounded function. Additionally,
$
\int x_i \phi(x) \,\mathrm{d}x = 0
$
for every $i \in\{1, \ldots, d_x\}$.

\item[a.2] The filter density $p_t$ has continuous and bounded
derivatives up to order 2, that is, $D^\alpha p_t \in C_b(\Real
^{d_x})$ for every $\alpha$ such that $|\alpha| \le2$.

\item[a.3] The number of particles is selected to guarantee that
$N=N(k)\ge k^{2(d_x+2)}$.
\end{itemize}
Then we have the following refinement of Theorem~\ref{thMISE} for
$\alpha=0$.
%
\begin{Teorema} \label{thMISE2}
If both $p_t(x)$ and the kernel $\phi(x)$ have a compact support set
$\mK\subset\Real^{d_x}$ and assumptions \textup{a.1}, \textup{a.2} and \textup{a.3} hold, then
\[
\mathit{MISE} \equiv\int_{\mK} E \bigl[ \bigl(
p_t^k(x) - p_t(x) \bigr)^2
\bigr]\, \mathrm{d}x \le\frac{
C_{\mK,t}
}{
k^4
},
\]
where $C_{\mK,t} < \infty$ is constant w.r.t. $k$.
\end{Teorema}
%
\begin{pf} Recall the deterministic approximation $\tilde
p_t^k(x) = (\phi_k^x, \pi_t)$ of $p_t(x)$. Using the multivariate
version of Taylor's theorem, the difference $\tilde p_t^k(x) - p_t(x)$
can be written as
%
\begin{eqnarray}\label{eqMVTaylor}
\tilde p_t^k(x) - p_t(x) &=& \int
\phi_k(z) \bigl( p_t(x-z) - p_t(x) \bigr)
\,\mathrm{d}z
\nonumber
\\[-8pt]\\[-8pt]
&=& \int\phi_k(z) \biggl( \sum_{\alpha:|\alpha|=1}
D^\alpha p_t(x) (-z)^\alpha+ \sum
_{\alpha
:|\alpha|=2} R_\alpha(x-z) (-z)^\alpha \biggr)\,\mathrm{d}z,
\nonumber
\end{eqnarray}
where $z^\alpha= z_1^{\alpha_1}\cdots z_{d_x}^{\alpha_{d_x}}$ and
the remainder terms, $R_\alpha$, satisfy
%
\begin{equation}\label{eqResto}
\bigl|R_\alpha(x-z)\bigr| \le\max_{\alpha:|\alpha|=2} \bigl\| D^\alpha
p_t \bigr\| _\infty.
\end{equation}
From assumption a.1, $\int\phi_k(z) z_i \,\mathrm{d}z = 0$ for any $1 \le i \le
d_x$, hence
%
\begin{equation}\label{eqEliminaLineal}
\sum_{\alpha:|\alpha|=1} D^\alpha p_t(x)
\int\phi_k(z) (-z)^\alpha \,\mathrm{d}z = - \sum
_{i=1}^{d_x} \frac{\partial p_t}{\partial x_i}(x) \int
\phi_k(z) z_i \,\mathrm{d}z = 0.
\end{equation}
Substituting (\ref{eqEliminaLineal}) and (\ref{eqResto}) into (\ref
{eqMVTaylor}) and taking the absolute value of the difference yields
\[
\bigl| \tilde p_t^k(x) - p_t(x) \bigr| \le \Bigl(
\max_{\alpha:|\alpha|=2} \bigl\| D^\alpha p_t
\bigr\|_\infty \Bigr) \sum_{i,j \in\{1, \ldots, d_x\}} \int
\phi_k(z) | z_i z_j | \,\mathrm{d}z.
\]
However, $\max_{\alpha:|\alpha|=2} \| D^\alpha p_t \|_\infty<
\infty$ from assumption a.2 and $\int\phi_k(z) |z_i z_j| \,\mathrm{d}z \le
\frac{C_2}{k^2}$ from assumption a.1. Therefore, we obtain
%
\begin{equation}\label{eqNewSquareBound}
\bigl| \tilde p_t^k(x) - p_t(x) \bigr| \le
\frac{
C_{2,t}
}{
k^2
},
\end{equation}
where the constant $C_{2,t} = \max_{\alpha:|\alpha|=2} \| D^\alpha
p_t \|_\infty d_x^2 C_2 < \infty$ is independent of $k$. Combining
(\ref{eqNewSquareBound}) with the inequalities \eqref{eqApplicLemma}
(for $\alpha=0$) and \eqref{eqTriangleIneqDp} yields
\[
\bigl\| p_t^k(x) - p_t(x) \bigr\|_p
\le\frac{
\bar c_t k^{d_x} \| \phi\|_\infty
}{
\sqrt{N(k)}
} + \frac{
C_{2,t}
}{
k^2
},
\]
where $\bar c_t$ is constant w.r.t. to $k$ (and $N(k)$). From
assumption a.3, $N(k) \ge k^{2(d_x+2)}$, we arrive at
%
\begin{equation}\label{eqConvergence_p_k2}
\bigl\| p_t^k(x) - p_t(x) \bigr\|_p
\le\frac{
\bar C_{2,t}
}{
k^2
},
\end{equation}
where $\bar C_{2,t} = \bar c_t \| \phi\|_\infty+ C_{2,t} < \infty$
is a constant.

Similarly to the proof of Theorem~\ref{thMISE}, we choose $\mK=
[-M,+M] \times\cdots\times[-M,+M]$ without loss of generality. Using
the inequality \eqref{eqConvergence_p_k2} with $p=2$, we readily obtain
\[
\int_\mK E \bigl[ \bigl( p_t^k(x)
- p_t(x) \bigr)^2 \bigr] \,\mathrm{d}x \le\frac{
\bar C_{2,t}^2
}{
k^4
} \mL(
\mK) \le\frac{
C_{\mK,t}
}{
k^4
},
\]
where $\mL(\mK) = (2M)^{d_x}$ is the Lebesgue measure of $\mK$ and
$C_{\mK,t} = (2M)^{d_x} \bar C_{2,t}^2$ is constant w.r.t.~$k$.
\end{pf}

Note that the improvement of the convergence rate in Thorem \ref
{thMISE2} ($k^{-4}$ versus $k^{-2}$ in Theorem~\ref{thMISE}) is
obtained at the expense of slightly increasing the computational cost
of the particle filter ($N(k) \ge k^{2(d_x+2)}$ are needed, versus
$N(k) \ge k^{2(d_x+1)}$ in Theorem~\ref{thMISE} for $\alpha=\mathbf{0}$).

\subsection{Convergence with the number of particles $N$} \label
{ssConvergence_with_N}

The results stated in this section are given in terms of the index $k$
because this leads to concise expressions for the upper bounds of the
approximation errors and it also yields a straightforward connection
with classical kernel density estimation results in terms of the kernel
bandwidth (recall that $h=1/k$), as explicitly exploited in Section~\ref{ssISE}.

However, for the use of numerical schemes it may be useful to re-state,
or at least interpret, some of these results in terms of the number
particles, $N$, in the particle filter, since it is this parameter that
determines the computational complexity of the algorithm. Fortunately,
there is a straightforward (and deterministic) connection between the
values of $N$ and $k$, as already discussed in Section~\ref{ssSummary}. Here, we elaborate on this issue and provide versions of
Theorems \ref{thSupK} (uniform convergence over the state space), \ref
{thConvTV_1} (convergence in total variation distance of the continuous
particle approximation of $\pi_t$) and \ref{thISE} (convergence of
the ISE) with rates given in terms of $N$. They are given as
corollaries, as their proofs are straightforward from the original theorems.

Under the standard conditions in Remark~\ref{rmStandardAss}, the
number of particles $N$ and the inverse bandwidth $k$ satisfy the
inequality $N \ge k^{2(d_x+1)}$, and they are both integer quantities.
Therefore, given $N$, the largest inverse bandwidth that we can choose is
%
\begin{equation}\label{eqMapNK}
k(N) = \bigl\lfloor N^{\afrac{ 1 } { 2(d_x+1) } }\bigr
\rfloor,
\end{equation}
where $\lfloor z \rfloor= \sup\{ m \in\mathbb{Z} \dvtx  m \le z \}$. It
is apparent that $\lim_{N\rightarrow\infty} k(N) = \infty$. For conciseness
in the notation, let us write
\[
\hat p_t^N(x) = p_t^{k(N)}(x) =
\bigl(\phi_{k(N)}^x,\pi_t^N\bigr)
\]
for the kernel approximation of $p_t$ with $N$ particles determined by
the map \eqref{eqMapNK}. Similarly, consider the sequence of hypercubes
\[
\hat\mK_N = [-\hat M_N, +\hat M_N] \times
\cdots\times[-\hat M_N,+\hat M_N],
\]
where $\hat M_N = \frac{1}{2} k(N)^{\afrac{\beta}{d_x p}}$, with
positive constants $p \ge2$ and $0 \le\beta< 1$. This is the
counterpart of the sequence $\mK_k$ in Section~\ref{ssAS_convergence}. Then, the next result follows readily from Theorem~\ref{thSupK}.
%
\begin{Corolario} \label{coSupK}
If the standard conditions are satisfied, then
\[
\sup_{ x \in\hat\mK_N } \bigl\llvert \hat p_t^N(x)
- p_t(x) \bigr\rrvert \le\frac{
U^\varepsilon
}{
k(N)^{1-\varepsilon}
},
\]
where $k(N) = \lfloor
N^{\afrac{
1
}{
2(d_x+1)
}}
\rfloor$, $U^\varepsilon\ge0$ is an a.s. finite random variable and
$0<\varepsilon<1$ is a constant, both of them independent of $N$ and
$x$. In particular,
\[
\lim_{N\rightarrow\infty} \sup_{x \in\hat\mK_N} \bigl\llvert \hat
p_t^N(x) - p_t(x) \bigr\rrvert = 0\qquad
\mbox{a.s.}
\]
\end{Corolario}
%

If we write $\breve\pi_t^N(\mathrm{d}x) = \hat p_t^N(x)\,\mathrm{d}x$ for the continuous
approximation of $\pi_t(\mathrm{d}x)$ constructed from the approximate density
for a given number of particles $N$, then we have the corollary below,
that follows immediately from Theorem~\ref{thConvTV_1}.
%
\begin{Corolario} \label{coConvTV_1}
If the standard conditions are satisfied and $\pi_t(\hat\mK_N^\compl
) \le\frac{b}{2}k(N)^{-\gamma}$, where $k(N) = \lfloor
N^{\afrac{
1
}{
2(d_x+1)
}}
\rfloor$ and $b>0$ and $\gamma>0$ are constants independent of $N$, then
\[
\int\bigl\llvert \hat p_t^N(x) - p_t(x)
\bigr\rrvert \,\mathrm{d}x < \frac{
Q^\varepsilon
}{
k(N)^{\min\{ 1-\varepsilon, \gamma\}}
},
\]
where $Q^\varepsilon$ is an a.s. finite random variable and $0 <
\varepsilon< 1$ is a constant, both of them independent of $N$. In particular,
\[
\lim_{N \rightarrow\infty} \int\bigl\llvert \hat p_t^N(x)
- p_t(x) \bigr\rrvert \,\mathrm{d}x = 0 \qquad \mbox{a.s.}
\]
and, as a consequence,
\[
\lim_{N \rightarrow\infty} d_{\mathrm{TV}}\bigl(\breve\pi_t^N,
\pi_t\bigr) = 0 \qquad \mbox{a.s.}
\]
\end{Corolario}
%

We can also give a version of Theorem~\ref{thISE} with the error bound
explicitly given in terms of the number of particles, $N$. To write it,
let $\hat p_t^{\top,N}(x) = p_t^{\top,k(N)}(x) = I_{\hat\mK_N}(x)
\hat p_t^N(x)$ be the\vspace*{-2pt} truncation of the approximate density within the
compact hypercube $\hat\mK_N$. Then we have the corollary below,
which is proved in a trivial way from Theorem~\ref{thISE}.
%
%
\begin{Corolario} \label{coISE}
If the standard conditions are satisfied, $p_t \in B(\Real^{d_x})$ and
$\pi_t(\hat\mK_N^\compl) \le bk(N)^{-\gamma}$, where $k(N) =
\lfloor
N^{\afrac{
1
}{
2(d_x+1)
}}
\rfloor$ and $b>0$ and $\gamma>0$ are constants independent of $N$, then
\[
\mathit{ISE}\equiv\int \bigl( \hat p_t^{\top,N}(x) -
p_t(x) \bigr)^2 \,\mathrm{d}x \le\frac{
U^\varepsilon
}{
k(N)^{\min\{ 2-\varepsilon, \gamma\}}
},
\]
where $U^\varepsilon\ge0$ is an a.s. finite random variable,
independent of $N$, and $0<\varepsilon<2$ is an arbitrarily small
constant. In particular,
\[
\lim_{N\rightarrow\infty} \int \bigl( \hat p_t^{\top,N}(x)
- p_t(x) \bigr)^2 \,\mathrm{d}x = 0 \qquad \mbox{a.s.}
\]
\end{Corolario}
%

\subsection{A simple example} \label{ssExample}

There are several possible choices for the kernel function $\phi(x)$
that comply with assumptions A.\ref{asPositivePdf} and A.\ref
{asSecondOrder}. In particular, the standard multivariate Gaussian
density with unit covariance,
\[
\phi_\mathrm{G}(x) = \frac{
1
}{
(
2\uppi
)^{\sfrac{d_x}{2}}
} \exp \Biggl\{ -\frac{1}{2}
\sum_{j=1}^{d_x} x_j^2
\Biggr\},
\]
the $d_x$-dimensional Laplacian pdf,
\[
\phi_\mathrm{L}(x) = \biggl( \frac{1}{2b} \biggr)^{d_x} \exp
\Biggl\{ -\frac{1}{b} \sum_{j=1}^{d_x}
|x_j| \Biggr\},
\]
where $b=\sqrt{\frac{1}{2d_x}}$, and the Epanechnikov kernel $\phi
_\mathrm{E}(x)$ of equation \eqref{eqEpanechnikov} are densities with bounded second
order moment.

It is also straightforward to check assumption A.\ref{asDPhi_k_in_Cb}
for $\alpha=\mathbf{0}$ and $\alpha=\mathbf{1}$. In particular, for
$\alpha=\mathbf{0}$, it is apparent that $\phi_\mathrm{G},\phi_\mathrm{L},\phi_\mathrm{E} \in
C_b(\Real^{d_x})$. For $\alpha=\mathbf{1}$, the partial derivatives
of the Gaussian and Laplacian kernels yield
\begin{eqnarray}
D^\mathbf{1} \phi_\mathrm{G}(x) &=& \frac{
(-1)^{d_x}
}{
(2\uppi)^{\sfrac{d_x}{2}}
} \prod
_{l=1}^{d_x} x_l \exp \Biggl\{
-\frac{1}{2} \sum_{j=1}^{d_x}
x_j^2 \Biggr\}\quad  \mbox{and}
\nonumber
\\
D^\mathbf{1} \phi_\mathrm{L}(x) &=& \frac{
(-1)^{n^+}
}{
2^{d_x} b^{2d_x}
} \exp \Biggl
\{ -\frac{1}{b} \sum_{j=1}^{d_x}
|x_j| \Biggr\},\qquad  x \ne0,
\nonumber
\end{eqnarray}
respectively, where $n^+ = \llvert
\{
l \in\{1,\ldots,d_x\} \dvtx  x_l > 0
\}
\rrvert $ is the number of positive elements of $x \in\Real^{d_x}$.
It is not hard to verify that $D^\mathbf{1}\phi_\mathrm{G} \in C_b(\Real
^{d_x})$, while $D^\mathbf{1}\phi_\mathrm{L} \in B(\Real^{d_x})$. As for the
Epanechnikov kernel, it is easy to show that $D^\mathbf{1} \phi_\mathrm{E}(x)
= 0\ \forall x\in\Real^{d_x}$.


In the sequel, we consider a simple example consisting in the
approximation of a Gaussian filtering density using the Epanechnikov kernel.
%
%
\begin{Ejemplo} \label{exGauss0}
Consider the state-space system
%
\begin{equation}\label{eqModelLinearGauss}
p_0(x_0) = N(x_0; 0,
\mI_2),\qquad  X_t = AX_{t-1} + U_t,\qquad
Y_t = BX_t + V_t,\qquad  t=1,2,\ldots
,\end{equation}
where $N(x_0; 0,\mI_2)$ is the bivariate Gaussian pdf with mean $0$
and $2 \times2$ identity covariance matrix, $\mI_2$; the matrices
$A,B \in\Real^{2 \times2}$ are
\[
A=\lleft[ %
\begin{array} {c@{\quad}c} 0.50 &-0.35
\\
0.39 &-0.45
\\
\end{array} %
\rright],\qquad  B=\lleft[ %
\begin{array} {c@{\quad}c} 0.50
&0.30
\\
-0.80 &0.20
\\
\end{array} %
\rright],
\]
and $U_t$, $V_t$, $t = 1,2,\ldots\, $, are sequences of independent and
identically distributed $2 \times1$ Gaussian vectors with zero mean
and covariance $\mI_2$. For this class of (linear and Gaussian),
models the filtering pdf $p_t$, $t \ge1$, can be computed exactly
using the Kalman filter \cite{Kalman60} and, therefore, we have a
reference for comparison with the approximations $p_t^{k}$ produced by
the particle filter with $N(k) = k^{2(d_x+1)} = k^6$ samples.

For the simulation, we generated two sequences, $x_0, x_1, \ldots,
x_T$ and $y_1, \ldots, y_T$ for $T=50$, according to the model \eqref
{eqModelLinearGauss}. Then, using the fixed data $y_{1:T}$, we run a
Kalman filter to compute the Gaussian pdf $p_T(x) = N(x; \bar x_T,
\Sigma_T)$ exactly,\vadjust{\goodbreak} where $\bar x_T$ and $\Sigma_T$ are the posterior
mean and covariance at time $T$, respectively. For the same sequence
$y_{1:T}$, we run independent particle filters with various values of
$k$ and $N(k)=k^6$ particles each.\looseness=1
\begin{figure}
\begin{tabular}{cc}

\includegraphics{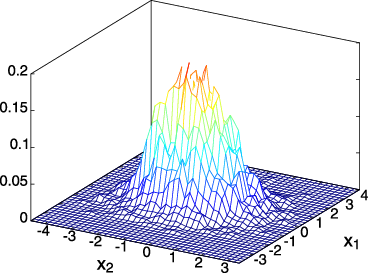}
& \includegraphics{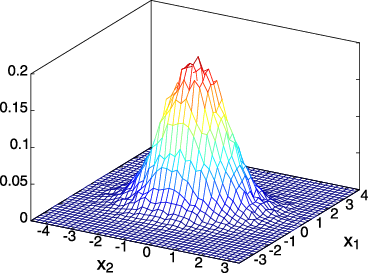}\\
\multicolumn{1}{c}{\scriptsize{(a) $k=4$ and $N(k)=4096$}}&
\multicolumn{1}{c}{\scriptsize{(b) $k=7$ and $N(k)=117\,649$}}\\[6pt]

\includegraphics{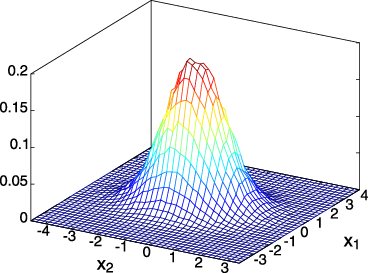}
&\includegraphics{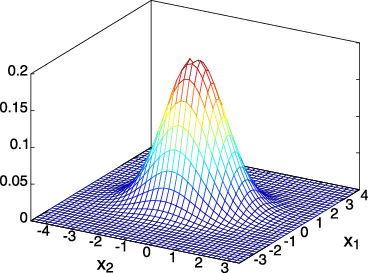}\\
\multicolumn{1}{c}{\scriptsize{(c) $k=10$ and $N(k)= 1\,000\,000$}}& \multicolumn{1}{c}{\scriptsize{(d) $p_T(x)$ (exact)}}
\end{tabular}
\caption{Plots (a)--(c) display the approximations of the filtering
density produced by the particle filter, $p_T^{k}(x)$, with an
increasing number of particles $N(k)=k^6$, and an Epanechnikov kernel,
$\phi_\mathrm{E}$. The true pdf, $p_T(x)$, is shown in plot (d) for comparison.
The plots correspond to the discrete grid $G_T$ in equation \protect\eqref{eqTheGrid}.}
\label{fBells}
\end{figure}

Figure~\ref{fBells} shows plots of the approximations $p_T^k(x)$ for
$k=4, 7, 10$ (constructed using the Epanechnikov kernel, $\phi_\mathrm{E}$) and
the true pdf $p_T(x)$. The plots are drawn from a regular grid of
points in $\Real^2$, namely
%
\begin{equation}\label{eqTheGrid}
x \in G_T = \bigl\{ (x_1,x_2)
\dvtx  x_1 = -2.92 + 0.2n,\ x_2 = -3.54 + 0.2n,\ 1 \le n \le42
\bigr\}
\end{equation}
(the offsets $-2.92$ and $-3.54$ correspond, approximately, to the true
posterior mean of $X_t$). We can see that there is an obvious error for
small $k$, while for $k=10$ the difference between $p_T(x)$ and its
approximation is negligible.
\end{Ejemplo}



%

\section{Applications} \label{sApplications}

We illustrate the use of the convergence results in Section~\ref{sApproximation} by addressing two application problems: the
computation of maximum a posteriori (MAP) estimators and the
approximation of functionals of the filtering density, $p_t$. All
through this section, we implicitly assume that the standard conditions
of Remark~\ref{rmStandardAss} are satisfied.

\subsection{MAP estimation}

We tackle the problem of approximating the maximum a posteriori (MAP)
estimator of the r.v. $X_t$. In particular, we address the numerical
search of elements of the set
%
\begin{equation}\label{eqSt}
S_t = \arg\max_{x\in\Real^{d_x}} p_t(x),
\end{equation}
where $s \in S_t$ if, and only if, $p_t(s)=\max_{x \in\Real^{d_x}}
p_t(x)$. Note that this is a relevant problem since MAP estimates are
often used, for example, in signal processing and engineering applications
(see, e.g., \cite{Gauvain92,Logothetis99,Frenkel99}), and the density
$p_t(x)$ cannot be analytically found in general.

Let
%
\begin{equation}\label{eqStk}
S_t^k = \arg\max_{x \in\Real^{d_x}}
p_t^{k}(x)
\end{equation}
be the set of MAP estimates for the approximation density $p_t^{k}(x)$
and note that $\hat x_k \in S_t^k$ if, and only if, $p_t^{k}(\hat x_k)
= \max_{x\in\Real^{d_x}} p_t^{k}(x)$. We can build a sequence of
approximate estimates, denoted $\{ \hat x_k \}_{k\ge1}$, by taking one
element from each set $S_t^k$, $k=1, 2, \ldots\, $, at time $t$. If $S_t$ is
nonempty, then any convergent subsequence of $\{ \hat x_k \}_{k \ge1}$
yields an arbitrarily accurate approximation of a true MAP estimator $s
\in S_t$, as stated below.
%
\begin{Teorema} \label{thMAP}
Assume that $S_t \ne\emptyset$ and take any convergent subsequence of
$\{ \hat x_k \}_{k\ge1}$, denoted $\{ \hat x_{k_i} \}_{i \ge1}$. Let
$\hat x = \lim_{i \rightarrow\infty} \hat x_{k_i}$ be the limit of such
subsequence. If $p_t \in C_b(\Real^{d_x})$, then $p_t(\hat x) = \max_{x\in\Real^{d_x}} p_t(x)$. In particular, if $p_t(x)$ has a unique
maximum, then $S_t$ is a singleton and $\lim_{i \rightarrow\infty}
\hat
x_{k_i} = \arg\max_{x\in\Real^{d_x}} p_t(x)$.
\end{Teorema}
%
\begin{pf} We prove the theorem by contradiction.
Specifically, assume that $p_t(\hat x) <\linebreak[4]  \max_{x\in\Real^{d_x}}
p_t(x)$. Then, choose some $s \in S_t$, so that $p_t(s)=\max_{x\in
\Real^{d_x}}p_t(x)$ and $p_t(\hat x) < p_t(s)$, and let
%
\begin{equation}\label{eq3eps}
\epsilon\triangleq\frac{
p_t(s)-p_t(\hat x)
}{
3
} > 0.
\end{equation}
Now, choose a compact subset $\mK\subset\Real^{d_x}$ that contains
$s$, $\{ \hat x_{k_i} \}_{i \ge1}$ and $\hat x$. From Remark~\ref
{rmFixedK}, $\lim_{k \rightarrow\infty} \sup_{x \in\mK} |
p_t^{k}(x)-p_t(x) | = 0$ a.s., hence there exists $m$ such that for all
$k \ge m$
%
\begin{equation}\label{eqEps1}
\sup_{x\in\mK} \bigl\llvert p_t^{k}(x)-p_t(x)
\bigr\rrvert < \epsilon.
\end{equation}
Moreover, since $p_t(x)$ is continuous at every point $x \in\mK$, we
can choose an integer $i_0$ such that for all $i \ge i_0$ we obtain
%
\begin{equation}\label{eqEps2}
\bigl| p_t(\hat x_{k_i})-p_t(\hat x) \bigr| <
\epsilon.
\end{equation}
Now, choose an index $\ell$ such that $\ell\ge i_0$ and $\ell\ge m$.
Then, for every $i, k_i > \ell$, we have
%
\begin{eqnarray}
p_t^{k_i}(\hat x_{k_i}) - p_t^{k_i}(s)
&=& \overbrace{ p_t^{k_i}(\hat x_{k_i}) -
p_t(\hat x_{k_i}) }^{< \epsilon} + \overbrace{
p_t(\hat x_{k_i}) - p_t(\hat x)
}^{ < \epsilon}
\nonumber
\\[-8pt]\\[-8pt]
&& {}+ \overbrace{ p_t(\hat x) - p_t(s)
}^{=-3\epsilon} + \overbrace{ p_t(s) - p_t^{k_i}(s)
}^{<\epsilon} < 0,\nonumber
\end{eqnarray}
where the first term on the right-hand side, $p_t^{k_i}(\hat x_{k_i}) -
p_t(\hat x_{k_i}) < \epsilon$, follows from inequality (\ref
{eqEps1}), the second term, $p_t(\hat x_{k_i}) - p_t(\hat x) < \epsilon
$, follows from inequality (\ref{eqEps2}), the third term, $p_t(\hat
x) - p_t(s) = -3\epsilon$, is due to the definition in (\ref{eq3eps})
and for the fourth term, $p_t(s) - p_t^{k_i}(s) <\epsilon$, is
obtained from the inequality (\ref{eqEps1}). Therefore, $\hat x_{k_i}
\notin\arg\max_{x \in\Real^{d_x}} p_t^{k_i}(x)$ and we arrive at a
contradiction. Hence, $p_t(\hat x)=\max_{x\in\Real^{d_x}} p_t(x)$.
\end{pf}
%
\begin{Nota}
Note that the whole sequence $\{ \hat x_k \}$ may not converge to a MAP
estimate since it may, for example, alternate between different
elements of $S_t$.
\end{Nota}
%

Many global optimization algorithms, such as simulated annealing \cite
{Corana87,Hedar06} or accelerated random search \cite{Appel03}, rely
only on the evaluation of the objective function and Theorem~\ref
{thMAP} justifies their use with the approximation $p_t^{k}(x)$. Many
other optimization procedures are based on the evaluation of derivates
of the objective function. For example, we may want to use a gradient
search method to find a local maximum of $p_t(x)$, that is, to find a
solution of the equation
%
\begin{equation}\label{eqGradient}
\nabla_x p_t(x) = 0,
\end{equation}
where $x=(x_1,\ldots,x_{d_x})$ and
\[
\nabla_x p_t(x) = \lleft[ %
\begin{array} {c} \frac{ \partial p_t }{ \partial x_1 }
\\
\vdots
\\\noalign{\vspace*{2pt}}
\frac{ \partial p_t }{ \partial x_{d_x} }
\\
\end{array} %
\rright](x) = \lleft[ %
\begin{array} {c} D^{\alpha_1} p_t
\\
\vdots
\\\noalign{\vspace*{2pt}}
D^{\alpha_{d_x}} p_t
\\
\end{array} %
\rright](x),
\]
with $\alpha_i = (0, \ldots, \overbrace{1}^{i\mathrm{th}}, \ldots,
0)$. Let $x^*$ be a solution of (\ref{eqGradient}), that is, $\nabla
_x p_t(x^*)=0$. Under the assumptions of Theorem~\ref
{thConvergenceDp}, for every $\epsilon>0$ there exists $k_\epsilon$
such that, $\forall k > k_\epsilon$,
\[
-\epsilon< D^{\alpha_i} p_t^k \bigl(x^*\bigr) <
\epsilon\qquad \mbox{a.s.}
\]
Therefore,
\[
\bigl\| \nabla_x p_t^{k}\bigl(x^*\bigr) \bigr\| =
\sqrt{ \sum_{i=1}^{d_x} \bigl( {
D^{\alpha_i} p_t }^{k}\bigl(x^*\bigr)
\bigr)^2 } < \epsilon\sqrt{d_x}\qquad  \forall k <
k_\epsilon,
\]
and, since $\epsilon$ can be chosen as small as we wish,
\[
\lim_{k\rightarrow\infty} \bigl\| \nabla_x p_t^{k}
\bigl(x^*\bigr) \bigr\| = 0 \qquad \mbox{a.s.},
\]
which justifies the application of a gradient search procedure using
the approximation of the filtering pdf.
%
\begin{Ejemplo}\label{exGradient}
We illustrate the application of a gradient search procedure using the
same example as in Section~\ref{ssExample}. In particular, we consider
the approximation of the maximum of the Gaussian filtering pdf
$p_T(x)$, $T=50$, using a steepest descent method. Given an
approximation $p_T^{k}(x)$ of the filtering density constructed with
the Gaussian kernel $\phi_\mathrm{G}$, we run the iterative algorithm
%
\begin{equation}\label{eqGradientApprox}
\hat x_T(i+1)^k = \hat x_T(i)^k
+ a  \nabla_x p_T^{k}(x) \bigl\vert
_{x = \hat x_T(i)^k}, \qquad i=0,1,2,\ldots
\end{equation}
with initial condition $\hat x_T(0)^k = (-2,-2)^\top$ and step-size
parameter $a=0.1$. This procedure yields a sequence of approximations
$\hat x_T(1)^k, \ldots, \hat x_T(i)^k, \ldots$ of the MAP estimator
$\hat x_T$. Since for the model of equation \eqref{eqModelLinearGauss} it
is possible to obtain $p_T(x)$ exactly, we have also run a steepest
descent search over the true filtering pdf, namely,
%
\begin{equation}\label{eqGradientTrue}
\hat x_T(i+1) = \hat x_T(i) + a
\nabla_x p_T(x) \rrvert _{x = \hat x_T(i)},\qquad  i=0,1,2,
\ldots,
\end{equation}
that generates the estimates $\hat x_T(1), \ldots, \hat x_T(i), \ldots
$ for the same initial condition and step size.
%
\begin{figure}\begin{tabular}{cc}

\includegraphics{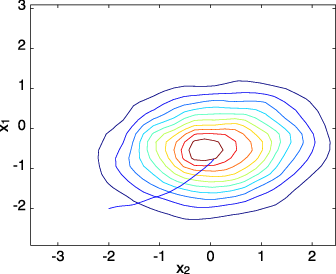}
& \includegraphics{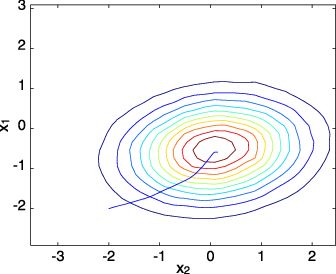}\\
\multicolumn{1}{c}{\scriptsize{(a) $k=5$, $N(k)=15\,625$}}&
\multicolumn{1}{c}{\scriptsize{(b) $k=9$, $N(k)=531\,441$}}\\[6pt]
\multicolumn{2}{c}{
\includegraphics{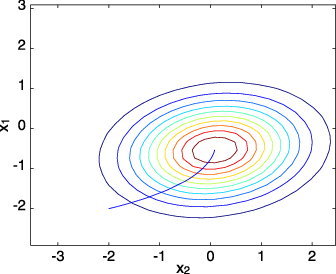}
}\\
\multicolumn{2}{c}{\scriptsize{(c) True pdf $p_t$}}
\end{tabular}
\caption{Trajectories of the gradient search algorithms. Plot (a)
shows the estimates produced by the gradient search algorithm of equation
\protect\eqref{eqGradientApprox} superimposed over a contour representation of
$p_T^{k}(x)$\vspace*{-1pt} for $k=5$. Plot (b) displays the estimates and contour
graph for $p_T^{k}(x)$ for $k=9$. Plot (c) shows the estimates produced
by the gradient search algorithm of equation \protect\eqref{eqGradientTrue}
superimposed over a contour representation of $p_T(x)$, for comparison.}
\label{fMAP}
\end{figure}

The results, using the same sequence of observations as in Section~\ref{ssExample},
are shown in Figure~\ref{fMAP}. Specifically, Figures~\ref{fMAP}(a) and \ref{fMAP}(b) show the trajectories described by
the estimates $\hat x_T(1)^k, \ldots, \hat x_T(i)^k, \ldots$
superimposed over the contour plots of the approximate pdf $p_T^{k}(x)$
for $k=5$ and $k=9$, respectively (and $N(k)=k^6$). For comparison,
Figure~\ref{fMAP}(c) depicts the sequence $\hat x_T(1), \ldots, \hat
x_T(i), \ldots$ obtained from the search over the true density
$p_T(x)$, together with the corresponding contour plot. We observe that
both the pdfs and the trajectories described by the search algorithms
are very similar.
\end{Ejemplo}

%

%

In practice, problem \eqref{eqStk} may turn out difficult to solve
because the approximation $p_t^k(x)$ can be rough, plagued with local
maxima, when the number of particles $N(k)$ is not sufficiently large
(see, e.g., Figure~\ref{fBells}(a)). In such cases, one may have to
resort to computationally expensive global optimization methods instead
of (simpler) gradient-based techniques. A computationally less
demanding approach consists in performing the search of the maximum of
$p_t^k(x)$ over the discrete set of particles $\Omega_t^{N(k)} = \{
x_t^{(n)} \}_{n=1, \ldots, N(k)}$ (where $N(k) \ge k^{2(2d_x+1)}$) instead
of over the complete (continuous) state space\footnote{This
alternative approximation of the MAP estimator of $X_t$ was pointed out
to us by one of the anonymous reviewers of the original manuscript.}
\cite{Abraham04}. To be specific, it is straightforward (e.g., by a
linear search) to obtain the set of particle values for which the
approximate density is maximum, namely
%
\begin{equation}\label{eqTildeS_k}
\tilde S_t^k = \arg\max_{ x \in\Omega_t^{N(k)} }
p_t^k(x).
\end{equation}
In the classical setup, when the target density is approximated using
i.i.d. samples drawn directly from the desired distribution, it can be
shown that the elements of $\tilde S_t^k$ become arbitrarily close to
the elements of $S_t^k$ as $k\rightarrow\infty$ (and, hence, as $N(k)
\rightarrow
\infty$) \cite{Abraham04}. The following theorem\vspace*{-1pt} yields a similar
asymptotic result when $\Omega_t^{N(k)}$ is generated by the standard
particle filter.
%
\begin{Teorema}
Assume that $S_t \neq\emptyset$ and $p_t \in C_b(\Real^{d_x})$. If
$s_t \in S_t$, $s_t^k \in S_t^k$ and $\tilde s_t^k \in\tilde S_t^k$, then,
%
\begin{equation}\label{eqVariousConv}
\lim_{k \rightarrow\infty} p_t\bigl(\tilde s_t^k
\bigr) = \lim_{k \rightarrow
\infty} p_t\bigl(s_t^k
\bigr) = p_t(s_t) 
\qquad \mbox{a.s.}
\end{equation}
\end{Teorema}
%
\begin{pf} Let us introduce the additional approximation
of the MAP estimator
\[
\check S_t^k = \arg\max_{x \in\Omega_t^{N(k)}}
p_t(x).
\]
The set $\check S_t^k$ cannot be computed in practice because $p_t(x)$
cannot be evaluated, but it will be auxiliary in proving that equation
\eqref{eqVariousConv} holds. Specifically, we first show (using an
argument taken from \cite{Miguez13b}) that the sequence $\{ p_t(\check
s_t^k)\dvtx  \check s_t^k \in\check S_t^k, k \ge1 \}$ converges to
$p_t(s_t)$ a.s. when $k \rightarrow\infty$. Then, we use the latter
result to
show that \eqref{eqVariousConv} holds.

We proceed to prove that $\lim_{k\rightarrow\infty} p_t(\check
s_t^k) =
p_t(s_t)$ a.s. Choose any MAP estimate $s_t \in S_t$ and define the
open ball
\[
B_m(s_t) = \biggl\{ x \in\Real^{d_x}\dvtx  \| x -
s_t \| < \frac{1}{m} \biggr\},
\]
where $m$ is a positive integer. From Proposition~\ref{propConvPF},
the integral $(I_{B_m(s_t)},\pi_t)$ (where $I_A(x)=1$ if $x \in A$ and
0 otherwise) can be approximated with asymptotically vanishing error.
Specifically, since $k \rightarrow\infty$ implies that $N(k)
\rightarrow\infty$, we have
\[
\lim_{k \rightarrow\infty} \bigl(I_{B_m(s_t)},\pi_t^{N(k)}
\bigr) = \lim_{k\rightarrow\infty
} \frac{
| B_m(s_t) \cap\Omega_t^{N(k)} |
}{
N(k)
} = (I_{B_m(s_t)},
\pi_t)\qquad  \mbox{a.s.},
\]
where $| B_m(s_t) \cap\Omega_t^{N(k)} |$ yields the number of
particles inside the ball $B_m(s_t)$. Since $p_t$ is continuous and
positive at $s_t$, then $(I_{B_m(s_t)},\pi_t) > 0$, hence
%
\begin{equation}\label{eqInfinitePoints}
\lim_{k\rightarrow\infty} \frac{| B_m(s_t) \cap\Omega_t^{N(k)}
|}{N(k)} > 0 \qquad \mbox{a.s.}
\end{equation}
for any integer $m$.

The inequality \eqref{eqInfinitePoints} means that the set $B_m(s_t)
\cap\Omega_t^{N(k)}$, consisting of particles which are ``close'' to
$s_t$ (namely, at a distance lesser than $1/m$), is asymptotically
nonempty, with probability 1, no matter how large we choose $m$.
Therefore, let us choose a point $s_t^{k,m} \in B_m(s_t) \cap\Omega
_t^{N(k)}$. Obviously, $p_t(s_t^{k,m}) \le p_t(s_t)$, but also
$p_t(s_t^{k,m}) \le p_t(\check s_t^k)$ by construction, hence
%
\begin{equation}\label{eqOrdenados}
p_t\bigl(s_t^{k,m}\bigr) \le
p_t\bigl(\check s_t^k\bigr) \le
p_t(s_t).
\end{equation}
Since $p_t$ is continuous at $s_t$, for any arbitrarily small $\epsilon
>0$ we can choose $m>0$ such that if $x \in B_m(s_t)$ then
$p_t(s_t)-p_t(x) < \epsilon$. However,\vadjust{\goodbreak} for every $m$ there exists
$k_m$ such that when $k>k_m$ the intersection $B_m(s_t) \cap\Omega
_t^{N(k)}$ is a.s. nonempty, hence there exists a particle $s_t^{k,m}
\in B_m(s_t) \cap\Omega_t^{N(k)}$ and the inequality \eqref
{eqOrdenados} yields $0 \le p_t(s_t)-p_t(\check s_t^k) \le p_t(s_t) -
p_t(s_t^{k,m}) < \epsilon$. Therefore,
%
\begin{equation}\label{eqLimitPart-1}
\lim_{k\rightarrow\infty} p_t\bigl(\check s_t^k
\bigr) = p_t(s_t) \qquad \mbox{a.s.}
\end{equation}

Now we prove the convergence of $p_t(\tilde s_t^k)$ and $p_t(s_t^k)$
toward $p_t(s_t) = \max_{x\in\Real^{d_x}} p_t(x)$. Consider first
the nonnegative difference
%
\begin{eqnarray}\label{eqLonga}
0 &\le& p_t(s_t)-p_t\bigl(\tilde
s_t^k\bigr) \nonumber\\
&=& \bigl( p_t(s_t)
- p_t\bigl(\check s_t^k\bigr) \bigr) +
\bigl( p_t\bigl(\check s_t^k\bigr) -
p_t^k\bigl(\check s_t^k\bigr)
\bigr)\\
&&{} + \bigl( p_t^k\bigl(\check s_t^k
\bigr) - p_t^k\bigl(\tilde s_t^k
\bigr) \bigr) + \bigl( p_t^k\bigl(\tilde
s_t^k\bigr) - p_t\bigl(\tilde
s_t^k\bigr) \bigr), \nonumber
\end{eqnarray}
where the inequality follows from the definition of $S_t$, and let us
look into each term on the right-hand side of \eqref{eqLonga} separately.

Choose any arbitrarily small $\epsilon>0$. From \eqref
{eqLimitPart-1}, there exists $k_1$ such that for every $k>k_1$,
%
\begin{equation}\label{eqItem-1}
0 \le p_t(s_t) - p_t\bigl(\check
s_t^k\bigr) < \frac{\epsilon}{6}.
\end{equation}
Let us now select, without loss of generality, a compact set $\mK
\supset S_t \cup S_t^k \cup\tilde S_t^k \cup\check S_t^k$. From
Remark~\ref{rmFixedK},
\[
\bigl| p_t\bigl(\check s_t^k\bigr) -
p_t^k\bigl(\check s_t^k\bigr)
\bigr| \le\sup_{x \in\mK} \bigl| p_t(x) - p_t^k(x)
\bigr| \le \frac{
\tilde U^\varepsilon
}{
k^{1-\varepsilon}
},
\]
where $\tilde U^\varepsilon$ is an a.s. finite random variable and $0
< \varepsilon< 1$ is arbitrary but constant. Hence, there exists $k_2$
such that, for every $k>k_2$,
%
\begin{equation}\label{eqItem-2}
-\frac{\epsilon}{6} < p_t\bigl(\check s_t^k
\bigr) - p_t^k\bigl(\check s_t^k
\bigr) < \frac
{\epsilon}{6}.
\end{equation}
By the same argument, there is some $k_3$ such that, for every $k>k_3$,
%
\begin{equation}\label{eqItem-3}
-\frac{\epsilon}{6} < p_t\bigl(\tilde s_t^k
\bigr) - p_t^k\bigl(\tilde s_t^k
\bigr) < \frac
{\epsilon}{6}.
\end{equation}
Since, by construction,
%
\begin{equation}\label{eqItem-4}
p_t^k\bigl(\check s_t^k
\bigr) - p_t^k\bigl(\tilde s_t^k
\bigr) \le0,
\end{equation}
substituting \eqref{eqItem-1}--\eqref{eqItem-4} into the inequality
\eqref{eqLonga} and solving for $p_t^k(\check s_t^k) - p_t^k(\tilde
s_t^k)$ yields
%
\begin{equation}\label{eqItem-5}
0 \ge p_t^k\bigl(\check s_t^k
\bigr) - p_t^k\bigl(\tilde s_t^k
\bigr) > -\frac{\epsilon}{2}
\end{equation}
for every $k > \max\{ k_1, k_2, k_3 \}$. However,
%
\begin{eqnarray}\label{eqLonga2}
\bigl\llvert p_t(s_t)-p_t\bigl(\tilde
s_t^k\bigr) \bigr\rrvert &\le&\bigl\llvert
p_t(s_t) - p_t\bigl(\check
s_t^k\bigr) \bigr\rrvert + \bigl\llvert
p_t\bigl(\check s_t^k\bigr) -
p_t^k\bigl(\check s_t^k\bigr)
\bigr\rrvert \nonumber\\[-8pt]\\[-8pt]
&&{}+ \bigl\llvert p_t^k\bigl(\check
s_t^k\bigr) - p_t^k\bigl(
\tilde s_t^k\bigr) \bigr\rrvert + \bigl\llvert
p_t^k\bigl(\tilde s_t^k\bigr)
- p_t\bigl(\tilde s_t^k\bigr) \bigr
\rrvert \nonumber
\end{eqnarray}
and substituting \eqref{eqItem-1}--\eqref{eqItem-3} and \eqref
{eqItem-5} into \eqref{eqLonga2} yields $| p_t(s_t)-p_t(\tilde s_t^k)
| < \epsilon$ a.s. for every $k > \max\{ k_1, k_2, k_3 \}$, hence
\[
\lim_{k\rightarrow\infty} p_t\bigl(\tilde s_t^k
\bigr) = p_t(s_t) \qquad \mbox{a.s.}
\]

A similar argument proves the convergence of $p_t(s_t^k) \rightarrow p_t(s_t)$.
In particular, if we choose a compact set $\mK\supset S_t \cup S_t^k$
we can again apply Remark~\ref{rmFixedK} to show that, for any
$\epsilon> 0$ there exists $k_4$ such that, for every $k > k_4$,
%
\begin{equation}\label{eqAnotherItem-1}
-\frac{\epsilon}{4} < p_t(s_t) -
p_t^k(s_t) < \frac{\epsilon}{4}
\end{equation}
and there exists $k_5$ such that, for every $k>k_5$,
%
\begin{equation}\label{eqAnotherItem-2}
-\frac{\epsilon}{4} < p_t^k\bigl(s_t^k
\bigr) - p_t\bigl(s_t^k\bigr) <
\frac{\epsilon}{4}.
\end{equation}
However,
%
\begin{equation}\label{eqOutraLonga}
0 \le p_t(s_t)-p_t
\bigl(s_t^k\bigr) = \bigl( p_t(s_t)
- p_t^k(s_t) \bigr) + \bigl(
p_t^k(s_t) - p_t^k
\bigl(s_t^k\bigr) \bigr) + \bigl( p_t^k
\bigl(s_t^k\bigr) - p_t
\bigl(s_t^k\bigr) \bigr)
\end{equation}
and, since $p_t^k(s_t) - p_t^k(s_t^k) \le0$ by definition of $S_t^k$,
substituting \eqref{eqAnotherItem-1} and \eqref{eqAnotherItem-2} into
\eqref{eqOutraLonga} and solving for $p_t^k(s_t) - p_t^k(s_t^k)$ yields
%
\begin{equation}\label{eqAnotherItem-3}
-\frac{\epsilon}{2} < p_t^k(s_t) -
p_t^k\bigl(s_t^k\bigr) \le0
\end{equation}
for every $k > \max\{ k_4, k_5 \}$. Finally, since
\[
\bigl\llvert p_t(s_t)-p_t^k
\bigl(s_t^k\bigr) \bigr\rrvert \le\bigl\llvert
p_t(s_t) - p_t^k
\bigl(s_t\bigr) \bigr\rrvert + \bigl\llvert
p_t^k(s_t) - p_t^k
\bigl(s_t^k\bigr) \bigr\rrvert ,
\]
we obtain that $| p_t(s_t)-p_t^k(s_t^k) | \le\epsilon$ for every $k >
\max\{ k_4, k_5 \}$, hence
\[
\lim_{k\rightarrow\infty} p_t^k
\bigl(s_t^k\bigr) = p_t(s_t)
\qquad \mbox{a.s.}
\]
\upqed
\end{pf}
%
\begin{Ejemplo}\label{exMAP2}
We consider, again, the Gaussian density $p_T(x)$, with $T=50$, of
Examples \ref{exGauss0} and \ref{exGradient} in order to compare
numerically the approximations $p_T(s_t^k)$ and $p_T(\tilde s_T^k)$
with the true maximum $p_T(s_T)$. The results are displayed in Table~\ref{tMAP}, which shows the maximum $p_T(s_T) = \max_{x \in\Real
^{d_x}} p_T(x)$ and the differences $p_T(s_T) - p_T(\tilde s_t^k)$ and
$p_T(s_T) - p_T(s_T^k)$ for $k=5$ ($N=15\,625$) and $k=9$ ($N=531\,441$).
\end{Ejemplo}
%
\begin{table}
\tablewidth=250pt
\tabcolsep=0pt
\caption{Approximation of the maximum posterior density $p_T(s_T) =
\max_{x\in\Real^{d_x}} p_T(x)$ by way of equations \protect\eqref{eqStk} and
\protect\eqref{eqTildeS_k} ($p_T(s_T^k)$ and $p_T(\tilde s_T^k)$,
respectively). The approximate MAP estimate $s_T^k$ has been computed
via the gradient search method of equation \protect\eqref{eqGradientApprox}}
\label{tMAP}
\begin{tabular*}{250pt}{@{\extracolsep{\fill}}llll@{}}
\hline
&$p_T(s_T)$ &$p_T(s_T)-p_T(s_T^k)$ &$p_T(s_T)-p_T(\tilde s_T^k)$\\
\hline
$k=5$ &0.201937 &0.005090 &0.004500\\
$k=9$ &0.201937 &0.001030 &0.002679\\
\hline
\end{tabular*}
\end{table}

%

\subsection{Functionals of $p_t$}

The result of Theorem~\ref{thConvTV_1} allows us to construct
(rigorous) approximations of functionals of the form $(f \circ p_t, \pi
_t)$, where $\circ$ denotes composition and $f$ is a
Lipschitz-continuous and bounded real function. In order to provide
rates for the convergence of the particle-kernel approximations $(f
\circ p_t^k, \pi_t^{N(k)})$, we again work with the sequence of
hypercubes $\mK_k = [-M_k,M_k] \times\cdots\times[-M_k,M_k] \subset
\Real^{d_x}$ where $M_k = \frac{1}{2} k^{\afrac{\beta}{d_x p}}$ and $0
< \beta< 1$, $p > 3$ are constants with respect to $k$. Specifically,
we have the following result.
%
\begin{Teorema} \label{thFunctionals}
Choose any bounded, Lipschitz continuous function $f$, that is, $f \in
B(\Real)$ and $\forall x,y \in\Real$
\[
\bigl\llvert f(x)-f(y) \bigr\rrvert \le c_f\llvert x-y \rrvert
\]
for some finite constant $c_f>0$. If $p_t \in B(\Real^{d_x})$ and $\pi
_t(\mK_k^\compl) \le\frac{b}{2}k^{-\gamma}$ for some constants
$\gamma,b>0$, then
%
\begin{equation}\label{eqConvergence_of_functionals}
\bigl\llvert \bigl(f \circ p_t^{k},
\pi_t^{N(k)}\bigr) - (f \circ p_t,
\pi_t) \bigr\rrvert \le\frac{ Q_f^\varepsilon}{ k^{ \min\{1-\varepsilon
,\gamma\} } } ,
\end{equation}
where $0 < \varepsilon< 1$ is an arbitrarily small constant and
$Q_f^\varepsilon$ is an a.s. finite random variable independent of
$k$. In particular,
%
\[
\lim_{k\rightarrow\infty} \bigl\llvert \bigl(f \circ p_t^{k},
\pi_t^{N(k)}\bigr) - (f \circ p_t,
\pi_t) \bigr\rrvert = 0\qquad  \mbox{a.s.}
\]
\end{Teorema}
%
\begin{pf} Consider first the absolute difference
%
\begin{eqnarray}\label{eqToBeBounded}
\bigl\llvert \bigl( f \circ p_t^{k},
\pi_t \bigr) - ( f \circ p_t, \pi_t )
\bigr\rrvert &=& \biggl\llvert \int \bigl[ \bigl( f\circ p_t^{k}
\bigr) (x) - ( f\circ p_t ) (x) \bigr] p_t(x)\,\mathrm{d}x
\biggr\rrvert
\nonumber
\\[-8pt]\\[-8pt]
&\le& \int\bigl\llvert \bigl( f\circ p_t^{k} \bigr) (x)
- ( f\circ p_t ) (x) \bigr\rrvert p_t(x)\,\mathrm{d}x,
\nonumber
\end{eqnarray}
where the inequality holds because $p_t(x) \ge0$. Using the Lipschitz
continuity of $f$ in the integral of equation \eqref{eqToBeBounded} yields
%
\begin{eqnarray}\label{eqError-I}
\bigl\llvert \bigl( f \circ p_t^{k},
\pi_t \bigr) - ( f \circ p_t, \pi_t )
\bigr\rrvert &\le& c_f \int\bigl\llvert p_t^{k}(x)
- p_t(x) \bigr\rrvert p_t(x)\,\mathrm{d}x
\nonumber
\\[-8pt]\\[-8pt]
&\le& c_f \| p_t \|_\infty\int\bigl\llvert
p_t^{k}(x) - p_t(x) \bigr\rrvert \,\mathrm{d}x,
\nonumber
\end{eqnarray}
where the second inequality follows from the assumption $p_t \in
B(\Real^{d_x})$ (hence $\| p_t \|_\infty< \infty$). Equation (\ref
{eqError-I}) together with Theorem~\ref{thConvTV_1} readily yields
%
\begin{equation}\label{eqPart-I}
\bigl\llvert \bigl( f \circ p_t^{k},
\pi_t \bigr) - ( f \circ p_t, \pi_t )
\bigr\rrvert \le\frac{
c_f\| p_t \|_\infty Q^\varepsilon
}{
k^{\min\{1-\varepsilon,\gamma\}}
} ,
\end{equation}
where $0<\varepsilon<1$ is a constant and $Q^\varepsilon$ is an a.s.
finite random variable. 

As a second step, consider the difference $\llvert
( f \circ p_t^{k}, \pi_t^{N(k)} ) - ( f \circ p_t^{k}, \pi_t )
\rrvert $. Since $f \in B(\Real)$, it follows that $\| f \circ p_t^{k}
\|_\infty\le\| f \|_\infty$ independently of $k$ and an application
of Proposition~\ref{propConvPF} yields
%
\[
E \bigl[ \bigl\llvert \bigl( f \circ p_t^{k},
\pi_t^{N(k)} \bigr) - \bigl( f \circ p_t^{k},
\pi_t \bigr) \bigr\rrvert ^q \bigr] \le
\frac{
c_t^q \| f \|^q_\infty
}{
N(k)^{\sfrac{q}{2}}
} \le\frac{
c_t^q\| f \|^q_\infty
}{
k^{q(2d_x+1)}
},
\]
where $q \ge1$ and the second inequality holds because $N(k) \ge
k^{2(d_x+| \mathbf{1} |+1)}$. Using Lemma~\ref{lmU} with $c = c_t^q\|
f \|^q_\infty$ and $\nu= 0$ (note that $q(2d_x+1) \ge2$ for any
$q,d_x \ge1$), we readily obtain the convergence rate for the absolute
error, that is,
%
\begin{equation}\label{eqPart-II}
\bigl\llvert \bigl( f \circ p_t^{k},
\pi_t^{N(k)} \bigr) - \bigl( f \circ p_t^{k},
\pi_t \bigr) \bigr\rrvert \le\frac{U^\varepsilon}{k^{1-\varepsilon}},
\end{equation}
where $0<\varepsilon<1$ is an arbitrarily small constant and
$U^\varepsilon\ge0$ is an a.s. finite random variable.

To conclude, consider the triangle inequality
%
\begin{eqnarray}\label{eqAnotherTriangle}
\bigl\llvert \bigl(f \circ p_t^{k},
\pi_t^{N(k)}\bigr) - (f \circ p_t,
\pi_t) \bigr\rrvert &\le& \bigl\llvert \bigl(f \circ
p_t^{k}, \pi_t^{N(k)}\bigr) -
\bigl(f \circ p_t^{k}, \pi_t\bigr) \bigr
\rrvert
\nonumber
\\[-8pt]\\[-8pt]
&&{} + \bigl\llvert \bigl(f \circ p_t^{k},
\pi_t\bigr) - (f \circ p_t, \pi_t) \bigr
\rrvert . \nonumber
\end{eqnarray}
Substituting \eqref{eqPart-I} and \eqref{eqPart-II} into \eqref
{eqAnotherTriangle} yields
%
\begin{equation}
\bigl\llvert \bigl(f \circ p_t^{k},
\pi_t^{N(k)}\bigr) - (f \circ p_t,
\pi_t) \bigr\rrvert \le\frac{
U^\varepsilon
}{
k^{1-\varepsilon}
} + \frac{
c_f \| p_t \|_\infty Q^\varepsilon
}{
k^{\min\{1-\varepsilon,\gamma\}}
} \le
\frac{
Q_f^\varepsilon
}{
k^{\min\{1-\varepsilon,\gamma\}}
},
\end{equation}
where the random variable $Q_f^\varepsilon= U^\varepsilon+ c_f \| p_t
\|_\infty Q^\varepsilon\ge0$ is a.s. finite and independent of~$k$.
\end{pf}

In statistical signal processing, machine learning and information
theory it is often of interest to evaluate the Shannon entropy of a
probability measure $\pi$ \cite{Beirlant97,Hulle05,Nilsson07}.
Assuming that $\pi$ has a density $p$ w.r.t. the Lebesgue measure, the
entropy of the probability distribution is
\[
\mH(\pi) = -(\log p,\pi) = -\int_\mS p(x) \log \bigl[ p(x)
\bigr] \,\mathrm{d}x,
\]
where $\mS$ is the support of $p$. In the case of the filtering
measure $\pi_t$, it is natural to think of a particle approximation of
the entropy $\mH(\pi_t)$ constructed as
\[
\mH(\pi_t)^{k} = -\bigl( \log p_t^{k},
\pi_t^{N(k)} \bigr) = -\frac{1}{N(k)} \sum
_{n=1}^{N(k)} \log p_t^{k}
\bigl( x_t^{(n)} \bigr).
\]
Unfortunately, the $\log$ function is neither bounded nor Lipschitz
continuous and, therefore, Theorem~\ref{thFunctionals} does not
guarantee the convergence $\mH(\pi_t)^{k} \rightarrow\mH(\pi_t)$.
Such a
result, however, can be obtained, with a more specific argument, if we
assume the support of the density $p_t$ to be compact.
%
\begin{Teorema} \label{thEntropy}
Let the sequence of observations $Y_{1:T}=y_{1:T}$ (for some large but
finite $T$) be fixed and assume that $g_t^{y_t}$ is positive and
bounded and $\log p_t \in\mathsf{F}_T^4$ for $1 \le t \le T$.
%
%
If there exists a compact set $\mS\subset\Real^{d_x}$ such that
$\int_\mS p_t(x)\,\mathrm{d}x = 1$ and $\inf_{x \in\mS} p_t(x)>0$, then
\[
\lim_{k\rightarrow\infty} \bigl\llvert \mH(\pi_t)^{k}
- \mH(\pi_t) \bigr\rrvert = 0 \qquad \mbox{a.s.}
\]
\end{Teorema}
%
\begin{pf} We apply the triangle inequality to obtain
%
\begin{eqnarray}\label{eqTriangle_entropy}
\bigl\llvert \bigl(-\log p_t^{k},\pi_t^{N(k)}
\bigr) - (-\log p_t,\pi_t) \bigr\rrvert &\le& \bigl
\llvert \bigl(-\log p_t^{k},\pi_t^{N(k)}
\bigr) - \bigl(-\log p_t,\pi_t^{N(k)}\bigr)
\bigr\rrvert
\nonumber
\\[-8pt]\\[-8pt]
&&{} + \bigl\llvert \bigl(-\log p_t,\pi_t^{N(k)}
\bigr) - (-\log p_t,\pi_t) \bigr\rrvert
\nonumber
\end{eqnarray}
and then analyze the two terms on the right-hand side of \eqref
{eqTriangle_entropy}.

The first one can be expanded to yield
%
\begin{eqnarray}\label{eqExpanded}
\bigl\llvert \bigl(-\log p_t^{k},\pi_t^{N(k)}
\bigr) - \bigl(-\log p_t,\pi_t^{N(k)}\bigr)
\bigr\rrvert 
&=& \Biggl\llvert
\frac{1}{N(k)} \sum_{i=1}^{N(k)} \log
\frac{
p_t(x_t^{(i)})
}{
p_t^{k}(x_t^{(i)})
} \Biggr\rrvert
\nonumber
\\[-8pt]\\[-8pt]
&\le& \frac{1}{N(k)} \sum_{i=1}^{N(k)}
\biggl\llvert \log\frac{
p_t(x_t^{(i)})
}{
p_t^{k}(x_t^{(i)})
} \biggr\rrvert . \nonumber
\end{eqnarray}
The logarithm of a ratio $x/y$ can be upper bounded as
%
\begin{equation}\label{eqBound_for_log}
\log\frac{x}{y} \le\frac{
\max\{ x,y \}
}{
\min\{ x,y \}
} - 1,
\end{equation}
hence applying \eqref{eqBound_for_log} into \eqref{eqExpanded} we
arrive at
%
\begin{equation}\label{eqExpanded2}
\bigl\llvert \bigl(-\log p_t^{k},\pi_t^{N(k)}
\bigr) - \bigl(-\log p_t,\pi_t^{N(k)}\bigr)
\bigr\rrvert \le\frac{1}{N(k)} \sum_{i=1}^{N(k)}
\biggl\llvert \frac{
\max \{
p_t^{k}(x_t^{(i)}), p_t(x_t^{(i)})
\}
}{
\min \{
p_t^{k}(x_t^{(i)}), p_t(x_t^{(i)})
\}
} - 1 \biggr\rrvert .
\end{equation}
However, from Theorem~\ref{thSupK} and Remark~\ref{rmFixedK},
\[
\lim_{k\rightarrow\infty} p_t^{k}(x) /
p_t(x) = \lim_{N\rightarrow
\infty} p_t(x) /
p_t^{k}(x) = 1\qquad  \mbox{a.s.}
\]
for every $x \in\mS$. Moreover, since we have assumed $\inf_{x \in
\mS} p_t(x) > 0$, it follows that for any $\epsilon>0$ there exists
$k_\epsilon$ independent of $x$ such that, for all $k>k_\epsilon$,
%
\begin{equation}\label{eqConvergence_of_ratio}
\frac{
\max \{
p_t^{k}(x_t^{(i)}), p_t(x_t^{(i)})
\}
}{
\min \{
p_t^{k}(x_t^{(i)}), p_t(x_t^{(i)})
\}
} \le1 + \epsilon.
\end{equation}
Substituting \eqref{eqConvergence_of_ratio} into \eqref{eqExpanded2}
yields, for all $k > k_\epsilon$,
\[
\bigl\llvert \bigl(-\log p_t^{k},\pi_t^{N(k)}
\bigr) - \bigl(-\log p_t,\pi_t^{N(k)}\bigr)
\bigr\rrvert \le\epsilon\qquad \mbox{a.s.}
\]
Since $\epsilon$ can be as small as we wish,
%
\begin{equation}\label{eqFirstTermConverges}
\lim_{k\rightarrow\infty} \bigl\llvert \bigl(-\log p_t^{k},
\pi_t^{N(k)}\bigr) - \bigl(-\log p_t,
\pi_t^{N(k)}\bigr) \bigr\rrvert = 0 \qquad \mbox{a.s.}
\end{equation}

The second term in \eqref{eqTriangle_entropy} converges to 0 because
of Proposition~\ref{propConvPF}, part (b), that is,
%
\begin{equation}\label{eqSecondTermConverges}
\lim_{k\rightarrow\infty} \bigl\llvert \bigl(-\log p_t,
\pi_t^{N(k)}\bigr) - (-\log p_t,
\pi_t) \bigr\rrvert = 0 \qquad \mbox{a.s.}
\end{equation}
and taking together equations \eqref{eqFirstTermConverges}, \eqref
{eqSecondTermConverges} and \eqref{eqTriangle_entropy} we arrive at
\[
\lim_{k\rightarrow\infty} \bigl\llvert \bigl(-\log p_t^{k},
\pi_t^{N(k)}\bigr) - (-\log p_t,
\pi_t) \bigr\rrvert = 0 \qquad \mbox{a.s.}
\]
\upqed
\end{pf}
%
\begin{Ejemplo}
We continue to use the model of Section~\ref{ssExample} to numerically
illustrate the particle approximation of $\mH(\pi_t)$. Since the
densities $p_t$ for this example are Gaussian with known covariance
matrices $\Sigma_t$, $t=1,2,\ldots\, $, we can compute their associated
Shannon entropies exactly, namely
\[
\mH(\pi_t) = \frac{1}{2}\log \bigl( (2\uppi
\mathrm{e})^{d_x} | \Sigma_t | \bigr),
\]
where $| \Sigma_t |$ is the determinant of matrix $\Sigma_t$. Taking
$t=T=50$ and using the same sequence of observations $y_{1:50}$ as in
Section~\ref{ssExample}, the resulting entropy is $\mH(\pi
_T)=2.5998$ nats.

Let us point out that, obviously, the Gaussian distribution has an
infinite support and, therefore, the convergence result of Theorem~\ref
{thEntropy} cannot be rigorously applied. However, the Gaussian pdf is
light-tailed and, as can be observed from Figure~\ref{fBells}(d), it
can be truncated within a compact (rectangular) support and still yield
a faithful representation of the original distribution.

Table~\ref{tEntropies} displays the empirical mean and standard
deviation of the absolute error
$
\llvert
\mH(\pi_T) - \mH(\pi_T)^{k}
\rrvert
$
obtained through computer simulations for $N(k) = k^6$ and $k=3,4,5$.
To be specific, we carried out 30 independent simulation runs for each
value of $k$. We observe how both the mean error and its standard
deviation reduce quickly as $k$ is increased.
\end{Ejemplo}

%

\section{Summary} \label{sConclusions}

We have addressed the approximation of the sequence of filtering pdfs
of a Markov state-space model using a particle
filter. The numerical technique is conceptually simple. We collect the
$N$ particles generated by the sequential Monte Carlo algorithm and
approximate the desired density as the sum of $N$ scaled kernel
functions located at the particle positions. The main contribution of
the paper is the analysis of the convergence of such particle-kernel
approximations. In particular, we have first proved the point-wise
convergence of the approximation of the filtering density and its
derivatives as the number of particles is increased and the kernel
bandwidth is correspondingly decreased. Explicit convergence rates are
provided and they are sufficient to prove that the approximation errors
vanish a.s. Under mild additional assumptions on the chosen kernel, it
is possible to extend the latter result to prove that the approximation
error converges uniformly on the support of the filtering density
(rather than point-wise) and a.s. to 0. We have also found an explicit
convergence rate for the supremum of the approximation error. The
analysis establishes a connection between the complexity of the
particle filter and the bandwidth of the kernel function used for
estimating the filtering pdf. For a given number of particles $N$, this
relationship yields an optimal value of the bandwidth.

\begin{table}
\tablewidth=260pt
\tabcolsep=0pt
\caption{Empirical mean and standard deviation of the
entropy-approximation error, $\llvert  \mH(\pi_T) - \mH(\pi_T)^{k}
\rrvert $, averaged over 30 independent simulations. The entropies are
evaluated in nats} \label{tEntropies}
\begin{tabular*}{250pt}{@{\extracolsep{\fill}}llll@{}}
\hline
&$k=3$ &$k=4$ &$k=5$\\
\hline
Mean &0.0616 &0.0370 &0.0128\\
Std. &0.0453 &0.0249 &0.0091\\
\hline
\end{tabular*}
\end{table}

The uniform approximation result has a number of applications. We have
first exploited it to prove the convergence, in total variation
distance, of the continuous measure generated by the estimated density
toward the true filtering measure. In a similar vein, we have also
shown that the MISE of the sequence of approximate densities converges
(quadratically with the kernel bandwidth) toward 0 when the state space
is compact. For a truncated version of the density approximation, the
(random) ISE is also shown to converge a.s. toward 0 without assuming
compactness of the support. Although the convergence rate found for the
ISE is only quadratic (versus fourth order for the asymptotic
approximation of the MISE in classical kernel density estimation
theory), one should be aware that all the results obtained in this
paper remain valid whenever the density estimator is obtained by
smoothing a discrete random measure $\pi_t^N$ that is ``good enough''
to estimate integrals of bounded functions in such a way that the $L_p$
norms of the approximation error convergence as $\frac{ c }{ \sqrt{N}
}$ (in particular, we do not require to have samples from the target
density $p_t$). As a consequence, the results obtained here can be
applied to, for example, kernel density estimators built from importance
samples as in \cite{West93}, or to the analysis of bootstrapped
estimators as considered in \cite{Hall01}. Convergence of the MISE
with the fourth power of the bandwidth (i.e., the same as for the AMISE
in the classical theory) can also be obtained at the expense of a
slight increase in the computational load of the particle filter and
some additional assumptions on the kernel function and the smoothness
of the filter density.

We have also proved that the maxima of the approximate filtering
density converge a.s. toward the true ones. Therefore, MAP estimation
of the state at time $t$ can be carried out using, for example,
gradient search methods on the approximate filtering pdf. We remark
that it is sound to apply such methods on the approximate function,
since we have proved convergence also for its derivatives. The last
application we consider is the approximation of functionals of the
filtering pdf. We provide a general result that guarantees the
convergence of the particle-kernel approximations for general bounded
and Lipschitz continuous functionals of the filtering density. Finally,
we prove that it is also possible to use the proposed constructs to
approximate the Shannon entropy of densities with a compact support. In
order to arrive at this result, we have also proved the convergence of
the particle filter approximations of integrals of unbounded test
functions under very mild assumptions (essentially, the integrability
of the function up to fourth order). This is a departure from most
existing approaches, which assume bounded test functions.


\begin{appendix}
\section{Proof of Proposition \texorpdfstring{\protect\ref{propConvPF}}{2.1}} \label{apProofLemma 1}

Part (a) of Proposition~\ref{propConvPF} is a straightforward
consequence of \cite{Miguez13b}, Lemma~1, hence we focus here on part
(b). We start with the following lemma, which is used as an auxiliary
result in the proof of the proposition.
%
\begin{Lema} \label{lmIntegrable}
Let $\{ \theta_n; n=1, \ldots, N \}$ be a set of random variables,
assumed centered and i.i.d. conditionally on some $\sigma$-algebra
$\mG$. If $E[ \theta_n^4 ] < \infty$, $n = 1, \ldots, N$, then
%
\begin{equation}\label{eqIneqMZ}
E \Biggl[ \Biggl( \frac{1}{N}\sum_{n=1}^N
\theta_n \Biggr)^4 \Biggr] \le\frac{
cE [ \theta_1^4  ]
}{
N^2
},
\end{equation}
where $c$ is a constant independent of $N$. In particular,
%
\[
\lim_{N\rightarrow\infty} \Biggl\llvert \frac{1}{N}\sum
_{n=1}^N \theta _n \Biggr\rrvert = 0
\qquad \mbox{a.s.} 
\]
\end{Lema}
%
%
\begin{pf} Conditional on $\mG$, the variables are zero mean and
independent, hence it is straightforward to show that
%
\begin{equation}\label{eq1stDecompE4}
E \Biggl[ \Biggl( \frac{1}{N} \sum_{n=1}^N
\theta_n \Biggr)^4 \Bigl| \mG \Biggr] = \frac{1}{N^4}
\sum_{n=1}^N E\bigl[ \theta_n^4
| \mG\bigr] + \frac{
6
}{
N^4
} \sum_{1 \le j < k \le N} E\bigl[
\theta_j^2 | \mG\bigr] E\bigl[ \theta_k^2
| \mG\bigr].
\end{equation}
Since the conditional (on $\mG$) distributions of the $\theta_n$'s
are identical, we can rewrite \eqref{eq1stDecompE4} in terms of
$E[\theta_1^4|\mG]$ and $E[\theta_1^2|\mG]$ alone, namely
%
\begin{equation}\label{eq2ndDecompE4}
E \Biggl[ \Biggl( \frac{1}{N} \sum_{n=1}^N
\theta_n \Biggr)^4 \Bigl| \mG \Biggr] = \frac{1}{N^3} E
\bigl[ \theta_1^4 | \mG\bigr] + \frac{
6(N-1)
}{
N^3
}
E^2\bigl[ \theta_1^2 | \mG\bigr].
\end{equation}
However, $E[\theta_n^4|\mG] \ge E^2[\theta_n^2|\mG]$ (from Jensen's
inequality), which readily yields the bound
%
\begin{equation}\label{eqFinalBound}
E \Biggl[ \Biggl( \frac{1}{N} \sum_{n=1}^N
\theta_n \Biggr)^4 \Bigl| \mG \Biggr] \le E\bigl[
\theta_1^4 | \mG\bigr] \frac{
1+6(N-1)
}{
N^3
} \le
\frac{
c E[\theta_1^4 | \mG]
}{
N^2
}
\end{equation}
for any constant $c \ge6$. Taking unconditional expectations on the
right-hand and left-hand sides of \eqref{eqFinalBound} leads to the
desired inequality \eqref{eqIneqMZ}.

Finally, a standard Borel--Cantelli argument yields
$
\lim_{N\rightarrow\infty} E [
\llvert
\frac{1}{N} \sum_{n=1}^N \theta_n
\rrvert
] = 0
$ a.s.
\end{pf}


Lemma~\ref{lmIntegrable} enables us to prove the convergence of
particle approximations, $\lim_{N \rightarrow\infty} (f,\allowbreak \pi_t^N)
\rightarrow(f,\pi
_t)$ a.s., when $f \in\mathsf{F}_T^4$. We follow an induction
argument to prove the latter result. In particular, let
%
\begin{equation}\label{eqDefBarPi}
\bar\pi_{t}^N(\mathrm{d}x) = \sum_{n=1}^N
w_t^{(n)} \delta_{\bar x_t^{(n)}}(\mathrm{d}x)
\end{equation}
be the random measure resulting from assigning importance weights
$w_t^{(n)} = {
g_t^{y_t}(\bar x_t^{(n)})
}/\linebreak[4] ({
\sum_{k=1}^N g_t^{y_t}(\bar x_t^{(k)}))
}$ to the particles $\bar x_t^{(n)}$. We prove that:
\begin{enumerate}[2.]
\item[1.] At time $t=0$, $\lim_{N\rightarrow\infty} | (f,\pi
_0^N)-(f,\pi_0) |
= 0$ a.s. and, at time $t=1$,
\[
\lim_{N\rightarrow\infty} \bigl\llvert \bigl(f,\bar\pi_1^N
\bigr) - (f,\pi_1) \bigr\rrvert = 0 \qquad \mbox{a.s.}
\]

\item[2.] If $\lim_{N\rightarrow\infty} | (f,\bar\pi_t^N) - (f,\pi_t)
| = 0$
a.s. at some time $1\le t < T$, then
\[
\lim_{N\rightarrow\infty} \bigl\llvert \bigl(f,\bar\pi_{t+1}^N
\bigr) - (f,\pi_{t+1}) \bigr\rrvert = 0 \qquad \mbox{a.s.}
\]
\end{enumerate}
In the induction step, it is explicitly shown that $\lim_{N\rightarrow
\infty
} | (f,\bar\pi_t^N) - (f,\pi_t) | = 0$ a.s. implies $\lim_{N\rightarrow
\infty} | (f,\pi_t^N) - (f,\pi_t) | = 0$. The latter is the result
in the statement of Proposition~\ref{propConvPF}, hence the argument
above yields a complete proof.

\subsection{Base case (\texorpdfstring{$t\le1$}{t<=1})}

For $t=0$, the samples $x_0^{(n)}$, $n=1,\ldots,N$, are i.i.d. with common
distribution $\pi_0$, hence the approximation $\pi_0^N$ is
constructed as $\pi_0^N(\mathrm{d}x) = \frac{1}{N}\sum_{n=1}^N \delta
_{x_0^{(n)}}(\mathrm{d}x)$. Consider the random\vspace*{-3pt} variables $\theta_{n,0} =
f(x_0^{(n)}) - (f,\pi_0)$. It is apparent that they are i.i.d. and $E[
\theta_{n,0} ] = 0$. Also
\begin{eqnarray}
E\bigl[ \theta_{n,0}^4 \bigr] &=& E \bigl[ \bigl( f\bigl(
x_0^{(n)} \bigr) - (f,\pi_0)
\bigr)^4 \bigr]
\nonumber
\\
&\le& 2^4 \bigl( E\bigl[ f\bigl(x_0^{(n)}
\bigr)^4 \bigr] + (f,\pi_0)^4 \bigr) <
\infty,
\nonumber
\end{eqnarray}
where the last inequality follows from $E[ f(x_0^{(n)})^4] = (f^4,\pi
_0)$ and the assumption $f \in\mathsf{F}_T^4$. Since the variables
$\{ \theta_{n,0}; n=1, \ldots, N \}$ satisfy the assumptions of Lemma~\ref{lmIntegrable}, we readily obtain
%
\begin{equation}\label{eqBaseCase-1}
\lim_{N\rightarrow\infty} \bigl\llvert \bigl(f,\pi_0^N
\bigr) - (f,\pi_0) \bigr\rrvert = \lim_{N\rightarrow\infty}
\Biggl\llvert \frac{1}{N} \sum_{n=1}^N
f\bigl( x_0^{(n)} \bigr) - (f,\pi_0) \Biggr
\rrvert = 0 \qquad \mbox{a.s.}
\end{equation}

Consider the (predictive) measure $\xi_{t+1} = \tau_{t+1}\pi_t$ as
defined in equation \eqref{eqSuccessive}. After the sampling step, the
particle filter produces a random approximation
\[
\xi_{t+1}^N(\mathrm{d}x)= \frac{1}{N}\sum
_{n=1}^N \delta_{\bar x_{t+1}^{(n)}}(\mathrm{d}x),
\]
that is, $\xi_{t+1}^N = \tau_{t+1}\pi_t^N$. We look now into the
approximation error $| (f,\xi_1^N) - (f,\xi_1) |$.

Let $\mF_t=\sigma (x_{0:t}^{(n)},\bar x_{1:t}^{(n)}\dvtx  1 \le n \le
N  )$ denote the $\sigma$-algebra generated by the random\vspace*{-1pt}
variables $x_s^{(n)}$ and $\bar x_r^{(n)}$ for $n=1, \ldots, N$, $s=0,
\ldots, t$ and $r=1, \ldots, t$. It is apparent\vspace*{-1pt} that $E[ f(\bar
x_{t+1}^{(n)}) | \mF_t ] = \tau_{t+1}(f)(x_t^{(n)})$, hence the
conditional mean an of $(f,\xi_{t+1}^N)$ is
\begin{eqnarray}
E\bigl[ \bigl(f,\xi_{t+1}^N\bigr) | \mF_t
\bigr] &=& \frac{1}{N} \sum E\bigl[ f\bigl(\bar
x_{t+1}^{(n)}\bigr) | \mF_t \bigr]
\nonumber
\\
&=& \frac{1}{N} \sum_{n=1}^N
\tau_{t+1}(f) \bigl(x_t^{(n)}\bigr) = \bigl(f,
\tau _{t+1}\pi_t^N\bigr),
\nonumber
\end{eqnarray}
and it is natural to use the triangular inequality
%
\begin{equation}\label{eqTriangle_xi}
\bigl\llvert \bigl(f,\xi_{t+1}^N\bigr) - (f,
\xi_{t+1}) \bigr\rrvert \le\bigl\llvert \bigl(f, \xi_{t+1}^N
\bigr) - \bigl(f,\tau_{t+1}\pi_t^N\bigr) \bigr
\rrvert + \bigl\llvert \bigl(f,\tau_{t+1}\pi_t^N
\bigr) - (f,\tau_{t+1}\pi_t) \bigr\rrvert
\end{equation}
to analyze the approximation error $| (f,\xi_{t+1}^N) - (f,\xi_{t+1}) |$.

We proceed with the case $t=0$. If we look into the  second term on
the right-hand side of \eqref{eqTriangle_xi} we observe that
\[
(f,\tau_{1}\pi_0) = \int\tau_{1}(f)
(x_0) \pi_0(\mathrm{d}x_0) = \bigl(\tau
_{1}(f),\pi_0\bigr)
\]
and, similarly, $(f,\tau_{1}\pi_0^N) = (\tau_{t+1}(f),\pi_t^N)$.
Since $f \in\mathsf{F}_T^4$, it follows that $\tau_{t+1}(f) \in
\mathsf{F}_T^4$ as well and, from equation \eqref{eqBaseCase-1}, we
readily see that
%
\begin{equation}\label{eqTriangle_xi_2nd_part}
\lim_{N\rightarrow\infty} \bigl\llvert \bigl(f,\tau_{1}
\pi_0^N\bigr) - (f,\tau_{1}
\pi_0) \bigr\rrvert = 0 \qquad \mbox{a.s.}
\end{equation}

In order to analyze the first term on the right-hand side of \eqref
{eqTriangle_xi} (for $t=0$), let us choose the random variables
\[
\bar\theta_{1,n} = f\bigl(\bar x_1^{(n)}\bigr)
- \tau_1(f) \bigl(x_0^{(n)}\bigr),\qquad  n = 1,
\ldots, N.
\]
It is straightforward to check that they are unconditionally i.i.d. %
To see that they are centered, simply observe that, for every $n = 1,
\ldots, N$,
\[
E[\bar\theta_{1,n}] = E \bigl[ E[ \bar\theta_{1,n} |
\mF_0 ] \bigr] = 0,
\]
since $E[ \bar\theta_{1,n} | \mF_0 ] = E[ f(\bar x_1^{(n)}) | \mF_0
] - \tau_1(f)(x_0^{(n)})$ and $E[ f(\bar x_1^{(n)}) | \mF_0 ] = \tau
_1(f)(x_0^{(n)})$. Moreover,
\[
E \bigl[ \bar\theta_{1,n}^4 | \mF_0 \bigr]
\le2^4 \bigl( \tau_1\bigl(f^4\bigr)
\bigl(x_0^{(n)}\bigr) + \bigl(\tau_1(f)^4,
\pi_0\bigr) \bigr),
\]
hence
\[
E \bigl[ \bar\theta_{1,n}^4 \bigr] = E \bigl[ E \bigl[
\bar\theta_{1,n}^4 | \mF_0 \bigr] \bigr]
\le2^4 \bigl( \bigl(\tau_1\bigl(f^4\bigr),
\pi_0\bigr) + \bigl(\tau_1(f)^4,
\pi_0\bigr) \bigr) < \infty,
\]
where the last inequality holds because $f \in\mathsf{F}_T^4$ and, as
a consequence $(\tau_1(f)^4,\pi_0) \le(\tau_1(f^4), \pi_0) <
\infty$. As the variables $\bar\theta_{1,n}=f(\bar x_1^{(n)})-\tau
_1(f)(x_0^{(n)})$ satisfy the assumptions of Lemma~\ref{lmIntegrable},
we readily obtain that
%
\begin{equation}\label{eqTriangle_xi_1st_part_t1}
\lim_{N\rightarrow\infty} \bigl\llvert \bigl(f,\xi_1^N
\bigr) - \bigl(f,\tau_1\pi_0^N\bigr) \bigr
\rrvert = 0 \qquad \mbox{a.s.}
\end{equation}
Taking together \eqref{eqTriangle_xi}, \eqref
{eqTriangle_xi_1st_part_t1} and \eqref{eqTriangle_xi_2nd_part}, we obtain
%
\begin{equation}\label{eqBaseCase-2}
\lim_{N\rightarrow\infty} \bigl\llvert \bigl(f,\xi_1^N
\bigr) - (f,\xi_1) \bigr\rrvert = 0\qquad  \mbox{a.s.}
\end{equation}

After computing the importance weights, we obtain the random measure
$\bar\pi_1^N(\mathrm{d}x)$ defined in equation \eqref{eqDefBarPi} (with $t=1$).
Integrals w.r.t. $\bar\pi_{1}^N$ can be written in terms of
$g_{1}^{y_{1}}$ and $\xi_{1}^N$, namely
\[
\bigl(f,\bar\pi_{1}^N\bigr) = \sum
_{n=1}^N \frac{
g_1^{y_1}(\bar x_1^{(n)})
}{
\sum_{k=1}^N g_1^{y_1}(\bar x_1^{(k)})
} f\bigl(\bar
x_1^{(n)}\bigr) = \frac{
(fg_{1}^{y_{1}},\xi_{1}^N)
}{
(g_{1}^{y_{1}},\xi_{1}^N)
}.
\]
Similarly, for $\pi_{1}$ and $\xi_{1}$ Bayes' theorem yields
\[
(f,\pi_{1}) = \frac{
(fg_{1}^{y_{1}},\xi_{1})
}{
(g_{1}^{y_{1}},\xi_{1})
},
\]
and the difference $(f,\bar\pi_{1}^N) - (f,\pi_{1})$ can be written as
%
\begin{eqnarray}\label{eqSubtraction}
&&\bigl(f,\bar\pi_{1}^N\bigr) - (f,\pi_{1})\nonumber\\
&&\quad = \frac{
(fg_{1}^{y_{1}},\xi_{1}^N)
}{
(g_{1}^{y_{1}},\xi_{1}^N)
} - \frac{
(fg_{1}^{y_{1}},\xi_{1})
}{
(g_{1}^{y_{1}},\xi_{1})
} \pm\frac{
(fg_{1}^{y_{1}},\xi_{1}^N)
}{
(g_{1}^{y_{1}},\xi_{1})
}
\\
&&\quad = \frac{
(fg_{1}^{y_{1}},\xi_{1}^N) - (fg_{1}^{y_{1}},\xi_{1})
}{
(g_{1}^{y_{1}},\xi_{1})
} 
+ \frac{
(fg_{1}^{y_{1}},\xi_{1}^N)
}{
(g_{1}^{y_{1}},\xi_{1}^N)
} \times
\frac{
(g_{1}^{y_{1}},\xi_{1}) - (g_{1},\xi_{1}^N)
}{
(g_{1}^{y_{1}},\xi_{1})
}.\nonumber
\end{eqnarray}
Note that, since $g_{1}^{y_{1}} \in B(\Real^{d_x})$ (hence
$g_{1}^{y_{1}} \in\mathsf{F}_T^4$), equation \eqref{eqBaseCase-2} yields
%
\begin{equation}\label{eqZeroNumerator1}
\lim_{N\rightarrow\infty} \bigl\llvert \bigl(g_{1}^{y_{1}},
\xi_{1}\bigr) - \bigl(g_{1},\xi_{1}^N
\bigr) \bigr\rrvert = 0\qquad  \mbox{a.s.}
\end{equation}
and, since we have assumed
%
\begin{equation}\label{eqBoundedDenom1}
\bigl(g_{1}^{y_{1}},\xi_{1}\bigr)>0,
\end{equation}
it follows that
%
\begin{equation}\label{eqBoundedDenom2}
\lim_{N\rightarrow\infty} \bigl(g_{1},\xi_{1}^N
\bigr) > 0\qquad  \mbox{a.s.}
\end{equation}
Also as a consequence of the likelihood $g_1^{y_1}$ being bounded, we
have $fg_{1}^{y_{1}} \in\mathsf{F}_T^4$ and \eqref{eqBaseCase-2}
guarantees that
%
\begin{equation}\label{eqZeroNumerator2}
\lim_{N\rightarrow\infty} \bigl\llvert \bigl(fg_{1}^{y_{1}},
\xi_{1}\bigr) - \bigl(fg_{1},\xi_{1}^N
\bigr) \bigr\rrvert = 0\qquad  \mbox{a.s.}
\end{equation}
Taking equations \eqref{eqBoundedDenom1} and \eqref{eqBoundedDenom2}
together, we deduce that
$\lim_{N\rightarrow\infty} \frac{
(fg_1^{y_1},\xi_1^N)
}{
(g_1^{y_1},\xi_1^N)
} < \infty$ a.s.\vspace*{-2pt} This result, combined with \eqref{eqZeroNumerator1}
and \eqref{eqZeroNumerator2} yields
%
\begin{equation}\label{eqBar_pi_converges}
\lim_{N\rightarrow\infty} \bigl\llvert \bigl(f,\bar\pi_{1}^N
\bigr) - (f,\pi_{1}) \bigr\rrvert = 0 \qquad \mbox{a.s.}
\end{equation}

\subsection{Induction step ($t>1$)}

Let us assume that $\lim_{N\rightarrow\infty} \llvert
(f,\bar\pi_{t}^N) - (f,\pi_{t})
\rrvert  = 0 \mbox{ a.s.}$ for some $1 \le t < T$.

We first show that the difference $|(f,\pi_{t}^N)-(f,\bar\pi_{t})|$
converges to 0 a.s. Recall that $\pi_t^N$ is obtained from the
equally-weighted particles after the resampling step. Let us introduce
the generated $\sigma$-algebra $\bar\mF_t = \sigma (
x_{0:t-1}^{(n)}, \bar x_{1:t}^{(n)};\ 1 \le n \le N
)$ and the random variables
\[
\theta_{t,n} = f\bigl(x_{t}^{(n)}\bigr) -
\bigl(f,\bar\pi_{t}^N\bigr),\qquad  n= 1, \ldots, N.
\]
It is simple to check that
\[
E \bigl[ f\bigl(x_{t}^{(n)}\bigr) | \bar\mF_{t}
\bigr] = \bigl(f,\bar\pi_{t}^N\bigr),\qquad  n=1, \ldots, N,
\]
hence $\theta_{t,n}$, $n = 1, \ldots, N$, are centered (and obviously
i.i.d.) given $\bar\mF_{t}$. Also, $E[\theta_{t,n}^4|\bar\mF
_{t}]<\infty$. Specifically, $(f^4,\bar\pi_{t}^N)$ is $\bar\mF
_{t}$-measurable hence
\[
E \bigl[ f\bigl(x_{t}^{(n)}\bigr)^4 | \bar
\mF_{t} \bigr] = \bigl(f^4,\bar\pi_{t}^N
\bigr),
\]
and, from the induction hypothesis and $f \in\mathsf{F}_T^4$,
\[
\lim_{N\rightarrow\infty} \bigl(f^4,\bar\pi_{t}^N
\bigr) = \bigl(f^4, \pi_{t}\bigr) < \infty \qquad \mbox{a.s.}
\]
Therefore, for sufficiently large $N$, $(f^4,\bar\pi_{t}^N) < \infty
$. However,
\[
E\bigl[\theta_{t,n}^4|\bar\mF_{t}\bigr]
\le2^4 \bigl( \bigl(f^4,\bar\pi_t^N
\bigr) + \bigl(f,\bar\pi_t^N\bigr)^4 \bigr),
\]
hence $E[ \theta_{t,n}^4 ] = E [ E[ \theta_{t,n}^4 | \bar\mF_t
]  ] < \infty$ for sufficiently large $N$. As the conditions of
Lemma~\ref{lmIntegrable} are satisfied for $\theta_{t,n}$, $n=1, \ldots,
N$, we obtain
%
\begin{equation}\label{eqPi_converges_to_bar_pi}
\lim_{N\rightarrow\infty} \bigl\llvert \bigl(f,\pi_{t}^N
\bigr) - \bigl(f,\bar\pi_{t}^N\bigr) \bigr\rrvert = \lim
_{N\rightarrow\infty} \Biggl\llvert \frac{1}{N}\sum
_{n=1}^N \theta_{t,n} \Biggr\rrvert = 0
\qquad \mbox{a.s.}
\end{equation}
Finally, taking together the induction hypothesis, \eqref
{eqPi_converges_to_bar_pi} and the triangle inequality
\[
\bigl\llvert \bigl(f,\pi_{t}^N\bigr) - (f,
\pi_{t}) \bigr\rrvert \le\bigl\llvert \bigl(f,\pi_{t}^N
\bigr) - \bigl(f,\bar\pi_{t}^N\bigr) \bigr\rrvert + \bigl
\llvert \bigl(f,\bar\pi_{t}^N\bigr) - (f,
\pi_{t}) \bigr\rrvert ,
\]
readily yields
%
\begin{equation}\label{eqPi_N_converges}
\lim_{N\rightarrow\infty} \bigl\llvert \bigl(f,\pi_{t}^N
\bigr) - (f,\pi_{t}) \bigr\rrvert = 0 \qquad \mbox{a.s.}
\end{equation}

Next, we prove that $\lim_{N\rightarrow\infty} | (f,\xi_{t+1}^N) -
(f,\xi
_{t+1})| = 0$ a.s. We resort again to the triangular inequality \eqref
{eqTriangle_xi}. Since $(f,\tau_{t+1}\pi_t) = (\tau_{t+1}(f),\pi
_t)$, $(f,\tau_{t+1}\pi_t^N) = (\tau_{t+1}(f),\pi_t^N)$ and $\tau
_{t+1}(f) \in\mathsf{F}_T^4$, it is a straightforward consequence of
\eqref{eqPi_N_converges} that
%
\begin{equation}\label{eqInduction_2nd_term}
\lim_{N\rightarrow\infty} \bigl\llvert \bigl(f,\tau_{t+1}
\pi_t^N\bigr) - (f,\tau_{t+1}
\pi_t) \bigr\rrvert = 0 \qquad \mbox{a.s.}
\end{equation}

To show that the error $(f,\xi_{t+1}^N) - (f,\tau_{t+1}\pi_t^N)$
also vanishes, let us choose the random variables $\bar\theta_{t+1,n}
= f(\bar x_{t+1}^{(n)}) - \tau_{t+1}(f)(x_t^{(n)})$. These are
i.i.d.\footnote{In particular, note that
\begin{itemize}
\item$\{ \bar x_{t+1}^{(n)} \}_{n=1, \ldots, N}$ can be viewed as i.i.d.
samples from the probability measure $m_{t+1}(\mathrm{d}x) = \sum_{n=1}^N
w_t^{(n)}\tau_{t+1}(\mathrm{d}x|\allowbreak \bar x_t^{(n)})$, where both $w_t^{(n)}$ and
$\bar x_t^{(n)}$, $1\le n\le N$, are $\bar\mF_t$-measurable, and
\item$\{ x_t^{(n)} \}_{n=1, \ldots, N}$ are also i.i.d. given $\bar\mF_t$.
\end{itemize}
} conditional on $\bar\mF_t$. They are also centered, since $E[ \bar
\theta_{t+1,n} | \mF_t ] = E[ f(\bar x_{t+1}^{(n)}) | \mF_t ] - \tau
_{t+1}(f)(x_t^{(n)})= 0$ and $\bar\mF_t \subset\mF_t$. Therefore,
we just need to check that $E[ \bar\theta_{t+1,n}^4 ] < \infty$ in
order to apply Lemma~\ref{lmIntegrable}. We note that
\[
E\bigl[ \bar\theta_{t+1,n}^4 | \mF_t \bigr]
\le2^4 \bigl( \tau_{t+1}\bigl(f^4\bigr)
\bigl(x_t^{(n)}\bigr) + \tau_{t+1}(f)^4
\bigl(x_t^{(n)}\bigr) \bigr)
\]
and then readily obtain
%
\begin{eqnarray}\label{eqChain3}
E\bigl[ \bar\theta_{t+1,n}^4 | \bar\mF_t
\bigr] &=& E \bigl[ E[\bar\theta _{t+1,n} | \mF_t] | \bar
\mF_t \bigr]
\nonumber
\\[-8pt]\\[-8pt]
&\le& 2^4 \bigl( \bigl(\tau_{t+1}\bigl(f^4
\bigr), \bar\pi_t^N\bigr) + \bigl(\tau_{t+1}(f)^4,
\bar\pi_t^N\bigr) \bigr). \nonumber
\end{eqnarray}
However, $\tau_{t+1}(f)^4 \le\tau_{t+1}(f^4)$ and $f \in\mathsf
{F}_T^4$ implies that $(\tau_{t+1}(f^4), \pi_t) < \infty$. Moreover,
the induction hypothesis yields
%
\[
\lim_{N\rightarrow\infty} \bigl\llvert \bigl(\tau_{t+1}
\bigl(f^4\bigr),\bar\pi_t^N\bigr) - \bigl(
\tau_{t+1}\bigl(f^4\bigr),\bar\pi_t\bigr)
\bigr\rrvert = 0 \qquad \mbox{a.s.}
\]
hence $(\tau_{t+1}(f)^4,\bar\pi_t^N) \le(\tau_{t+1}(f^4),\bar\pi
_t^N) < \infty$ for sufficiently large $N$. As a consequence,
$E[ \bar\theta_{t+1,n}^4 ] < \infty$ and the conditions of
Lemma~\ref{lmIntegrable} are satisfied for the random\vspace*{-1pt} variables $\bar\theta
_{t+1,n}$ and the $\sigma$-algebra $\bar\mF_t$. In particular, we have
%
\begin{equation}\label{eqTriangle_xi_1st_part_t+1}
\lim_{N\rightarrow\infty} \bigl\llvert \bigl(f,\xi_{t+1}^N
\bigr) - \bigl(f,\tau_{t+1}\pi_t^N\bigr) \bigr
\rrvert = \lim_{N\rightarrow\infty} \Biggl\llvert \frac{1}{N} \sum
_{n=1}^N \bar\theta_{t+1,n} \Biggr
\rrvert = 0 \qquad \mbox{a.s.}
\end{equation}
Taking together \eqref{eqTriangle_xi_1st_part_t+1}, \eqref
{eqInduction_2nd_term} and \eqref{eqTriangle_xi} yields
%
\begin{equation}\label{eqConvergence_of_xi}
\lim_{N\rightarrow\infty} \bigl\llvert \bigl(f,\xi_{t+1}^N
\bigr) - (f,\xi_{t+1}) \bigr\rrvert = 0 \qquad \mbox{a.s.}
\end{equation}

Finally, given \eqref{eqConvergence_of_xi}, it is straightforward to
prove that $\lim_{N\rightarrow\infty} | (f,\bar\pi_{t+1}^N) -
(f,\pi
_{t+1})| = 0$ a.s. using the same argument as in the base case for
$\bar\pi_1^N$.

\section{Proof of Lemma \texorpdfstring{\protect\ref{lmU}}{4.1}} \label{apLema_U}

Choose a constant $\beta$ such that $\nu< \beta< p-1$ and define
\[
U^{\beta,p} = \sum_{m=1}^\infty
m^{p-1-\beta} \bigl(\theta^m\bigr)^p.
\]
The random variable $U^{\beta,p}$ is obviously nonnegative and,
additionally, it has a finite mean, $E[U^{\beta,p}]<\infty$. Indeed,
from Fatou's lemma
\begin{eqnarray}
E \bigl[ U^{\beta,p} \bigr] &\le& \sum_{m=1}^\infty
m^{p-1-\beta} E\bigl[\bigl(\theta^m\bigr)^p
\bigr]%
\le c \sum
_{m=1}^\infty m^{-1-\beta+\nu},
\nonumber
\end{eqnarray}
where the second inequality follows from equation \eqref{eqAss_lema_U}.
Since $\beta-\nu>0$, it follows that $\sum_{m=1}^\infty m^{-1-(\beta
-\nu)} < \infty$, hence $E[U^{\beta,p}]<\infty$.

We use the so-defined random variable $U^{\beta,p}$ in order to
determine the convergence rate of $\theta^k$. Obviously, $k^{p-1-\beta
} (\theta^k)^p \le U^{\beta,p}$ and solving for $\theta^k$ yields
\[
\theta^k \le\frac{
(
U^{\beta,p}
)^{\sfrac{1}{p}}
}{
k^{1 - \vfrac{1+\beta}{p}}
}.
\]
If we define $\varepsilon= \frac{1+\beta}{p}$ and $U^\varepsilon=
( U^{\beta,p}  )^{\sfrac{1}{p}}$, the we obtain the inequality
\[
\theta^k \le\frac{U^\varepsilon}{k^{1-\varepsilon}}.
\]
Since $E[U^{\beta,p}] < \infty$, it follows that $E [  (
U^\varepsilon)^p  )  ] < \infty$, hence $U^\varepsilon$
is a.s. finite. Also, we recall that $\nu< \beta< p-1$, therefore
$\frac{1+\nu}{p} < \varepsilon< 1$.
\end{appendix}
\section*{Acknowledgements}

The work of D. Crisan was partially supported by the EPSRC Grant No:
EP/H0005500/1. The work of J. M\'{\i}guez was partially supported by
\emph{Ministerio de Econom\'{\i}a y Competitividad} of Spain (program
Consolider-Ingenio 2010 CSD2008-00010 COMONSENS and project
TEC2012-38883-C02-01 COMPREHENSION) and \emph{Ministerio de Educaci\'
on, Cultura y Deporte} of Spain (\emph{Programa Nacional de Movilidad
de Recursos Humanos} PRX12/00690).

%




\printhistory


\begin{thebibliography}{1}

\bibitem{Crisan14a}
D.~Crisan and J.~Miguez.
\newblock Particle-kernel estimation of the filter density in state-space
  models.
\newblock {\em Bernoulli}, 20(4):1879--1929, 2014.

\end{thebibliography}


\begin{thebibliography}{48}

\bibitem{Abraham04}
\begin{barticle}[mr]
\bauthor{\bsnm{Abraham},~\bfnm{Christophe}\binits{C.}},
  \bauthor{\bsnm{Biau},~\bfnm{G{\'e}rard}\binits{G.}} \AND
  \bauthor{\bsnm{Cadre},~\bfnm{Beno{\^{\i}}t}\binits{B.}}
(\byear{2004}).
\btitle{On the asymptotic properties of a simple estimate of the mode}.
\bjournal{ESAIM Probab. Stat.}
\bvolume{8}
\bpages{1--11 (electronic)}.
\bid{doi={10.1051/ps:2003015}, issn={1292-8100}, mr={2085601}}
\bptok{imsref}%
\end{barticle}
\endbibitem

\bibitem{Appel03}
\begin{barticle}[mr]
\bauthor{\bsnm{Appel},~\bfnm{M.~J.}\binits{M.J.}},
  \bauthor{\bsnm{LaBarre},~\bfnm{R.}\binits{R.}} \AND
  \bauthor{\bsnm{Radulovi{\'c}},~\bfnm{D.}\binits{D.}}
(\byear{2003}).
\btitle{On accelerated random search}.
\bjournal{SIAM J. Optim.}
\bvolume{14}
\bpages{708--731 (electronic)}.
\bid{doi={10.1137/S105262340240063X}, issn={1052-6234}, mr={2085938}}
\bptok{imsref}%
\end{barticle}
\endbibitem

\bibitem{Bain08}
\begin{bbook}[mr]
\bauthor{\bsnm{Bain},~\bfnm{Alan}\binits{A.}} \AND
  \bauthor{\bsnm{Crisan},~\bfnm{Dan}\binits{D.}}
(\byear{2009}).
\btitle{Fundamentals of Stochastic Filtering}.
\bseries{Stochastic Modelling and Applied Probability}
\bvolume{60}.
\blocation{New York}: \bpublisher{Springer}.
\bid{mr={2454694}}
\bptnote{check year}%
\bptok{imsref}%
\end{bbook}
\endbibitem

\bibitem{Beirlant97}
\begin{barticle}[mr]
\bauthor{\bsnm{Beirlant},~\bfnm{J.}\binits{J.}},
  \bauthor{\bsnm{Dudewicz},~\bfnm{E.~J.}\binits{E.J.}},
  \bauthor{\bsnm{Gy{\"o}rfi},~\bfnm{L.}\binits{L.}} \AND
  \bauthor{\bparticle{van~der} \bsnm{Meulen},~\bfnm{E.~C.}\binits{E.C.}}
(\byear{1997}).
\btitle{Nonparametric entropy estimation: An overview}.
\bjournal{Int. J. Math. Stat. Sci.}
\bvolume{6}
\bpages{17--39}.
\bid{issn={1055-7490}, mr={1471870}}
\bptok{imsref}%
\end{barticle}
\endbibitem

\bibitem{Brewer00}
\begin{barticle}[auto:STB|2013/12/09|07:59:19]
\bauthor{\bsnm{Brewer},~\bfnm{M.~J.}\binits{M.J.}}
(\byear{2000}).
\btitle{A Bayesian model for local smoothing in kernel density estimation}.
\bjournal{Statist. Comput.}
\bvolume{10}
\bpages{299--309}.
\bptok{imsref}%
\end{barticle}
\endbibitem

\bibitem{Corana87}
\begin{barticle}[mr]
\bauthor{\bsnm{Corana},~\bfnm{A.}\binits{A.}},
  \bauthor{\bsnm{Marchesi},~\bfnm{M.}\binits{M.}},
  \bauthor{\bsnm{Martini},~\bfnm{C.}\binits{C.}} \AND
  \bauthor{\bsnm{Ridella},~\bfnm{S.}\binits{S.}}
(\byear{1987}).
\btitle{Minimizing multimodal functions of continuous variables with the
  ``simulated annealing'' algorithm}.
\bjournal{ACM Trans. Math. Software}
\bvolume{13}
\bpages{262--280}.
\bid{doi={10.1145/29380.29864}, issn={0098-3500}, mr={0918580}}
\bptok{imsref}%
\end{barticle}
\endbibitem

\bibitem{Cover91}
\begin{bbook}[mr]
\bauthor{\bsnm{Cover},~\bfnm{Thomas~M.}\binits{T.M.}} \AND
  \bauthor{\bsnm{Thomas},~\bfnm{Joy~A.}\binits{J.A.}}
(\byear{1991}).
\btitle{Elements of Information Theory}.
\bseries{Wiley Series in Telecommunications}.
\blocation{New York}: \bpublisher{Wiley}.
\bid{doi={10.1002/0471200611}, mr={1122806}}
\bptok{imsref}%
\end{bbook}
\endbibitem

\bibitem{Crisan01}
\begin{bincollection}[mr]
\bauthor{\bsnm{Crisan},~\bfnm{Dan}\binits{D.}}
(\byear{2001}).
\btitle{Particle filters -- a theoretical perspective}.
In \bbooktitle{Sequential {M}onte {C}arlo Methods in Practice}
(\beditor{A. Doucet}, \beditor{N. de Freitas} and \beditor{N. Gordon}, eds.).
\bseries{Stat. Eng. Inf. Sci.}
\bpages{17--41}.
\blocation{New York}: \bpublisher{Springer}.
\bid{mr={1847785}}
\bptok{imsref}%
\end{bincollection}
\endbibitem

\bibitem{Crisan99}
\begin{barticle}[mr]
\bauthor{\bsnm{Crisan},~\bfnm{D.}\binits{D.}},
  \bauthor{\bsnm{Del~Moral},~\bfnm{P.}\binits{P.}} \AND
  \bauthor{\bsnm{Lyons},~\bfnm{T.}\binits{T.}}
(\byear{1999}).
\btitle{Discrete filtering using branching and interacting particle systems}.
\bjournal{Markov Process. Related Fields}
\bvolume{5}
\bpages{293--318}.
\bid{issn={1024-2953}, mr={1710982}}
\bptok{imsref}%
\end{barticle}
\endbibitem

\bibitem{Crisan00}
\begin{bmisc}[auto:STB|2013/12/09|07:59:19]
\bauthor{\bsnm{Crisan},~\bfnm{D.}\binits{D.}} \AND
  \bauthor{\bsnm{Doucet},~\bfnm{A.}\binits{A.}}
(\byear{2000}).
\bhowpublished{Convergence of sequential Monte Carlo methods. Technical
  Report {CUED/FINFENG/TR381}, Cambridge University}.
\bptok{imsref}%
\end{bmisc}
\endbibitem

\bibitem{Crisan02}
\begin{barticle}[mr]
\bauthor{\bsnm{Crisan},~\bfnm{Dan}\binits{D.}} \AND
  \bauthor{\bsnm{Doucet},~\bfnm{Arnaud}\binits{A.}}
(\byear{2002}).
\btitle{A survey of convergence results on particle filtering methods for
  practitioners}.
\bjournal{IEEE Trans. Signal Process.}
\bvolume{50}
\bpages{736--746}.
\bid{doi={10.1109/78.984773}, issn={1053-587X}, mr={1895071}}
\bptok{imsref}%
\end{barticle}
\endbibitem

\bibitem{Dean11}
\begin{bmisc}[auto:STB|2013/12/09|07:59:19]
\bauthor{\bsnm{Dean},~\bfnm{T.~A.}\binits{T.A.}},
  \bauthor{\bsnm{Singh},~\bfnm{S.~S.}\binits{S.S.}},
  \bauthor{\bsnm{Jasra},~\bfnm{A.}\binits{A.}} \AND
  \bauthor{\bsnm{Peters},~\bfnm{G.~W.}\binits{G.W.}}
(\byear{2011}).
 \bhowpublished{Parameter estimation for hidden Markov models with
  intractable likelihoods. Available at \arxivurl{arXiv:1103.5399v1}
  [math.ST]}.
\bptok{imsref}%
\end{bmisc}
\endbibitem

\bibitem{DelMoral96b}
\begin{barticle}[mr]
\bauthor{\bsnm{Del~Moral},~\bfnm{P.}\binits{P.}}
(\byear{1996}).
\btitle{Non-linear filtering using random particles}.
\bjournal{Theory Probab. Appl.}
\bvolume{40}
\bpages{690--701}.
\bptnote{check year}%
\bptok{imsref}%
\end{barticle}
\endbibitem

\bibitem{DelMoral96a}
\begin{barticle}[mr]
\bauthor{\bsnm{Del~Moral},~\bfnm{P.}\binits{P.}}
(\byear{1996}).
\btitle{Nonlinear filtering: Interacting particle solution}.
\bjournal{Markov Process. Related Fields}
\bvolume{2}
\bpages{555--579}.
\bid{issn={1024-2953}, mr={1431187}}
\bptok{imsref}%
\end{barticle}
\endbibitem

\bibitem{DelMoral04}
\begin{bbook}[mr]
\bauthor{\bsnm{Del~Moral},~\bfnm{Pierre}\binits{P.}}
(\byear{2004}).
\btitle{Feynman--{K}ac Formulae: Genealogical and Interacting Particle Systems with Applications}.
\bseries{Probability and Its Applications (New York)}.
\blocation{New York}: \bpublisher{Springer}.
\bid{doi={10.1007/978-1-4684-9393-1}, mr={2044973}}
\bptok{imsref}%
\end{bbook}
\endbibitem

\bibitem{DelMoral00}
\begin{bincollection}[mr]
\bauthor{\bsnm{Del~Moral},~\bfnm{P.}\binits{P.}} \AND
  \bauthor{\bsnm{Miclo},~\bfnm{L.}\binits{L.}}
(\byear{2000}).
\btitle{Branching and interacting particle systems approximations of
  {F}eynman--{K}ac formulae with applications to non-linear filtering}.
In \bbooktitle{S\'eminaire de {P}robabilit\'es, {XXXIV}}.
\bseries{Lecture Notes in Math.}
\bvolume{1729}
\bpages{1--145}.
\blocation{Berlin}: \bpublisher{Springer}.
\bid{doi={10.1007/BFb0103798}, mr={1768060}}
\bptok{imsref}%
\end{bincollection}
\endbibitem

\bibitem{DelMoral11}
\begin{bmisc}[auto:STB|2013/12/09|07:59:19]
\bauthor{\bsnm{Del Moral},~\bfnm{P.}\binits{P.}},
  \bauthor{\bsnm{Doucet},~\bfnm{A.}\binits{A.}} \AND
  \bauthor{\bsnm{Singh},~\bfnm{S.}\binits{S.}}
(\byear{2011}).
\bhowpublished{Uniform
  stability of a particle approximation of the optimal filter derivative.
  Available at \arxivurl{arXiv:1106.2525v1} [math.ST]}.
\bptok{imsref}%
\end{bmisc}
\endbibitem

\bibitem{Devroye85}
\begin{bbook}[mr]
\bauthor{\bsnm{Devroye},~\bfnm{Luc}\binits{L.}} \AND
  \bauthor{\bsnm{Gy{\"o}rfi},~\bfnm{L{\'a}szl{\'o}}\binits{L.}}
(\byear{1985}).
\btitle{Nonparametric Density Estimation: The $L{\sb{1}}$ View}.
\bseries{Wiley Series in Probability and Mathematical Statistics: Tracts on
  Probability and Statistics}.
\blocation{New York}: \bpublisher{Wiley}.
\bid{mr={0780746}}
\bptok{imsref}%
\end{bbook}
\endbibitem

\bibitem{Douc05}
\begin{bincollection}[auto:STB|2013/12/09|07:59:19]
\bauthor{\bsnm{Douc},~\bfnm{R.}\binits{R.}},
  \bauthor{\bsnm{Capp{\'e}},~\bfnm{O.}\binits{O.}} \AND
  \bauthor{\bsnm{Moulines},~\bfnm{E.}\binits{E.}}
(\byear{2005}).
\btitle{Comparison of resampling schemes for particle filtering}.
In \bbooktitle{Proceedings of the 4th International Symposium on Image and
  Signal Processing and Analysis}
\bpages{64--69}.
\bptok{imsref}%
\end{bincollection}
\endbibitem


\bibitem{Doucet01}
\begin{bincollection}[mr]
\bauthor{\bsnm{Doucet},~\bfnm{A.}\binits{A.}},
  \bauthor{\bsnm{de~Freitas},~\bfnm{N.}\binits{N.}} \AND
  \bauthor{\bsnm{Gordon},~\bfnm{N.}\binits{N.}}
(\byear{2001}).
\btitle{An introduction to sequential Monte Carlo
methods}.
In
\bbooktitle{Sequential {M}onte {C}arlo Methods in Practice}
(\beditor{A. Doucet}, \beditor{N. de Freitas} and \beditor{N. Gordon}, eds.).
\bseries{Statistics for Engineering and Information Science}.
\blocation{New York}: \bpublisher{Springer}.
\bid{mr={1847783}}
\bptok{imsref}%
\end{bincollection}
\endbibitem

\bibitem{Doucet01b}
\begin{bbook}[mr]
\bauthor{\bsnm{Doucet},~\bfnm{A.}\binits{A.}},
  \bauthor{\bsnm{de~Freitas},~\bfnm{N.}\binits{N.}} \AND
  \bauthor{\bsnm{Gordon},~\bfnm{N.}\binits{N.}}
(\byear{2001}).
\btitle{Sequential {M}onte {C}arlo Methods in Practice}
(\beditor{A.~Doucet}, \beditor{N. de Freitas} and \beditor{N. Gordon}, eds.).
\bseries{Statistics for Engineering and Information Science}.
\blocation{New York}: \bpublisher{Springer}.
\bid{mr={1847783}}
\bptok{imsref}%
\end{bbook}
\endbibitem


\bibitem{Doucet00}
\begin{barticle}[auto:STB|2013/12/09|07:59:19]
\bauthor{\bsnm{Doucet},~\bfnm{A.}\binits{A.}},
  \bauthor{\bsnm{Godsill},~\bfnm{S.}\binits{S.}} \AND
  \bauthor{\bsnm{Andrieu},~\bfnm{C.}\binits{C.}}
(\byear{2000}).
\btitle{On sequential Monte Carlo sampling methods for Bayesian filtering}.
\bjournal{Statist. Comput.}
\bvolume{10}
\bpages{197--208}.
\bptok{imsref}%
\end{barticle}
\endbibitem

\bibitem{Duong05}
\begin{barticle}[mr]
\bauthor{\bsnm{Duong},~\bfnm{Tarn}\binits{T.}} \AND
  \bauthor{\bsnm{Hazelton},~\bfnm{Martin~L.}\binits{M.L.}}
(\byear{2005}).
\btitle{Cross-validation bandwidth matrices for multivariate kernel density
  estimation}.
\bjournal{Scand. J. Stat.}
\bvolume{32}
\bpages{485--506}.
\bid{doi={10.1111/j.1467-9469.2005.00445.x}, issn={0303-6898}, mr={2204631}}
\bptok{imsref}%
\end{barticle}
\endbibitem

\bibitem{Frenkel99}
\begin{barticle}[auto:STB|2013/12/09|07:59:19]
\bauthor{\bsnm{Frenkel},~\bfnm{L.}\binits{L.}} \AND
  \bauthor{\bsnm{Feder},~\bfnm{M.}\binits{M.}}
(\byear{1999}).
\btitle{Recursive expectation--maximization (EM) algorithms for time-varying
  parameters with applications to multiple target tracking}.
\bjournal{IEEE Trans. Signal Process.}
\bvolume{47}
\bpages{306--320}.
\bptok{imsref}%
\end{barticle}
\endbibitem

\bibitem{Gauvain92}
\begin{barticle}[auto:STB|2013/12/09|07:59:19]
\bauthor{\bsnm{Gauvain},~\bfnm{J.~L.}\binits{J.L.}} \AND
  \bauthor{\bsnm{Lee},~\bfnm{C.~H.}\binits{C.H.}}
(\byear{1992}).
\btitle{Bayesian learning for hidden Markov model with Gaussian mixture state
  observation densities}.
\bjournal{Speech Commun.}
\bvolume{11}
\bpages{205--213}.
\bptok{imsref}%
\end{barticle}
\endbibitem

\bibitem{Godsill01}
\begin{barticle}[mr]
\bauthor{\bsnm{Godsill},~\bfnm{Simon}\binits{S.}},
  \bauthor{\bsnm{Doucet},~\bfnm{Arnaud}\binits{A.}} \AND
  \bauthor{\bsnm{West},~\bfnm{Mike}\binits{M.}}
(\byear{2001}).
\btitle{Maximum a posteriori sequence estimation using {M}onte {C}arlo particle
  filters}.
\bjournal{Ann. Inst. Statist. Math.}
\bvolume{53}
\bpages{82--96}.
\bid{doi={10.1023/A:1017968404964}, issn={0020-3157}, mr={1777255}}
\bptok{imsref}%
\end{barticle}
\endbibitem

\bibitem{Gordon93}
\begin{barticle}[auto:STB|2013/12/09|07:59:19]
\bauthor{\bsnm{Gordon},~\bfnm{N.}\binits{N.}},
  \bauthor{\bsnm{Salmond},~\bfnm{D.}\binits{D.}} \AND
  \bauthor{\bsnm{Smith},~\bfnm{A.~F.~M.}\binits{A.F.M.}}
(\byear{1993}).
\btitle{Novel approach to nonlinear and non-{G}aussian Bayesian state
  estimation}.
\bjournal{IEE Proc. F}
\bvolume{140}
\bpages{107--113}.
\bptok{imsref}%
\end{barticle}
\endbibitem

\bibitem{Hall01}
\begin{barticle}[mr]
\bauthor{\bsnm{Hall},~\bfnm{Peter}\binits{P.}} \AND
  \bauthor{\bsnm{Kang},~\bfnm{Kee-Hoon}\binits{K.H.}}
(\byear{2001}).
\btitle{Bootstrapping nonparametric density estimators with empirically chosen
  bandwidths}.
\bjournal{Ann. Statist.}
\bvolume{29}
\bpages{1443--1468}.
\bid{doi={10.1214/aos/1013203461}, issn={0090-5364}, mr={1873338}}
\bptok{imsref}%
\end{barticle}
\endbibitem

\bibitem{Hedar06}
\begin{barticle}[mr]
\bauthor{\bsnm{Hedar},~\bfnm{Abdel-Rahman}\binits{A.R.}} \AND
  \bauthor{\bsnm{Fukushima},~\bfnm{Masao}\binits{M.}}
(\byear{2006}).
\btitle{Derivative-free filter simulated annealing method for constrained
  continuous global optimization}.
\bjournal{J. Global Optim.}
\bvolume{35}
\bpages{521--549}.
\bid{doi={10.1007/s10898-005-3693-z}, issn={0925-5001}, mr={2249547}}
\bptok{imsref}%
\end{barticle}
\endbibitem

\bibitem{Heine08}
\begin{barticle}[mr]
\bauthor{\bsnm{Heine},~\bfnm{Kari}\binits{K.}} \AND
  \bauthor{\bsnm{Crisan},~\bfnm{Dan}\binits{D.}}
(\byear{2008}).
\btitle{Uniform approximations of discrete-time filters}.
\bjournal{Adv. in Appl. Probab.}
\bvolume{40}
\bpages{979--1001}.
\bid{issn={0001-8678}, mr={2488529}}
\bptok{imsref}%
\end{barticle}
\endbibitem

\bibitem{Hu08}
\begin{barticle}[mr]
\bauthor{\bsnm{Hu},~\bfnm{Xiao-Li}\binits{X.L.}},
  \bauthor{\bsnm{Sch{\"o}n},~\bfnm{Thomas~B.}\binits{T.B.}} \AND
  \bauthor{\bsnm{Ljung},~\bfnm{Lennart}\binits{L.}}
(\byear{2008}).
\btitle{A basic convergence result for particle filtering}.
\bjournal{IEEE Trans. Signal Process.}
\bvolume{56}
\bpages{1337--1348}.
\bid{doi={10.1109/TSP.2007.911295}, issn={1053-587X}, mr={2512468}}
\bptok{imsref}%
\end{barticle}
\endbibitem

\bibitem{Kalman60}
\begin{barticle}[auto:STB|2013/12/09|07:59:19]
\bauthor{\bsnm{Kalman},~\bfnm{R.~E.}\binits{R.E.}}
(\byear{1960}).
\btitle{A new approach to linear filtering and prediction problems}.
\bjournal{J. Basic Eng.}
\bvolume{82}
\bpages{35--45}.
\bptok{imsref}%
\end{barticle}
\endbibitem

\bibitem{Kitagawa96}
\begin{barticle}[mr]
\bauthor{\bsnm{Kitagawa},~\bfnm{Genshiro}\binits{G.}}
(\byear{1996}).
\btitle{Monte {C}arlo filter and smoother for non-{G}aussian nonlinear state
  space models}.
\bjournal{J. Comput. Graph. Statist.}
\bvolume{5}
\bpages{1--25}.
\bid{doi={10.2307/1390750}, issn={1061-8600}, mr={1380850}}
\bptok{imsref}%
\end{barticle}
\endbibitem

\bibitem{Kunsch05}
\begin{barticle}[mr]
\bauthor{\bsnm{K{\"u}nsch},~\bfnm{Hans~R.}\binits{H.R.}}
(\byear{2005}).
\btitle{Recursive {M}onte {C}arlo filters: Algorithms and theoretical
  analysis}.
\bjournal{Ann. Statist.}
\bvolume{33}
\bpages{1983--2021}.
\bid{doi={10.1214/009053605000000426}, issn={0090-5364}, mr={2211077}}
\bptok{imsref}%
\end{barticle}
\endbibitem

\bibitem{LeGland04}
\begin{barticle}[mr]
\bauthor{\bsnm{Le~Gland},~\bfnm{Fran{\c{c}}ois}\binits{F.}} \AND
  \bauthor{\bsnm{Oudjane},~\bfnm{Nadia}\binits{N.}}
(\byear{2004}).
\btitle{Stability and uniform approximation of nonlinear filters using the
  {H}ilbert metric and application to particle filters}.
\bjournal{Ann. Appl. Probab.}
\bvolume{14}
\bpages{144--187}.
\bid{doi={10.1214/aoap/1075828050}, issn={1050-5164}, mr={2023019}}
\bptok{imsref}%
\end{barticle}
\endbibitem

\bibitem{Liu98}
\begin{barticle}[mr]
\bauthor{\bsnm{Liu},~\bfnm{Jun~S.}\binits{J.S.}} \AND
  \bauthor{\bsnm{Chen},~\bfnm{Rong}\binits{R.}}
(\byear{1998}).
\btitle{Sequential {M}onte {C}arlo methods for dynamic systems}.
\bjournal{J. Amer. Statist. Assoc.}
\bvolume{93}
\bpages{1032--1044}.
\bid{doi={10.2307/2669847}, issn={0162-1459}, mr={1649198}}
\bptok{imsref}%
\end{barticle}
\endbibitem

\bibitem{Logothetis99}
\begin{barticle}[auto:STB|2013/12/09|07:59:19]
\bauthor{\bsnm{Logothetis},~\bfnm{A.}\binits{A.}} \AND
  \bauthor{\bsnm{Krishnamurthy},~\bfnm{V.}\binits{V.}}
(\byear{1999}).
\btitle{Expectation maximization algorithms for MAP estimation of jump Markov
  linear systems}.
\bjournal{IEEE Trans. Signal Process.}
\bvolume{47}
\bpages{2139--2156}.
\bptok{imsref}%
\end{barticle}
\endbibitem

\bibitem{Miguez13b}
\begin{barticle}[mr]
\bauthor{\bsnm{M{\'{\i}}guez},~\bfnm{Joaqu{\'{\i}}n}\binits{J.}},
  \bauthor{\bsnm{Crisan},~\bfnm{Dan}\binits{D.}} \AND
  \bauthor{\bsnm{Djuri{\'c}},~\bfnm{Petar~M.}\binits{P.M.}}
(\byear{2013}).
\btitle{On the convergence of two sequential {M}onte {C}arlo methods for
  maximum a posteriori sequence estimation and stochastic global optimization}.
\bjournal{Stat. Comput.}
\bvolume{23}
\bpages{91--107}.
\bid{doi={10.1007/s11222-011-9294-4}, issn={0960-3174}, mr={3018352}}
\bptok{imsref}%
\end{barticle}
\endbibitem

\bibitem{Musso01}
\begin{bincollection}[mr]
\bauthor{\bsnm{Musso},~\bfnm{Christian}\binits{C.}},
  \bauthor{\bsnm{Oudjane},~\bfnm{Nadia}\binits{N.}} \AND
  \bauthor{\bsnm{Le~Gland},~\bfnm{Francois}\binits{F.}}
(\byear{2001}).
\btitle{Improving regularised particle filters}.
In \bbooktitle{Sequential {M}onte {C}arlo Methods in Practice}
(\beditor{A. Doucet}, \beditor{N. de Freitas} and \beditor{N. Gordon}, eds.).
\bseries{Stat. Eng. Inf. Sci.}
\bpages{247--271}.
\blocation{New York}: \bpublisher{Springer}.
\bid{mr={1847795}}
\bptok{imsref}%
\end{bincollection}
\endbibitem

\bibitem{Najim06}
\begin{barticle}[auto:STB|2013/12/09|07:59:19]
\bauthor{\bsnm{Najim},~\bfnm{K.}\binits{K.}},
  \bauthor{\bsnm{Ikonen},~\bfnm{E.}\binits{E.}} \AND
  \bauthor{\bsnm{Del Moral},~\bfnm{P.}\binits{P.}}
(\byear{2006}).
\btitle{Open-loop regulation and tracking control based on a
  genealogical decision tree}.
\bjournal{Neural Comput. Appl.}
\bvolume{15}
\bpages{339--349}.
\bptok{imsref}%
\end{barticle}
\endbibitem

\bibitem{Nilsson07}
\begin{barticle}[mr]
\bauthor{\bsnm{Nilsson},~\bfnm{Mattias}\binits{M.}} \AND
  \bauthor{\bsnm{Kleijn},~\bfnm{W.~Bastiaan}\binits{W.B.}}
(\byear{2007}).
\btitle{On the estimation of differential entropy from data located on embedded
  manifolds}.
\bjournal{IEEE Trans. Inform. Theory}
\bvolume{53}
\bpages{2330--2341}.
\bid{doi={10.1109/TIT.2007.899533}, issn={0018-9448}, mr={2319377}}
\bptok{imsref}%
\end{barticle}
\endbibitem

\bibitem{Silverman86}
\begin{bbook}[mr]
\bauthor{\bsnm{Silverman},~\bfnm{B.~W.}\binits{B.W.}}
(\byear{1986}).
\btitle{Density Estimation for Statistics and Data Analysis}.
\bseries{Monographs on Statistics and Applied Probability}.
\blocation{London}: \bpublisher{Chapman \& Hall}.
\bid{mr={0848134}}
\bptok{imsref}%
\end{bbook}
\endbibitem

\bibitem{Simonoff96}
\begin{bbook}[mr]
\bauthor{\bsnm{Simonoff},~\bfnm{Jeffrey~S.}\binits{J.S.}}
(\byear{1996}).
\btitle{Smoothing Methods in Statistics}.
\bseries{Springer Series in Statistics}.
\blocation{New York}: \bpublisher{Springer}.
\bid{doi={10.1007/978-1-4612-4026-6}, mr={1391963}}
\bptok{imsref}%
\end{bbook}
\endbibitem

\bibitem{Hulle05}
\begin{barticle}[auto:STB|2013/12/09|07:59:19]
\bauthor{\bsnm{Van~Hulle},~\bfnm{M.~M.}\binits{M.M.}}
(\byear{2005}).
\btitle{Edgeworth approximation of multivariate differential entropy}.
\bjournal{Neural Comput.}
\bvolume{17}
\bpages{1903--1910}.
\bptok{imsref}%
\end{barticle}
\endbibitem

\bibitem{Wand95}
\begin{bbook}[mr]
\bauthor{\bsnm{Wand},~\bfnm{M.~P.}\binits{M.P.}} \AND
  \bauthor{\bsnm{Jones},~\bfnm{M.~C.}\binits{M.C.}}
(\byear{1995}).
\btitle{Kernel Smoothing}.
\bseries{Monographs on Statistics and Applied Probability}
\bvolume{60}.
\blocation{London}: \bpublisher{Chapman \& Hall}.
\bid{mr={1319818}}
\bptok{imsref}%
\end{bbook}
\endbibitem

\bibitem{West93}
\begin{barticle}[mr]
\bauthor{\bsnm{West},~\bfnm{Mike}\binits{M.}}
(\byear{1993}).
\btitle{Approximating posterior distributions by mixtures}.
\bjournal{J. R. Stat. Soc. Ser. B Stat. Methodol.}
\bvolume{55}
\bpages{409--422}.
\bid{issn={0035-9246}, mr={1224405}}
\bptok{imsref}%
\end{barticle}
\endbibitem





\bibitem{Zhang06}
\begin{barticle}[mr]
\bauthor{\bsnm{Zhang},~\bfnm{Xibin}\binits{X.}},
  \bauthor{\bsnm{King},~\bfnm{Maxwell~L.}\binits{M.L.}} \AND
  \bauthor{\bsnm{Hyndman},~\bfnm{Rob~J.}\binits{R.J.}}
(\byear{2006}).
\btitle{A {B}ayesian approach to bandwidth selection for multivariate kernel
  density estimation}.
\bjournal{Comput. Statist. Data Anal.}
\bvolume{50}
\bpages{3009--3031}.
\bid{doi={10.1016/j.csda.2005.06.019}, issn={0167-9473}, mr={2239655}}
\bptok{imsref}%
\end{barticle}
\endbibitem

\end{thebibliography}
\end{document}


\maketitle

\abstract{
This is an erratum to the article in {\em Bernoulli} 20, no. 4 (2014), pp. 1879--1929. There is a gap in the proof of Theorem 4.2 of the latter paper, as it relies on an inequality that does not necessarily hold under the assumptions of that theorem. In this note we fill this gap by providing an extended proof. The statement and assumptions of the theorem are exactly the same as in the original paper.
}

\vspace{+0.5cm}






\section{Introduction}

We have found a gap in the published proof of \cite[Theorem 4.2]{Crisan14a}. Specifically, the upper bound for the error $| p_t^k(x) - p_t(x) |$ in the expression (4.16) of \cite{Crisan14a} does not necessarily hold true unless additional assumptions are imposed on the density $p_t(x)$ (see Remark \ref{rem0} at the end of this note). The argument for the proof provided in the original article is, therefore, incomplete.  The statement of Theorem 4.2 remains valid, however, and a complete proof is given in Section \ref{sCorrection} below.

Unless otherwise specified, we adopt the same notation as in \cite{Crisan14a}. The ``standard conditions'' in the statement of Theorem 4.2 are detailed in Remark 4.5 of \cite{Crisan14a}.

\section{Correction to the proof of \cite[Theorem 4.2]{Crisan14a}} \label{sCorrection}

Let us recall that the posterior density $p_t(x)$ is estimated as $p_t^k(x) = (\phi_k^x,\pi_t^{N(k)})$, where $\phi_k^x(x')=k^{d_x}\phi(kx')$, $\phi$ is a bounded kernel with bounded derivatives, and $\pi_t^{N(k)}$ is the particle approximation of the posterior measure $\pi_t$. The integer index $k$ determines the properties of the kernel $\phi_k^x$ (see \cite[Remark 4.5]{Crisan14a}), including its bandwidth, and the number of particles, $N(k)$, used for the approximation of $\pi_t$. 

We also recall the sequence of hypercubes 
$
\mK_k=[-M_k,+M_k]^{d_x} \subset \Real^{d_x}
$
introduced in \cite[p. 1895]{Crisan14a}, where, for any given $p \ge 2$, $M_k = \frac{1}{2}k^{\frac{\beta}{d_x p}}$, the integer $d_x \ge 1$ is the dimension of the state space and $0 \le \beta < 1$ is a constant independent of $k$.

The following class of function-valued random variables is instrumental to our analysis.
\begin{Definicion} \label{defH}
Let $d \ge 1$ be a positive integer. A function-valued random variable (r.v.) $\sfh^k : \Real^d \rw \Real$ belongs to the family $\sfH^k(d)$ if, and only if, for every $x \in \Real^d$ we can express $\sfh^k(x)$ as
\begin{equation}
\sfh^k(x) = (a_k^x,\pi_t^N) - (a_k^x,\pi_t),
\nonumber
\end{equation}
where $N=N(k) \ge k^{2(2d_x+1)}$ and $a_k^x(x')=a_k(x,x')$ is a bounded function, $a_k:\Real^d \times \Real^{d_x} \rw \Real$, with bounded derivatives up to order $d$ w.r.t. the variable $x$, such that
\begin{equation}
\| a_k \|_\infty = \sup_{(x,x') \in \Real^d \times \Real^{d_x}} \left| 
	a_k^x(x') 
\right| \le k^{d_x} C_a < \infty 
\quad \mbox{and} \quad 
\| D^\alpha a_k \|_\infty = \sup_{(x,x') \in \Real^d \times \Real^{d_x}} \left| 
	D^\alpha a_k^x(x') 
\right| \le k^{d_x + |\alpha|} C_a < \infty
\nonumber 
\end{equation}
for some constant $C_a < \infty$ independent of $k$, where $\alpha=(\alpha_1, \alpha_2, \ldots, \alpha_d)$ is a multi-index and $|\alpha|=\sum_{i=1}^d \alpha_i \le d$.
\end{Definicion}

\begin{Nota}
The partial derivative operator acts on $x$, i.e., if $x=(x_1, \ldots, x_d)$ and $x'=(x_1', \ldots, x_{d_x}')$, then $D^\alpha a_k^x(x') = \frac{ \partial^{\alpha_1} \cdots \partial^{\alpha_d} a_k^x }{ \partial x_1 \cdots \partial x_d}(x_1', \ldots, x_{d_x}')$. 
\end{Nota}

\begin{Teorema}[Theorem 4.2 in \cite{Crisan14a}]
If the standard conditions are satisfied, then
\begin{equation}
\sup_{x \in \mK_k} \left|
        p_t^k(x) - p_t(x)
\right| \le \frac{
        U^{\varepsilon}
}{
        k^{1-\varepsilon}
}, \label{eqStatement}
\end{equation}
where $U^\varepsilon \ge 0$ is an a.s. finite random variable and $0<\varepsilon<1$ is  a constant, both of them independent of $k$ and $x$. In particular,
\begin{equation}
\lim_{k\rw\infty} \sup_{x \in \mK_k} \left|
        p_t^k(x)-p_t(x)
\right| = 0 \quad \mbox{a.s.}
\nonumber 
\end{equation}
\end{Teorema}

\noindent \textbf{Proof:} 
We are going to prove, more generally, that for any $1 \le d \le d_x$, $\sfh^k \in \sfH^k(d)$ and any $0 < \varepsilon < 1$, there exists an a.s. finite random variable $\bar U^\varepsilon$, independent of $x$ and $k$, such that
\begin{equation}
\sup_{x \in \left[ -M_k, M_k \right]^d} \left| \sfh^k(x) \right| \le \frac{
        \bar U^{\varepsilon}
}{
        k^{1-\varepsilon}
}, \label{eqConvergenceSupP}
\end{equation}
where $M_k = \frac{1}{2}k^{\frac{\beta}{dp}}$, with $0< \beta <1$ and $p \ge 2$. Note that, for $d=d_x$, $\left[ -M_k, M_k \right]^d = \mK_k$. We prove that the inequality \eqref{eqConvergenceSupP} holds by induction in the dimension $d$.

We start with the case $d=1$, hence $x \in \Real$. From Definition \ref{defH}, any $\sfh^k \in \sfH^k(1)$ is differentiable in every interval $[-M_k,M_k]$, hence we can apply the fundamental theorem of calculus (FTC) to express $\sfh^k(x)$, for $-M_k \le x \le M_k$, as
\begin{equation}
\sfh^k(x) = \sfh^k(0) + \int_0^x D^1\sfh^k(z)dz.  
\nonumber
\end{equation}
As a consequence, we obtain a simple upper bound for the magnitude of $\sfh^k(x)$, namely
\begin{equation}
\sup_{x \in [-M_k,M_k]} |\sfh^k(x)| \le |\sfh^k(0)| + A^k, \label{eqSupsff}
\end{equation}
where 
\begin{equation}
A^k = \int_{-M_k}^{M_k} \left| D^1 \sfh^k(z) \right |dz.
\end{equation}

In order to find an upper bound for the term $A^k$, we apply Jensen's inequality, which yields, for $p \ge 1$, 
\begin{equation}
\left(
        \frac{1}{2 M_k} A^k
\right)^p \le
\frac{1}{2 M_k} \int_{-M_k}^{M_k}  \left| D^1 \sfh^k(z) \right |^p dz \label{eqJensenIneq}
\end{equation}
and the inequality \eqref{eqJensenIneq} above readily leads to
\begin{equation}
\left(
        A^k
\right)^p \le 2^{p-1} M_k^{p-1} \int_{-M_k}^{M_k}  \left| D^1 \sfh^k(z) \right |^p dz.
\label{eqBoundBigA}
\end{equation}
However, since $\sfh^k \in \sfH^k(1)$, there exists some function $a_k^x(x')$, with $x \in \Real$ and $x'\in\Real^{d_x}$ such that
\begin{equation}
D^1 \sfh^k(x) = \left(
	D^1 a_k^x, \pi_t^N
\right) - \left(
	D^1 a_k^x, \pi_t
\right),
\nonumber
\end{equation}
where $N=N(k) \ge k^{2(2d_x+1)}$ and, from Definition \ref{defH}, 
\begin{equation}
\| D^1 a_k \|_\infty = \sup_{(x,x')\in \Real \times \Real^{d_x}} |D^1 a_k^x(x')| \le k^{d_x+1} C_a < \infty.
\label{eqCa}
\end{equation}
Since  $D^1 a_k^x \in B(\Real)$, we can apply \cite[Proposition 2.1]{Crisan14a} together with \eqref{eqCa} to arrive at
\begin{equation}
E\left[
	\left| 
		D^1 \sfh^k(x)
	\right|^p
\right] = E\left[
	\left|
		\left(
			D^1 a_k^x, \pi_t^N(k)
		\right) - \left(
			D^1 a_k^x, \pi_t
		\right)
	\right|^p
\right] \le \frac{
        c_{1,t}^p k^{(d_x+1)p} C_a^p
}{
       \left( N(k) \right)^\frac{p}{2}
},
\label{eqSmallBoundAk}
\end{equation}
which holds true for every $x \in [-M_k,M_k]$ and where the constants $c_{1,t}, C_a < \infty$ are independent of $k$ (and $x$). We can combine \eqref{eqSmallBoundAk} and \eqref{eqBoundBigA} to obtain
\begin{equation}
E\left[
        (A^k)^p
\right] \le \frac{
        2^p M_k^p c_{1,t}^p k^{(d_x+1)p} C_a^p 
}{
        \left( N(k) \right)^\frac{p}{2}
} \le \frac{
        c_{1,t}^p C_a^p 
}{
        k^{d_xp-\beta}
} \le \frac{
        c_{1,t}^p C_a^p 
}{
        k^{p-\beta}
}, \nonumber
\end{equation}
where the second inequality follows from the relationships $M_k = \frac{1}{2}k^\frac{\beta}{p}$ and $N(k) \ge k^{2(2d_x+1)}$, and the third inequality holds because $d_x\ge 1$.

If we now apply \cite[Lemma 4.1]{Crisan14a} with $\theta^k = A^k$, $p \ge 2$, $\nu=\beta$ and $c=c_t^p C_a^p$, then we obtain a constant $\varepsilon_1 \in \left( \frac{1+\beta}{p}, 1 \right)$ and a non-negative and a.s. finite random variable $V^{A,\varepsilon_1}$, both of them independent of $k$, such that
\begin{equation}
A^k \le \frac{
        V^{A,\varepsilon_1}
}{
        k^{1-\varepsilon_1}
}. \label{eqRate_part1}
\end{equation} 

Moreover, from Definition \ref{defH}, $\| a_k \|_\infty = \sup_{x \in \Real} \| a_k^x \|_\infty \le k^{d_x}C_a$ and this bound combined with \cite[Proposition 2.1]{Crisan14a} yields 
\begin{equation}
E\left[
        \left|
                \sfh^k(0)
        \right|^p
\right] = E\left[
        \left|
                (a_k^0,\pi_t^N(k)) - (a_k^0,\pi_t)
        \right|^p
\right] \le \frac{
        c_{0,t}^p k^{d_x p} C_a^p
}{
        ( N(k) )^{\frac{p}{2}}
}, \nonumber
\end{equation}
where $c_{0,t}<\infty$ is a constant independent of $k$ (and $x$). Since $N(k) \ge k^{2(2d_x+1)}$, the inequality above implies that 
\begin{equation}
E\left[
        \left|
                \sfh^k(0)
        \right|^p
\right] \le \frac{
        c_{0,t}^p C_a^p
}{
        k^{p(d_x+1)}
} \le \frac{
        c_{0,t}^p C_a^p
}{
        k^{2p}
},
\nonumber
\end{equation}
where the second inequality holds because $d_x \ge 1$. Now we can apply \cite[Lemma 4.1]{Crisan14a} again, with $\theta^k = | \sfh^k(0) |$, $p \ge 2$, $\nu=0$ and $c=c_{0,t}^p C_a^p$ to obtain the relationship
\begin{equation}
\left|
        \sfh^k(0)
\right| \le \frac{
        V^{0,\varepsilon_2}
}{
        k^{1-\varepsilon_2}
}, \label{eqRate_part2}
\end{equation}
where $\varepsilon_2 \in (0,1)$ is an arbitrary constant and $V^{0,\varepsilon_2}$ is a non-negative and a.s. finite r.v., both of them independent of $k$. 

If we choose $\varepsilon=\varepsilon_1=\varepsilon_2 \in \left( \frac{1+\beta}{p}, 1 \right)$ and define $\bar U^\varepsilon = V^{A,\varepsilon_1} + V^{0,\varepsilon_2}$, then the combination of Eqs. \eqref{eqSupsff}, \eqref{eqRate_part1} and \eqref{eqRate_part2} yields 
\begin{equation}
\sup_{x \in [-M_k,M_k]} \left|
        \sfh^k(x)
\right| \le \frac{\bar U^\varepsilon}{k^{1-\varepsilon}},
\nonumber
\end{equation}
where $\bar U^\varepsilon$ is a.s. finite. Note that $\bar U^\varepsilon$ and $\varepsilon$ are independent of $k$. Moreover, we can choose $p$ as large as we wish and $\beta>0$ as small as needed, hence we can effectively select $\varepsilon \in (0,1)$ as small as we wish. This completes the analysis for $d=1$.

Next, we assume that the inequality \eqref{eqConvergenceSupP} holds for members of the class $\sfH^k(d-1)$, with $1 \le d-1 < d_x$, and show that, in such case, it also holds for $\sfH^k(d)$, with $d \le d_x$.

Choose any $x=(x_1, \ldots, x_d)\in [-M_k, M_k]^d$. Using the FTC we obtain
\begin{equation}
\sfh^k(x_1, \ldots, x_d) = \sfh^k(x_1, \ldots, x_{d-1}, 0) + \int_0^{x_d} D^{\alpha_1} \sfh^k(x_1, \ldots, x_{d-1}, z_d ) dz_d,
\label{eqX0}
\end{equation}
where $\alpha_1=(0, ..., 0, 1)$. The function in the integral of the right hand side (rhs) of \eqref{eqX0} can be expanded, using the FTC again, as 
\begin{equation}
D^{\alpha_1} \sfh^k(x_1, \ldots, x_{d-1}, z_d ) = 
D^{\alpha_1} \sfh^k(x_1, \ldots, x_{d-2}, 0, z_d ) + 
\int_0^{x_{d-1}} D^{\alpha_2} \sfh^k(x_1, \ldots, x_{d-2}, z_{d-1},z_d) dz_d dz_{d-1},
\label{eqX1}
\end{equation}
where $\alpha_2 = (\overbrace{0, ..., 0}^{d-2}, 1, 1)$, and substituting \eqref{eqX1} into \eqref{eqX0} yields
\begin{eqnarray}
\sfh^k(x_1, \ldots, x_d) &=& \sfh^k(x_1, \ldots, x_{d-1}, 0) + \int_0^{x_d} D^{\alpha_1} \sfh^k(x_1, \ldots, x_{d-2}, 0, z_d ) dz_d \nonumber\\
&& + \int_0^{x_d} \int_0^{x_{d-1}} D^{\alpha_2} \sfh^k(x_1, \ldots, x_{d-2}, z_{d-1},z_d) dz_d dz_{d-1}.
\nonumber 
\end{eqnarray}
It is straightforward to see that, by successively applying the FTC $d$ times, we arrive at the expression
\begin{equation}
\sfh^k(x_1, \ldots, x_d) = \sum_{i=0}^{d-1} \tilde \sfh_i^k(x_{(d-i)}) + \int_0^{x_d} \cdots \int_0^{x_1} D^{\bf 1} \sfh^k(z_1, \ldots, z_d) dz_d \cdots dz_1,
\label{eqX3}
\end{equation}
where $x_{(d-i)} = (x_1, \ldots, x_{d-i-1}, x_{d-i+1}, x_d) \in [-M_k,M_k]^{d-1}$,
\begin{eqnarray}
\tilde \sfh_0^k(x_{(d)}) &\dfn& \sfh^k(x_1, \ldots, x_{d-1}, 0), \label{eqX4} \\
\tilde \sfh_i^k(x_{(d-i)}) &\dfn& \int_0^{x_d} \cdots \int_0^{x_{d-i+1}} D^{\alpha_i} \sfh^k(x_1, \ldots, x_{d-i-1}, 0, z_{d-i+1}, \ldots, z_d) dz_d \cdots dz_{d-i+1}. \label{eqX5}
\end{eqnarray}
and $\alpha_i=(\overbrace{0, ..., 0}^{d-i}, \overbrace{1, ..., 1}^{i})$. From Eq. \eqref{eqX3} we readily obtain the bound
\begin{equation}
\left| 
	\sfh^k(x_1, \ldots, x_d) 
\right| \le \sum_{i=0}^{d-1} \left| 
	\tilde \sfh_i^k(x_{(d-i)}) 
\right| + \int_{-M_k}^{M_k} \cdots \int_{-M_k}^{M_k} \left|
	D^{\bf 1} \sfh^k(z_1, \ldots, z_d) 
\right| dz_d \cdots dz_1,
\label{eqX6}
\end{equation}
that holds for every $-M_k \le x_i \le M_k$, $i=1, 2, ..., d$.

By inspecting \eqref{eqX4} and \eqref{eqX5} we realise that if $\sfh^k \in \sfH^k(d)$, then $\tilde \sfh_i^k \in \sfH^k(d-1)$ for $i=0, 1, ..., d-1$. Therefore, from the induction hypothesis (and the fact that $k^{\frac{\beta}{(d-1)p}} \ge k^{\frac{\beta}{dp}}$) we deduce that for any $\varepsilon_3 \in (0,1)$ there exist a.s. finite random variables $\tilde V_i^{\varepsilon_3}$, $i=0,1, ..., d-1$, such that
\begin{equation}
\sup_{z \in \left[ -M_k, M_k \right]^{d-1}}
\left|
	\tilde \sfh_i^k(z)
\right| \le \frac{\tilde V_i^{\varepsilon_3}}{k^{1-\varepsilon_3}}.
\label{eqX7}
\end{equation}

As for the $d$-dimensional integral on the rhs of \eqref{eqX6}, we can find a suitable upper bound by the same procedure as in the base case, as shown below. Let $z = (z_1, \ldots, z_d)$ and denote, for $d>1$,
$$
A_d^k = \int_{M_k}^{M_k}\cdots\int_{-M_k}^{M_k} \left|
        D^{\bf 1} \sfh^k(z)
\right| dz.
$$
An application of Jensen's inequality yields, for $p \ge 1$, 
\begin{equation}
\left(
        \frac{1}{2^d M_k^d} A_d^k
\right)^p \le \frac{1}{2^d M_k^d} \int_{-M_k}^{M_k} \cdots \int_{-M_k}^{M_k} \left|
        D^{\bf 1} \sfh^k(z)
\right|^p dz, \nonumber
\end{equation}
which leads to
\begin{equation}
\left(
        A_d^k
\right)^p \le 2^{d(p-1)} M_k^{d(p-1)} \int_{-M_k}^{M_k} \cdots \int_{-M_k}^{M_k} \left|
        D^{\bf 1} \sfh^k(z) 
\right|^p dz. 
\label{eqBoundBigA_d}
\end{equation}
From Definition \ref{defH}, $\sfh^k(z) = (a_k^z, \pi_t^{N(k)}) - (a_k^z, \pi_t)$ for some $a_k^z$ with bounded derivatives, i.e., $\| D^{\bf 1} a_k \|_\infty = \sup_{z \in \Real^d} \| D^{\bf 1} a_k^z \|_\infty \le k^{d_x+d} C_a < \infty$. Therefore,  from \cite[Proposition 2.1]{Crisan14a} we readily obtain
\begin{equation}
E\left[
        \left|
                D^{\bf 1} \sfh^k(z)
        \right|^p
\right] = E\left[
        \left|
                (D^{\bf 1} a_k^z, \pi_t^{N(k)}) - (D^{\bf 1} a_k^z, \pi_t)
        \right|^p
\right] \le \frac{
        c_t^p k^{(d_x+d)p} C_a^p
}{
        \left( N(k) \right)^\frac{p}{2}
},
\label{eqSmallBoundAk_d}
\end{equation}
where the finite constants $c_t$ and $C_a$ are independent of $k$ and $z$. We can combine \eqref{eqSmallBoundAk_d} and \eqref{eqBoundBigA_d} to arrive at
\begin{equation}
E\left[
        (A_d^k)^p
\right] \le \frac{
        2^{dp} M_k^{d p} c_t^p k^{(d_x+d)p} C_a^p 
}{
        \left( N(k) \right)^\frac{p}{2}
} \le \frac{
        c_t^p C_a^p 
}{
        k^{(d_x-d)p + p - \beta}
} \le \frac{
	c_t^p C_a^p
}{
	k^{p-\beta}
}, \nonumber
\end{equation}
where the second inequality follows from the relationships $M_k = \frac{1}{2} k^{\frac{\beta}{d p}}$ and $N(k) \ge k^{2(2d_x+1)}$ and the third inequality holds because $1 < d \le d_x$. 

If we now apply \cite[Lemma 4.1]{Crisan14a} with $\theta^k = A_d^k$, $p \ge 2$, $\nu=\beta$ and $c=c_t^p C_a^p$, then we conclude that for any constant $\varepsilon_4 \in \left( \frac{1+\beta}{p}, 1 \right)$ there exists a non-negative and a.s. finite random variable $\tilde V^{A,\varepsilon_4}$, independent of $k$, such that
\begin{equation}
A_d^k \le \frac{
        \tilde V^{A,\varepsilon_4}
}{
        k^{1-\varepsilon_4}
}. \label{eqRate_part1_d}
\end{equation} 
Taking the inequalities \eqref{eqX6}, \eqref{eqX7} and \eqref{eqRate_part1_d} together, and choosing $\varepsilon=\varepsilon_3=\varepsilon_4$, we arrive at
\begin{equation}
\sup_{x \in \left[ -M_k, M_k \right]^d} | \sfh^k(x) | \le \frac{ \bar U^\varepsilon}{ k^{1-\varepsilon} },
\label{eqX8}
\end{equation}
where $\bar U^\varepsilon = \tilde V^{A,\varepsilon} + \sum_{i=0}^{d-1} \tilde V_i^\varepsilon$ is an a.s. finite random variable. The inequality \eqref{eqX8} holds for any constant $\varepsilon \in \left( \frac{1+\beta}{p}, 1 \right)$; hence, since we can select $p$ as large as we need, then we can effectively choose $\varepsilon \in (0,1)$. We have now proved that the bound \eqref{eqConvergenceSupP} holds for every $\sfh^k \in \sfH^k(d)$ and any $d \in \{1, 2, \ldots, d \}$.

To conclude the proof, we note that, under the assumptions \Ai, \Aii~and \Aiv,
\begin{equation}
\sfh_*^k(x) \dfn ( \phi_k^x, \pi_t^{N(k)} ) - ( \phi_k^x, \pi_t ) = p_t^k(x) - \tilde p_t^k(x) \in \sfH^k(d_x),
\nonumber
\end{equation}
since 
\begin{equation}
\sup_{(x,x') \in \Real^{d_x} \times \Real^{d_x}} |\phi_k^x(x')| = \sup_{x \in \Real^{d_x}} |k^{d_x} \phi(kx)| = k^{d_x} \| \phi \|_\infty < \infty, 
\nonumber
\end{equation}
from the definition of $\phi_k^x(x')$ in \cite[Section 3.2]{Crisan14a}, and
\begin{equation}
\sup_{(x,x') \in \Real^{d_x} \times \Real^{d_x}} |D^{\bf 1} \phi_k^x(x')| = k^{2d_x} \| D^{\bf 1} \phi \|_\infty < \infty
\nonumber
\end{equation}
(see \cite[Remark 3.5]{Crisan14a}). Therefore, from the inequality \eqref{eqConvergenceSupP} we obtain the bound
\begin{equation}
\sup_{x \in \mK_k} | p_t^k(x) - \tilde p_t^k(x) | \le \frac{
	\bar U^\varepsilon
}{
	k^{1-\varepsilon}
}, \label{eqAlmostThere}
\end{equation}
where $\varepsilon \in (0,1)$ can be chosen arbitrarily small and $\bar U^\varepsilon$ is an a.s. finite random variable. Moreover, from the relation (4.9) in \cite{Crisan14a}, 
\begin{equation}
| \tilde p_t^k(x) - p_t(x) | \le \frac{c_{0,t}\sqrt{c_2}}{k},
\label{eqFinal}
\end{equation} 
where the constants $c_{0,t}$ and $c_2$ are finite and independent of $k$ and $x$. Taking together \eqref{eqAlmostThere} and \eqref{eqFinal}, a simple triangle inequality yields the desired bound in \eqref{eqStatement}, with $U^\varepsilon = \bar U^\varepsilon + c_{0,t}\sqrt{c_2}$.

\hfill \QED

\begin{Nota} \label{rem0}
The reference point $x=(0,0,\ldots,0)$ in the proof is arbitrarily chosen. If 
\begin{itemize}
\item the support of the measure $\pi_t$ is a compact set ${\sf S}$ and 
\item the method producing the particle approximation $\pi_t^{N(k)}$ keeps it entirely within $\sf S$, 
\end{itemize}
then by choosing a reference point $\tilde x=(\tilde x_1, \ldots, \tilde x_{d_x})$ outside the support and such that the hyperplanes $(x_1, \ldots, x_{d_x-i-1}, \tilde x_{d_x-i}, x_{d_x-i+1}, \ldots, x_{d_x})$ do not intersect with $\sf S$, expression \eqref{eqX3} becomes the original integral decomposition in \cite{Crisan14a} and the proof of the Theorem in the paper becomes valid. These additional assumptions are not needed when $d_x=1$.  
\end{Nota}

\bibliographystyle{plain}
\bibliography{bibliografia}